\documentclass[aps,prd,superscriptaddress,twocolumn]{revtex4}

\usepackage{amsfonts}
\usepackage{amsmath}
\usepackage{graphicx}
\usepackage{color}
\usepackage{amssymb}
\usepackage{dsfont}

\usepackage{soul}

\usepackage[normalem]{ulem}

\usepackage{subfig}
\usepackage{float}
\usepackage{subfloat}

\begin{document}

\title{Vavilov-Cherenkov radiation for parallel motion in three-dimensional topological insulators}

\author{O. J. Franca}
\email{uk081688@uni-kassel.de}
\affiliation{Theoretische Physik III, Universit\"at Kassel, Heinrich-Plett-Stra\ss e 40, 34132 Kassel, Germany}

\author{Stefan Yoshi Buhmann}
\email{stefan.buhmann@uni-kassel.de}
\affiliation{Theoretische Physik III, Universit\"at Kassel, Heinrich-Plett-Stra\ss e 40, 34132 Kassel, Germany}

%
\begin{abstract}
Our study delves into the modifications observed in Vavilov-Cherenkov radiation when its generating charged particle moves parallel to an interface formed by two generic magnetoelectric media, focusing on topological insulators. We compute the electromagnetic field through the Green's function. Applying the far-field approximation and the steepest descent method, we derive analytical expressions for the electric field, revealing contributions from spherical and lateral waves with topological origins. Subsequently, we analyze the angular distribution of the radiation, particularly focusing on parallel motions in close proximity to the interface. Our findings indicate that the radiation along the Vavilov-Cherenkov cone is inhomogeneous and asymmetric. We analyze the radiated energy at both sides of the interface. Finally, we discuss the particle's retarding force, which is notably enhanced in the ultrarelativistic regime. We illustrate these results for the topological insulator TlBiSe$_2$ and the magnetoelectric TbPO$_4$.
\end{abstract}

\maketitle

\section{Introduction}
Over 16 years have elapsed since the theoretical prediction \cite{TIs Prediction 1,TIs Prediction 2} and experimental validation of the existence \cite{Hsieh} of three-dimensional (3D) topological insulators (TIs), catalyzing the research into these materials leading to the detection of new varieties of TIs \cite{TIs variety}. Three-dimensional TIs constitute an interesting new state of matter exhibiting conducting surface states arising out of an unusual band structure that belongs to a different topological class different from that of ordinary semiconductors \cite{Anirban}, being protected by time-reversal symmetry and insulating bulk \cite{Hasan,Qi Review}. Originally, two types of TIs were established: weak three-dimensional TIs with an even number of surface states on all but a single surface, whereas strong ones exhibit an odd number of them on each surface \cite{TIs Prediction 1, TIs Prediction 2, Morgenstern review}. The introduction of time-reversal symmetry breaking at the interface between a conventional insulator and a 3D strong TI, accomplished through methods such as magnetic coating or doping the TI with transition metal elements, results in the opening of a gap in the surface states. This leads to a plethora of phenomena, including the quantum anomalous Hall effect \cite{QAH,Mogi}, the quantized magneto-optical effect \cite{Wu, Dziom}, the topological magnetoelectric effect \cite{Hasan, Qi PRB, Okada}, and the still to-be verified image magnetic monopole effect \cite{Qi Science}. Recently, TI-graphene heterostructures have been considered in order to enhance topological effects on the electric response through the interaction between a charged impurity and the material, leading to electronic transport influenced by the image magnetic monopole from the topological insulator's electromagnetic response \cite{Heterostructures1,Heterostructures2,Heterostructures3,Heterostructures4,Heterostructures5}. 

Despite the theoretical prediction of radiative effects in strong 3D TIs \cite{OJF-LFU-ORT, OJF-SYB, OJF-LFU}, their applications remain scarcely studied. Among the various types of radiation, the Vavilov-Cherenkov (VC) one has become an important tool in high-energy particle physics, high-power microwave sources and nuclear and cosmic-ray physics \cite{Jelley,Jelley1} from the theoretical and phenomenological sides. VC radiation arises when charged particles traveling through a dielectric medium with velocity $v$ higher than $c/\sqrt{\varepsilon \mu}$ \cite{cherenkov,Vavilov}. Here $c$ is the speed of light in vacuum, $\varepsilon$ denotes the permittivity of the medium, $\mu$ labels its permeability, and $n=\sqrt{\varepsilon\mu}$ stands for the refraction index of the material. Frank and Tamm were the first to explain this radiation theoretically and revealed its unique polarization and directional properties by means of classical electrodynamics \cite{FT}. In transparent dielectric media, VC radiation behaves as an extended mode and lies in a forward cone defined by the  angle $\theta^{C}= \cos^{-1}[c/vn] < \pi/2$ with respect to the direction of the generating charge. Multiple applications of VC radiation in nuclear and high-energy physics have been found for designing detectors \cite{Ypsilantis,PADG} and particle discoveries \cite{Chamberlain,Aubert}. VC radiation has also been actively studied beyond high-energy physics in other branches of physics ranging from metamaterials \cite{Tao}, magnetoelectric media \cite{ODELL,OJF-LFU-ORT}, and chiral media \cite{Kolomenskii,Galyamin 1}. Even medical applications like the Cherenkov luminescence imaging \cite{Spinelli} and the interplay between surface waves and VC radiation have recently been explored \cite{Hu-Lin-Wong}.

As predicted by Veselago it is possible to reverse the direction of VC radiation in left-handed materials \cite{Veselago} or, as the paper \cite{OJF-LFU-ORT} theorized, for strong three-dimensional TIs, which are right-handed materials. In the latter work, it was found that reversed VC radiation occurs in this class of materials when its generating particle travels with a trajectory perpendicular to the surface of the interface between the material and vacuum. 


Based on these ongoing developments, this paper will be devoted to 3D TIs and VC radiation but with a slightly different configuration. The aim of this work is to examine how VC radiation interacts with the interface between two arbitrary strong 3D TIs with different permeabilities and permeabilities when the charged particle moves parallel to this interface. Henceforth, we will refer to this configuration as parallel Vavilov-Cherenkov radiation to distinguish it from the perpendicular motion studied in Ref.~\cite{OJF-LFU-ORT}. We will compare with the results of Bolotovskii \cite{Bolotovskii} for an interface constituted by two ordinary media.

This paper is organized as follows. In Sec. \ref{EFT}, we describe the modified Maxwell equations which describe the electromagnetic response of three-dimensional TIs from an effective field theory point of view. Also in this section, we review the Green's function method to obtain the time-dependent electromagnetic field. Section \ref{CHARGE} is devoted to analyzing the electromagnetic field associated with parallel VC radiation. In Sec. \ref{ANGULAR}, we present angular distributions of the parallel VC radiation for the upper and lower hemispheres corresponding to the reflected and transmitted radiation, respectively. Section \ref{ENERGY} is devoted to obtaining an analytical expression for the radiated energy at both sides of the interface. In Section \ref{Friction Force} we study the retarding force experienced by the particle. Finally, a summary and concluding remarks of our results can be found in Sec. \ref{CONCLUSIONS}.

\section{Modified Maxwell Equations} \label{EFT}

The Maxwell equations and constitutive relations of the involved material constitute our natural starting point. For a three-dimensional TI, these are \cite{Qi PRB,Chang-Yang,Obukhov-Hehl}
\begin{eqnarray}
\nabla\cdot\mathbf{B}\left(\mathbf{r};\omega\right) &=& 0 \;, \label{Gauss B}\\
\nabla\times\mathbf{E}\left(\mathbf{r};\omega\right) &=& \mathrm{i}\omega \mathbf{B}\left(\mathbf{r};\omega\right) \;, \label{Faraday}\\
\nabla\cdot\mathbf{D}\left(\mathbf{r};\omega\right) &=& \varrho \left(\mathbf{r};\omega\right) \;, \label{Gauss D}\\
\nabla\times\mathbf{H}\left(\mathbf{r};\omega\right) + \mathrm{i} \omega \mathbf{D}\left(\mathbf{r};\omega\right) &=& \mathbf{j} \left(\mathbf{r};\omega\right) \;, \label{Ampere-Maxwell}
\end{eqnarray}
where $\varrho(\mathbf{r};\omega)$  and  $\mathbf{j} (\mathbf{r};\omega)$ denote the source term for electromagnetic waves generated by external charges and currents, respectively, $\mathbf{E}\left(\mathbf{r};\omega\right)$ stands for the electric field and $\mathbf{B}\left(\mathbf{r};\omega\right)$ is the induction field. For TIs, both fields are related with the displacement field $\mathbf{D}\left(\mathbf{r};\omega\right)$ and the magnetic field $\mathbf{H}\left(\mathbf{r};\omega\right)$ via
\begin{eqnarray}
\mathbf{D}\left(\mathbf{r};\omega\right) &=& \varepsilon_0\varepsilon\left(\mathbf{r};\omega\right) \mathbf{E}\left(\mathbf{r};\omega\right) + \frac{ \alpha \Theta\left(\mathbf{r};\omega\right) }{ \pi\mu_0 c }  \mathbf{B}\left(\mathbf{r};\omega\right) \nonumber\\
&& + \mathbf{P}_N\left(\mathbf{r};\omega\right) \;, \label{Constitutive 1} 
\end{eqnarray}
\begin{eqnarray}
\mathbf{H}\left(\mathbf{r};\omega\right) &=& \frac{ \mathbf{B}\left(\mathbf{r};\omega\right) }{ \mu_0\mu\left(\mathbf{r};\omega\right) } -  \frac{ \alpha \Theta\left(\mathbf{r};\omega\right) }{ \pi\mu_0 c }  \mathbf{E}\left(\mathbf{r};\omega\right) \nonumber\\
&& - \mathbf{M}_N\left(\mathbf{r};\omega\right) \;, \label{Constitutive 2}
\end{eqnarray}
where $\alpha$ is the fine structure constant and $\varepsilon\left(\mathbf{r};\omega\right)$, $\mu\left(\mathbf{r};\omega\right)$, and $\Theta\left(\mathbf{r};\omega\right)$ are the dielectric permittivity, magnetic permeability, and axion coupling, respectively. Here the $\mathbf{P}_N\left(\mathbf{r};\omega\right)$ and $\mathbf{M}_N\left(\mathbf{r};\omega\right)$ terms are the noise polarization and magnetization, respectively. These are Langevin noise terms that represent absorption within the material \cite{Scheel-Buhmann}.  

For weak three-dimensional TIs the axion coupling $\Theta\left(\mathbf{r};\omega\right)$ takes even multiples of $\pi$, whereas for strong three-dimensional TIs, it takes odd multiples of $\pi$, with the magnitude and sign of the multiple given by the strength and direction of the time-symmetry-breaking perturbation. However, when an effective field theory for a 3D TI is considered, it turns out that a suitable effective action transforming correctly under time-reversal symmetry can only be constructed for strong TIs while weak TIs are not suitable from the quantum mechanical point of view \cite{Qi Review,Qi PRB,Karch}. Possibly, such difficulties arise from the fact that lattice translation symmetry together with time-reversal symmetry protect the conducting states at the boundaries of a weak 3D TI, which requires to take into account also disorder and anisotropies Ref.~\cite{TIs Prediction 1,WTIs-Nature,WTIs}. Therefore, we will restrict to study only the strong TIs. Note that the modified Maxwell Eqs.~(\ref{Gauss B})-(\ref{Ampere-Maxwell}) can be derived from the Lagrangian density $\mathcal{L}=\mathcal{L}_0 + \mathcal{L}_{\Theta}$, where $\mathcal{L}_0$ is the usual Maxwell Lagrangian density and $\mathcal{L}_{\Theta}$ is given by
\begin{equation}
\mathcal{L}_{\Theta} =  \frac{ \alpha }{ 4\pi^2 } \frac{ \Theta\left(\mathbf{r};\omega\right) }{ \mu_0 c } \mathbf{E}\left(\mathbf{r};\omega\right) \cdot \mathbf{B}\left(\mathbf{r};\omega\right) \;.
\end{equation}
From a condensed matter perspective, $\Theta$ is referred as the (scalar) magnetoelectric polarizability \cite{Essin}, but in field theory, this quantity is called the axion field, which is the heart of axion electrodynamics \cite{Wilczek}. According to the Maxwell equations (\ref{Gauss B})--(\ref{Ampere-Maxwell}) and the modified constitutive relations (\ref{Constitutive 1}) and (\ref{Constitutive 2}), this additional $\Theta$ term induces field-dependent effective charge and current densities which model the topological magnetoelectric effect present in three-dimensional TIs \cite{Hasan,Qi PRB}. Let us briefly make contact with the broader perspective of bi-isotropic materials, which couple the electric and magnetic fields to each other by means of magnetoelectric parameters $\chi$ and $\kappa$ characterizing nonreciprocity and chirality, respectively. In this picture, we may rewrite the constitutive relations (\ref{Constitutive 1}) and (\ref{Constitutive 2}) in the following way:
\begin{eqnarray}
\mathbf{D} &=& \varepsilon_0\varepsilon \mathbf{E} + \frac{\alpha}{\pi}\mu \Theta \sqrt{\varepsilon_0\mu_0}\,\mathbf{H} \nonumber\\
&& + \frac{\alpha}{\pi}\mu \Theta \sqrt{\varepsilon_0\mu_0}\,\mathbf{M}_N + \mathbf{P}_N \;, \label{Constitutive 3}\\
%
%
%
\mathbf{B} &=& \mu_0\mu \mathbf{H} + \frac{\alpha}{\pi} \mu \Theta \sqrt{\varepsilon_0\mu_0}\,\mathbf{E} + \mu_0\mu \mathbf{M}_N \;,
\end{eqnarray}
where all spatial and frequency dependencies were omitted for brevity and higher-order terms in $\Theta$ were dropped in Eq.~(\ref{Constitutive 3}). They can be regarded as those for a bi-isotropic material $\mathbf{D}=\varepsilon_0\varepsilon\mathbf{E}+(\chi-\mathrm{i} \kappa)\sqrt{\varepsilon_0\mu_0}\,\mathbf{H}$ and $\mathbf{B}=\mu_0\mu\mathbf{H}+(\chi+\mathrm{i} \kappa)\sqrt{\varepsilon_0\mu_0}\,\mathbf{E}$, where noise terms are typically not considered, yielding the identifications $\chi=\alpha\mu\Theta/\pi$ and $\kappa=0$. Therefore, chirality is not present, which is the reason that TIs are classified as Tellegen media \cite{Jiang}. Furthermore, the axion coupling adds a nonreciprocal feature to the TIs which breaks time-reversal symmetry as mentioned in the Introduction. In the framework of Barron \cite{Barron 1, Barron 2}, the constitutive relationships for TIs are an instance of false chirality: the handedness of the medium may be inverted by either space inversion or time-reversal given that $\chi$ is a time-reversal-odd pseudoscalar \cite{Qi PRB}. 

Combining Eqs.~(\ref{Gauss B})-(\ref{Ampere-Maxwell}) together with the constitutive relations (\ref{Constitutive 1}) and (\ref{Constitutive 2}), one finds that the frequency components of the electric field obey the inhomogeneous Helmholtz equation 
\begin{equation}
\begin{aligned}
& \nabla \times \frac{ 1 }{ \mu(\mathbf{r};\omega) } \nabla \times \mathbf{E}(\mathbf{r};\omega) - \frac{ \omega^2 }{ c^2 }\varepsilon(\mathbf{r};\omega) \mathbf{E}(\mathbf{r};\omega) \\
&- \mathrm{i} \frac{ \omega \alpha }{ \pi c } \left[ \nabla \Theta(\mathbf{r};\omega) \right] \times \mathbf{E}(\mathbf{r};\omega) = \mathrm{i} \omega \mu_0 \left[\mathbf{j}(\mathbf{r};\omega)+ \mathbf{j}_N (\mathbf{r};\omega) \right]\;,  \label{inhomo Helmholtz}
\end{aligned}
\end{equation}
where $\mathbf{j}_N (\mathbf{r};\omega)=-\mathrm{i}\omega \mathbf{P}_N (\mathbf{r};\omega) + \nabla\times\mathbf{M}_N(\mathbf{r};\omega)$ is the source term for electromagnetic waves generated by noise fluctuations within the material. We observe that the last term on the left-hand side vanishes if the axion coupling is homogeneous, $\Theta(\mathbf{r};\omega)=\Theta(\omega)$, recovering the propagation of the electric field in a conventional magnetoelectric. As an immediate consequence, the propagation of electromagnetic waves will retain their usual properties within a homogeneous three-dimensional TI. Consequently, the effects due to the axion coupling will only be felt whenever $\Theta$ presents a spatial dependency, as is the case at the interface of two materials having different constant values of $\Theta$, for example. The so-called $\Theta$-electrodynamics \cite{OJF-LFU-ORT, AMR-LFU-MC 1, AMR-LFU-MC 2} considers $\Theta$ as a nondynamical field having the necessary topological features to model the electromagnetic behavior of strong three-dimensional TIs \cite{Qi PRB} as well as naturally existing magnetoelectric media \cite{Hehl,Landau-Lifshitz} such as Cr$_2$O$_3$ with $\Theta\simeq\pi/36$ \cite{Wu,Hehl-Obukov-Rivera}.

\begin{figure}[tbp]
\centering 
\includegraphics[width=8CM]{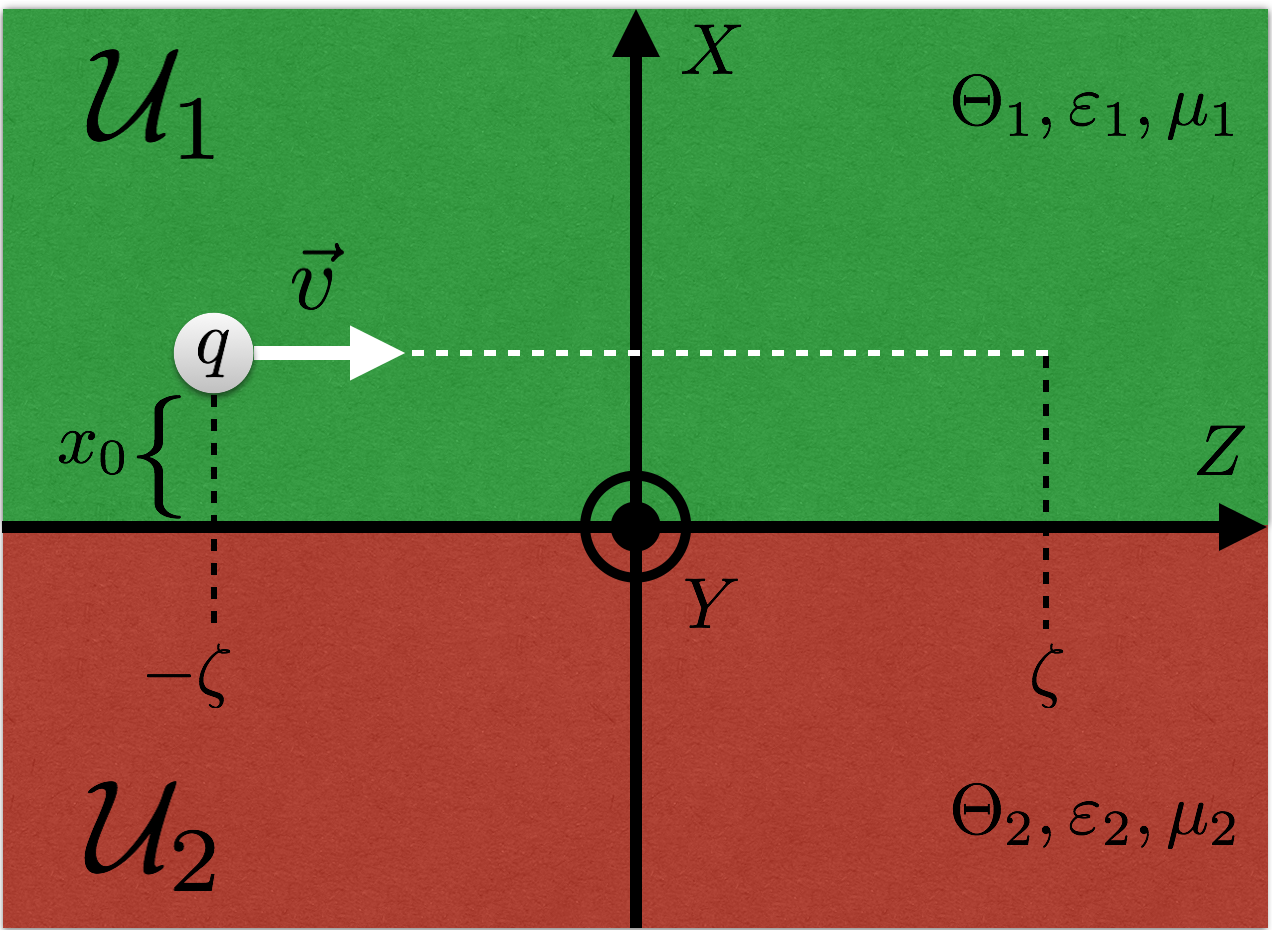} 
\caption{ Two semi-infinite magnetoelectric media with different magnetoelectric polarizabilities $\Theta_1$ and $\Theta_2$, having different permittivities $\varepsilon_1$ and $\varepsilon_2$, and different permeabilities $\mu_1$ and $\mu_2$, and separated by the planar interface. }
\label{REGIONS}
\end{figure}

With the help of the Green's function $\mathbb{G}(\mathbf{r},\mathbf{r}';\omega)$ associated with the Eq.~(\ref{inhomo Helmholtz}), which dictates the dynamics for the electric field, we are able to compute $\mathbf{E}(\mathbf{r};\omega)$ at any point from an arbitrary distribution of current sources. The formal solution to the differential equation (\ref{inhomo Helmholtz}) reads as
\begin{equation} \label{E GF j}
\mathbf{E}(\mathbf{r};\omega) = \mathrm{i}\omega\mu_0 \int d^{3}\mathbf{r}\, \mathbb{G}(\mathbf{r},\mathbf{r}';\omega) \cdot\left[\mathbf{j}(\mathbf{r};\omega)+ \mathbf{j}_N (\mathbf{r};\omega) \right]\;. 
\end{equation}
This Green's function carries all the information of the desired configuration, which for our case is made up of two semi-infinite magnetoelectric media separated by a planar interface located at $x=0$, filling the regions $\mathcal{U}_1$ and $\mathcal{U}_2$ of the space, as shown in Fig.~\ref{REGIONS}. At first sight, the choice of such interface at $x=0$ as opposed to $z=0$ seems to be uncommon but it provides analytical advantages as Refs.~\cite{Khan,Bolotovskii} show. On this work, the axion coupling $\Theta$ will be taken as a piecewise constant taking the values $\Theta=\Theta_1$ in the region $\mathcal{U}_1$ and $\Theta=\Theta_2$ in the region $\mathcal{U}_2$. This means that 
\begin{equation}\label{THETA}
\Theta(x) = \Theta_1 H(x) + \Theta_2 H(-x) \;,
\end{equation}
where $H(x)$ is the Heaviside function.

The required Green's function $\mathbb{G}(\mathbf{r},\mathbf{r}';\omega)$ for this configuration can be obtained from that deduced in Ref.~\cite{Crosse-Fuchs-Buhmann} by swapping $x\leftrightarrow z$ in all the indices and dependencies. An important feature of $\mathbb{G}(\mathbf{r},\mathbf{r}';\omega)$ that needs to be mentioned is its reflective and transmissive behaviors. From Fig.~\ref{REGIONS} we recognize two contributions for a field point at $x>0$ and if the source is in region $\mathcal{U}_1$ ($x'>0$): one from direct propagation from the source to the field point, which is described by the free-space Green's function $\mathbb{G}^{(0)}(\mathbf{r},\mathbf{r}';\omega)$, and an additional one from reflections from the surface, which is dictated by the reflective part of the Green's function $\mathbb{G}^{(1)}(\mathbf{r},\mathbf{r}';\omega)$. For $x<0$, the transmission at the surface is the only contribution that appears and is described by the transmissive part of the Green's function $\mathbb{G}^{(1)}(\mathbf{r},\mathbf{r}';\omega)$. This enables us to split the Green's function into two main parts,
\begin{equation} \label{GF split}
\mathbb{G}(\mathbf{r},\mathbf{r}';\omega) = \left\{
\begin{array}{cc}
\mathbb{G}^{(0)}(\mathbf{r},\mathbf{r}';\omega) + \mathbb{G}^{(1)}(\mathbf{r},\mathbf{r}';\omega) \;, & x>0 \;,     \\
\mathbb{G}^{(1)}(\mathbf{r},\mathbf{r}';\omega) \;, & x<0 \;,    
\end{array}
\right.
\end{equation}
each of which can be obtained separately as done in Ref.~\cite{Crosse-Fuchs-Buhmann}. A second feature of this Green's function that deserves attention is how it incorporates the nature of strong three-dimensional TIs or magnetoelectric media. As exposed in the Introduction, these materials break time-reversal-symmetry allowing their classification as nonreciprocal media \cite{Fuchs-Crosse-Buhmann}. Moreover, a nonreciprocal medium violates time-reversal symmetry and, hence, the Lorentz reciprocity principle for the Green's tensor \cite{Buhmann 1},
\begin{equation}
\mathbb{G}(\mathbf{r},\mathbf{r}';\omega) \neq \mathbb{G}^{\mathrm{T}}(\mathbf{r}',\mathbf{r};\omega) \;,
\end{equation}
where $\mathrm{T}$ denotes the transpose.

As an immediate consequence of the latter inequality, the symmetry relations \cite{Buhmann 1}
\begin{equation}\label{Usual Fresnel}
r_{\sigma}^{12} = -r_{\sigma}^{21} \;\; , \;\; \frac{ \mu_2(\omega) }{ k_{x,2} } t_{\sigma}^{21} = \frac{ \mu_1(\omega) }{ k_{x,1} } t_{\sigma}^{12} \;, (\sigma=\mathrm{TM}, \mathrm{TE})\;,
\end{equation}
for the standard Fresnel coefficients are no longer valid for our configuration. Instead of these, the four arising Fresnel coefficients satisfy the relations,
\begin{eqnarray}
R_{\mathrm{TE,TE}}^{12} + R_{\mathrm{TE,TE}}^{21} &=& \frac{\Delta_\Theta}{\mu_2 n_1} R_{\mathrm{TM,TE}}^{12} = \frac{\Delta_\Theta}{\mu_1 n_2} R_{\mathrm{TM,TE}}^{21} \,, \nonumber\\
R_{\mathrm{TM,TM}}^{12} + R_{\mathrm{TM,TM}}^{21} &=&  \frac{-\Delta_\Theta}{\mu_2 n_1} R_{\mathrm{TE,TM}}^{12}
 = \frac{-\Delta_\Theta}{\mu_1 n_2} R_{\mathrm{TE,TM}}^{21} \,, \nonumber\\
R_{\mathrm{TM,TE}}^{12} &=& R_{\mathrm{TE,TM}}^{21} \;, \nonumber\\
\frac{ \mu_1(\omega) }{ k_{x,1} } T_{\mathrm{TE,TE}}^{12} &=& \frac{ \mu_2(\omega) }{ k_{x,2} } T_{\mathrm{TE,TE}}^{21} \;, \label{TRS and Fresnel}\\
\frac{ \mu_1(\omega) }{ k_{x,1} } T_{\mathrm{TM,TM}}^{12} &=& \frac{ \mu_2(\omega) }{ k_{x,2} } T_{\mathrm{TM,TM}}^{21} \;, \nonumber\\
\frac{ k_{x,2} }{ n_2 } T_{\mathrm{TM,TE}}^{12} &=& - \frac{ k_{x,1} }{ n_1 } T_{\mathrm{TE,TM}}^{21} \;. \nonumber
\end{eqnarray}
For $T_{\mathrm{TE,TE}}^{12}$ and $T_{\mathrm{TM,TM}}^{12}$, we observe that they satisfy the same relation as in the  standard case when $\Delta_\Theta=0$. In Eqs.~(\ref{Usual Fresnel}), $r_{\sigma}^{12}$ denotes the Fresnel reflection coefficient and $t_{\sigma}^{12}$ is the transmission one with polarization $\sigma$, whose superscripts label the medium 1 as an upper layer and the medium 2 as the lower one or vice versa. Also, $k_{x,j}=\sqrt{k^2_j-\mathbf{k}_\parallel^2}$ stands for  the $x$ component of the wave vector $\mathbf{k}_{j\pm}=(\pm k_{x,j}, k_{y}, k_{z})$ with wave number $k_j=\sqrt{ \varepsilon_j(\omega) \mu_j(\omega) }\,\omega/c$ for $j=1,2$. From now on, every subscript on a physical quantity will denote that it is related to the medium 1 or medium 2.

The boundary conditions for the electric and magnetic fields upon assuming finite-time derivatives of the fields in the vicinity of the interface at $x=0$ as well as the noise terms, can be obtained from the modified Maxwell equations (\ref{Gauss B})--(\ref{Ampere-Maxwell}) and the constitutive relations (\ref{Constitutive 1}) and (\ref{Constitutive 2}) lead to the next  boundary conditions:
\begin{eqnarray}
&& \left[ \varepsilon_0\varepsilon \mathbf{E}_\perp \right]_{x=0^{-}}^{x=0^{+}} = \frac{ \Delta_\Theta \mathbf{B}_\perp|_{x=0} }{ c\mu_0\mu_1\mu_2 }\;,\; \left[\mathbf{B}_\perp\right]_{x=0^{-}}^{x=0^{+}} = \mathbf{ 0 } \;, \label{BC1}\\
&& \left[ \frac{ \mathbf{B}_{\parallel} }{ \mu_0\mu } \right]_{x=0^{-}}^{x=0^{+}} = -\frac{ \Delta_\Theta \mathbf{E}_{\parallel}|_{x=0} }{ c\mu_0\mu_1\mu_2 } \;,\; \left[\mathbf{E}_{\parallel}\right]_{x=0^{-}}^{x=0^{+}} = \mathbf{ 0 } \;, \label{BC2}
\end{eqnarray}
for vanishing external sources at the interface, where  
\begin{equation} \label{DELTA}
\Delta_\Theta=\alpha\mu_1\mu_2(\Theta_2-\Theta_1)/\pi \;.
\end{equation}
The notation is $\left[ \mathbf{V}\right]_{x=0^{-}}^{x=0^{+}}=\mathbf{V}(x=0^{+})-\mathbf{V}(x=0^{-})$, $\mathbf{V}\big|_{x=0}=\mathbf{V}(x=0)$, where $x=0^{\pm }$ indicates the limits $x=0\pm \epsilon $, with $\epsilon$ a positive real number such that $\; \epsilon\rightarrow 0$, and $\mathbf{V}$ an arbitrary vector field, respectively.

When an interface with a regular insulator ($\Theta_1=0$) in region $\mathcal{U}_1$ of Fig.~\ref{REGIONS} and a strong three-dimensional TI located in region $\mathcal{U}_2$ ($\Theta_2=\pi$), we have
\begin{equation}\label{DELTA_1}
\Delta_\Theta=\alpha(2m+1), 
\end{equation}
where $m$ is an integer depending on the details of the time-reversal-symmetry breaking at the interface and $\Delta_\Theta$ will be called as topological parameter.


\section{Electromagnetic fields} \label{CHARGE}
Without loss of generality and due to the azimuthal symmetry of our problem, let us consider a particle with charge $q$ and constant velocity $v\,\mathbf{e}_{z}$, parallel to the interface defined by the $yz$ plane ($x=0$) where $\mathbf{e}_{z}=(0,0,1)$ as shown in Fig.~\ref{REGIONS}. The external charge and current densities in frequency space are
\begin{eqnarray}
\varrho(\mathbf{r}';\omega) & = & \frac{ q }{ v } \delta(x'-x_0) \delta(y') e^{i\omega z' /v} \;, \\
\mathbf{j}(\mathbf{r}';\omega) & = &q \delta(x'-x_0) \delta(y') e^{i\omega z' /v} \mathbf{e}_{z} \;, \label{rho y j}
\end{eqnarray}
where we henceforth assume $x_0>0$. Instead of an infinite path for the charge, i.e., its movement will occur in the interval $z'\in(-\zeta,\zeta)$, with $\zeta\gg v/\omega$ and $\zeta$ is the half of the total length. This means that the particle travels from left to right parallel to the interface and never crosses it, as depicted in Fig.~\ref{REGIONS}, ruling out the presence of transition radiation studied in Ref.~\cite{OJF-SYB}. To ensure the emission of VC radiation, at least the particle velocity needs to be greater than the speed of light in one of the two media.

As the current density (\ref{rho y j}) has a single component, the components of the Green's function that matter for this configuration are $\mathbb{G}^{iz}(\mathbf{r},\mathbf{r}';\omega)$ with $i=x,y,z$, whose explicit forms are given in Appendix \ref{A}. Using these components, we perform the indicated convolution of Eq.~(\ref{E GF j}) by substituting the current density (\ref{rho y j}). The involved integrals can be simplified by converting to polar coordinates, after which the angular integrals are computed analytically through integral representations of the Bessel functions. The resulting Hankel transforms of the electric field components in the reflective region, upon carrying out the integral over the  parallel wave vector magnitude  $k_\parallel=\sqrt{ k_{y}^2 + k_{z}^2 }$ and the coordinate $z'$, are
\begin{eqnarray}
E^{\,x}_1(\mathbf{r};\omega) &=& \mathrm{i}q\omega\mu_0\mu_1(\omega) \nonumber\\
&& \times \int_{-\zeta}^{\zeta}dz' \frac{e^{\mathrm{i}k_1R}}{4\pi R} \left[ - \frac{ (x-x_0)(z-z') }{ R^2 }\right] e^{\mathrm{i}\frac{\omega z'}{v}} \nonumber\\
&& +\frac{\mathrm{i}q\omega\mu_0\mu_1(\omega)}{4\pi k_1^2 } \int_{-\zeta}^{\zeta}dz' e^{\mathrm{i}\frac{\omega z'}{v}} \frac{ (z-z') }{ R_\parallel } \frac{\partial \mathcal{I}_1 }{\partial R_\parallel} \nonumber\\
&& +\frac{\mathrm{i}q\omega\mu_0\mu_1(\omega)}{4\pi k_1 } \int_{-\zeta}^{\zeta}dz' e^{\mathrm{i}\frac{\omega z'}{v}} \frac{ y }{ R_\parallel } \frac{\partial \mathcal{I}_2 }{\partial R_\parallel} \;, \label{Ex1 int}
\end{eqnarray}
\begin{eqnarray}
E^{\,y}_1(\mathbf{r};\omega) &=& - \mathrm{i}q\omega\mu_0\mu_1(\omega) \int_{-\zeta}^{\zeta}dz' \frac{e^{\mathrm{i}k_1R}}{4\pi R}  \frac{ y(z-z') }{ R^2 }  e^{\mathrm{i}\frac{\omega z'}{v}} \nonumber\\
&& +\frac{q\omega\mu_0\mu_1(\omega)}{4\pi k_1 } \int_{-\zeta}^{\zeta}dz' e^{\mathrm{i}\frac{\omega z'}{v}} \mathcal{I}_3 \;, \label{Ey1 int}
\end{eqnarray}
\begin{eqnarray}
E^{\,z}_1(\mathbf{r};\omega) &=& - \mathrm{i}q\omega\mu_0\mu_1(\omega) \nonumber\\
&& \times \int_{-\zeta}^{\zeta}dz' \frac{e^{\mathrm{i}k_1R}}{4\pi R} \left[1 - \frac{ (z-z')^2 }{ R^2 }\right] e^{\mathrm{i}\frac{\omega x'}{v}} \nonumber\\
&& +\frac{q\omega\mu_0\mu_1(\omega)}{4\pi } \int_{-\zeta}^{\zeta}dz'  \frac{ e^{\mathrm{i}\frac{\omega z'}{v}} }{ R_\parallel } \frac{\partial \mathcal{I}_4 }{\partial R_\parallel} \nonumber\\
&& +\frac{q\omega\mu_0\mu_1(\omega)}{4\pi k_1^2 } \int_{-\zeta}^{\zeta}dz'  \frac{ e^{\mathrm{i}\frac{\omega z'}{v}} }{ R_\parallel } \frac{\partial \mathcal{I}_5 }{\partial R_\parallel} \nonumber\\   
&& +\frac{q\omega\mu_0\mu_1(\omega)}{4\pi k_1^2 } \int_{-\zeta}^{\zeta}dz' e^{\mathrm{i}\frac{\omega z'}{v}} \mathcal{I}_6 \;, \label{Ez1 int} 
\end{eqnarray}
where $R_\parallel=\sqrt{ y^2 + (z-z')^2 } $, $R^2= (x-x_0)^2 + R_\parallel^2 $ and the following integrals were defined
\begin{equation}
\begin{aligned}
\mathcal{I}_1  &=  \int_0^\infty dk_\parallel k_\parallel R_{\mathrm{TM,TM}}^{12}(k_\parallel) J_0(k_\parallel R_\parallel) e^{\mathrm{i}k_{x,1}(x+x_0)} \;, \label{integral I1 napp} 
\end{aligned}
\end{equation}
\begin{equation}
\mathcal{I}_2  =  \int_0^\infty \frac{ dk_\parallel k_\parallel }{ k_{x,1} }  R_{\mathrm{TM,TE}}^{12}(k_\parallel) J_0(k_\parallel R_\parallel) e^{\mathrm{i}k_{x,1}(x+x_0)} \;, \label{integral I2 napp} 
\end{equation}
\begin{equation}
\mathcal{I}_3  =  \int_0^\infty dk_\parallel k_\parallel R_{\mathrm{TE,TM}}^{12}(k_\parallel) J_0(k_\parallel R_\parallel) e^{\mathrm{i}k_{x,1}(x+x_0)} \;, \label{integral I3 napp} 
\end{equation}
\begin{equation}
\mathcal{I}_4  = \int_0^\infty \frac{ dk_\parallel }{ k_\parallel k_{x,1} }  R_{\mathrm{TE,TE}}^{12}(k_\parallel) J_0(k_\parallel R_\parallel) e^{\mathrm{i}k_{x,1}(x+x_0)}  \;, \label{integral I4 napp} 
\end{equation}
\begin{equation}
\mathcal{I}_5  = \int_0^\infty \frac{ dk_\parallel k_{x,1} }{ k_\parallel }  R_{\mathrm{TM,TM}}^{12}(k_\parallel) J_0(k_\parallel R_\parallel) e^{\mathrm{i}k_{x,1}(x+x_0)}  \;, \label{integral I5 napp} 
\end{equation}
\begin{equation}
\mathcal{I}_6  = \int_0^\infty dk_\parallel k_\parallel k_{x,1} R_{\mathrm{TM,TM}}^{12}(k_\parallel) J_0(k_\parallel R_\parallel) e^{\mathrm{i}k_{x,1}(x+x_0)}  \;. \label{integral I6 napp} 
\end{equation}
Analogously for the transmissive region, we have
\begin{eqnarray}
E^{\,x}_2(\mathbf{r};\omega) &=& \frac{\mathrm{i}q\omega\mu_0\mu_1(\omega)}{4\pi k_1 k_2 } \int_{-\zeta}^{\zeta}dz' e^{\mathrm{i}\frac{\omega z'}{v}} \frac{ (z-z') }{ R_\parallel } \frac{\partial \mathcal{I}_{7} }{\partial R_\parallel} \nonumber\\
&& +\frac{\mathrm{i}q\omega\mu_0\mu_1(\omega)}{4\pi k_2 } \int_{-\zeta}^{\zeta}dz' e^{\mathrm{i}\frac{\omega z'}{v}} \frac{ y }{ R_\parallel } \frac{\partial \mathcal{I}_{8} }{\partial R_\parallel} \;, \label{Ex2 int}
\end{eqnarray}
\begin{eqnarray}
E^{\,y}_2(\mathbf{r};\omega) &=& \frac{q\omega\mu_0\mu_1(\omega)}{4\pi k_1 } \int_{-\zeta}^{\zeta}dz'  \frac{ e^{\mathrm{i}\frac{\omega z'}{v}} }{ R_\parallel } \frac{\partial \mathcal{I}_{9} }{\partial R_\parallel} \nonumber\\
&& + \frac{q\omega\mu_0\mu_1(\omega)}{4\pi k_1 } \int_{-\zeta}^{\zeta}dz' e^{\mathrm{i}\frac{\omega z'}{v}} \mathcal{I}_{10} \nonumber\\
&& + \frac{q\omega\mu_0\mu_1(\omega)}{4\pi k_2 } \int_{-\zeta}^{\zeta}dz'  \frac{ e^{\mathrm{i}\frac{\omega z'}{v}} }{ R_\parallel } \frac{\partial \mathcal{I}_{11} }{\partial R_\parallel} \;, \label{Ey2 int}
\end{eqnarray}
\begin{eqnarray}
E^{\,z}_2(\mathbf{r};\omega) &=& \frac{q\omega\mu_0\mu_1(\omega)}{4\pi } \int_{-\zeta}^{\zeta}dz'  \frac{ e^{\mathrm{i}\frac{\omega z'}{v}} }{ R_\parallel } \frac{\partial \mathcal{I}_{12} }{\partial R_\parallel} \nonumber\\ 
&& - \frac{q\omega\mu_0\mu_1(\omega)}{4\pi k_1 k_2 } \int_{-\zeta}^{\zeta}dz'  \frac{ e^{\mathrm{i}\frac{\omega z'}{v}} }{ R_\parallel } \frac{\partial \mathcal{I}_{13} }{\partial R_\parallel} \nonumber\\   
&& -\frac{q\omega\mu_0\mu_1(\omega)}{4\pi k_1 k_2 } \int_{-\zeta}^{\zeta}dz' e^{\mathrm{i}\frac{\omega z'}{v}} \mathcal{I}_{14}  \;, \label{Ez2 int}
\end{eqnarray}
where the following integrals were also defined
\begin{equation}
\mathcal{I}_7  =  \int_0^\infty dk_\parallel k_\parallel T_{\mathrm{TM,TM}}^{12}(k_\parallel) J_0(k_\parallel R_\parallel) e^{\mathrm{i}k_{x,2}|x|+\mathrm{i}k_{x,1}x_0} \;, \label{integral I7 napp} 
\end{equation}
\begin{equation}
\mathcal{I}_8  =  \int_0^\infty \frac{ dk_\parallel k_\parallel }{ k_{x,1} }  T_{\mathrm{TM,TE}}^{12}(k_\parallel) J_0(k_\parallel R_\parallel) e^{\mathrm{i}k_{x,2}|x|+\mathrm{i}k_{x,1}x_0}  \;, \label{integral I8 napp} 
\end{equation}
\begin{equation}
\mathcal{I}_9  =  \int_0^\infty \frac{ dk_\parallel }{ k_\parallel }  T_{\mathrm{TE,TM}}^{12}(k_\parallel) J_0(k_\parallel R_\parallel) e^{\mathrm{i}k_{x,2}|x|+\mathrm{i}k_{x,1}x_0} \;, \label{integral I9 napp} 
\end{equation}
\begin{equation}
\mathcal{I}_{10}  =  \int_0^\infty dk_\parallel k_\parallel T_{\mathrm{TE,TM}}^{12}(k_\parallel) J_0(k_\parallel R_\parallel) e^{\mathrm{i}k_{x,2}|x|+\mathrm{i}k_{x,1}x_0} \;, \label{integral I10 napp} 
\end{equation}
\begin{equation}
\mathcal{I}_{11}  =  \int_0^\infty \frac{ dk_\parallel k_{x,2} }{ k_\parallel k_{x,1} }  T_{\mathrm{TM,TE}}^{12}(k_\parallel) J_0(k_\parallel R_\parallel) e^{\mathrm{i}k_{x,2}|x|+\mathrm{i}k_{x,1}x_0} \;, \label{integral I11 napp} 
\end{equation}
\begin{equation}
\mathcal{I}_{12}  = \int_0^\infty \frac{ dk_\parallel }{ k_\parallel k_{x,1} }  T_{\mathrm{TE,TE}}^{12}(k_\parallel) J_0(k_\parallel R_\parallel) e^{\mathrm{i}k_{x,2}|x|+\mathrm{i}k_{x,1}x_0} \;, \label{integral I12 napp} 
\end{equation}
\begin{equation}
\mathcal{I}_{13}  =  \int_0^\infty \frac{ dk_\parallel k_{x,2} }{ k_\parallel }  T_{\mathrm{TM,TM}}^{12}(k_\parallel) J_0(k_\parallel R_\parallel) e^{\mathrm{i}k_{x,2}|x|+\mathrm{i}k_{x,1}x_0} \;, \label{integral I13 napp} 
\end{equation}
\begin{equation}
\mathcal{I}_{14}  = \int_0^\infty dk_\parallel k_\parallel k_{x,2} T_{\mathrm{TM,TM}}^{12}(k_\parallel) J_0(k_\parallel R_\parallel) e^{\mathrm{i}k_{x,2}|x|+\mathrm{i}k_{x,1}x_0}  \;. \label{integral I14 napp} 
\end{equation}

In the components (\ref{Ex1 int})--(\ref{Ez1 int}) and (\ref{Ex2 int})--(\ref{Ez2 int}), terms of order higher than $R^{-2}$ were neglected and the eight modified Fresnel coefficients required for the electric field components are given as follows \cite{Crosse-Fuchs-Buhmann}:
\begin{eqnarray}
R_{\mathrm{TE,TE}}^{12} &=& \frac{(\mu_{2}k_{x,1}-\mu_{1}k_{x,2})\Omega_\varepsilon - \Delta_\Theta^{2}k_{x,1}k_{x,2}}{(\varepsilon_{2}k_{x,1}+\varepsilon_{1}k_{x,2})\Omega_\mu + \Delta_\Theta^{2}k_{x,1}k_{x,2}} \;, \label{Fresnel 1} \\
R_{\mathrm{TM,TE}}^{12} &=& \frac{ -2\mu_2n_1k_{x,1}k_{x,2}\Delta_\Theta}{(\varepsilon_{2}k_{x,1}+\varepsilon_{1}k_{x,2})\Omega_\mu+\Delta_\Theta^{2}k_{x,1}k_{x,2}} \;, \label{Fresnel 2} \\
T_{\mathrm{TE,TE}}^{12} &=& \frac{2\mu_{2} k_{x,1}\Omega_\varepsilon }{ (\varepsilon_{2}k_{x,1}+\varepsilon_{1}k_{x,2})\Omega_\mu + \Delta_\Theta^{2}k_{x,1}k_{x,2}} ,\;\; \label{Fresnel 3} \\
T_{\mathrm{TM,TE}}^{12} &=& \frac{ 2\mu_2n_2k_{x,1}^2\Delta_\Theta}{(\varepsilon_{2}k_{x,1}+\varepsilon_{1}k_{x,2})\Omega_\mu + \Delta_\Theta^{2}k_{x,1}k_{x,2}} \;,  \label{Fresnel 4} \\
R_{\mathrm{TM,TM}}^{12} &=& \frac{(\varepsilon_{2}k_{x,1}-\varepsilon_{1}k_{x,2})\Omega_\mu + \Delta_\Theta^{2}k_{x,1}k_{x,2}}{(\varepsilon_{2}k_{x,1}+\varepsilon_{1}k_{x,2})\Omega_\mu + \Delta_\Theta^{2}k_{x,1}k_{x,2}} \;, \label{Fresnel 5} \\
R_{\mathrm{TE,TM}}^{12} &=& R_{\mathrm{TM,TE}}^{12} \;, \label{Fresnel 6} \\
T_{\mathrm{TM,TM}}^{12} &=& \frac{n_2}{n_1} \frac{2\varepsilon_{1}k_{x,1}\Omega_\mu }{ (\varepsilon_{2}k_{x,1}+\varepsilon_{1}k_{x,2})\Omega_\mu + \Delta_\Theta^{2}k_{x,1}k_{x,2}} ,\;\;\;\; \label{Fresnel 7} \\
T_{\mathrm{TE,TM}}^{12} &=& - \frac{ n_1 k_{x,2} }{ n_2 k_{x,1} } T_{\mathrm{TM,TE}}^{12} \;, \label{Fresnel 8}
\end{eqnarray}
where $\Omega_\varepsilon=\mu_1 \mu_2 ( k_{x,1}\varepsilon_2 + k_{x,2}\varepsilon_1)$, $\Omega_\mu=\mu_1 \mu_2 ( k_{x,1}\mu_2 + k_{x,2}\mu_1)$, and recalling that $\Delta_\Theta$ was defined in Eq.~(\ref{DELTA}) and specified for strong TIs in Eq.~(\ref{DELTA_1}). For these modified Fresnel coefficients the notation for the superscripts described for the usual Fresnel coefficients (\ref{Usual Fresnel}) is employed, while the subscripts show explicitly a mixture between polarizations TE and TM \cite{Crosse-Fuchs-Buhmann}. Of course, the coefficients (\ref{Fresnel 1})--(\ref{Fresnel 8}) reduce properly to the standard ones when $\Delta_\Theta$ is turned off.

Our next task consists of solving the integrals over $k_\parallel$. Focusing on the radiation emitted by the particle, we will implement the far-field approximation retaining for the moment that $x_0$ could be of arbitrary size. To this aim we employ 
a steepest descent method \cite{Banhos,Chew,Mandel, Chew art,Wait}. 
 By means of this method, 
whose application with full details to the involved integrals is in Appendix \ref{B}, 
we obtain the following results after discarding terms of order higher than $R^{-2}$ for the reflected electric field 
%
%
%
%
%
\begin{eqnarray}
&& \mathbf{E}_{1} (\mathbf{r};\omega) = \frac{ \mathrm{i}q\omega \mu_0 \mu_1(\omega) \mathcal{K}_1(z/R_0,\omega) }{ 4\pi R_0 } e^{ \mathrm{i}k_1 R_0 } \nonumber\\ 
&& \times  \left(
\begin{array}{c}
  - (x-x_0)z/R^2_0  \\
   - yz/R^2_0 \\
   1 -z^2/R^2_0 
\end{array}
\right) \nonumber\\
&&+\frac{ \mathrm{i}q\omega \mu_0 \mu_1(\omega) \tilde{ \mathcal{K} }_1(z/R_1,\omega)  }{ 4\pi R_1 } e^{ \mathrm{i}k_1 R_1 }\nonumber\\
&& \times\left[ R_{\mathrm{TM,TM}}^{12}(k_1\rho/R_1,\Delta_\Theta) \left(
\begin{array}{c}
   (x+x_0)z/R_1^2 \\
    0 \\
   - (x+x_0)^2/R_1^2 
\end{array}
\right)  \right. \nonumber\\
&& \left. + R_{\mathrm{TE,TM}}^{12}(k_1\rho/R_1,\Delta_\Theta) \left(
\begin{array}{c}
  y/R_1 \\
  - (x+x_0)/R_1 \\
  0  
\end{array}
\right)  \right] \nonumber\\
&& +\frac{ q \omega \mu_0 \mu_1(\omega) }{ 4\pi k_1 } \sqrt{ \frac{ 2\pi\mathrm{i} \varkappa }{ \rho } }  e^{ \mathrm{i}\varkappa\rho + \mathrm{i}\sqrt{k_1^2 - \varkappa^2 }(x+x_0) } \mathcal{M}(z/\rho,\omega) \nonumber\\
&& \times \left[ \frac{ \mathrm{Res}\left(R_{\mathrm{TM,TM}}^{12};\varkappa\right) }{ k_1 } \left(
\begin{array}{c}
   -\varkappa z/\rho  \\
  0 \\
  \sqrt{k_1^2 - \varkappa^2 }  
\end{array}
\right) \right. \nonumber\\
&& + \left. \mathrm{Res}\left(R_{\mathrm{TE,TM}}^{12};\varkappa\right) \left(
\begin{array}{c}
   -\varkappa y/ (\rho \sqrt{k_1^2 - \varkappa^2 } )   \\
   1 \\
   0 
\end{array}
\right) \right] \;, \label{E reflex} 
\end{eqnarray}
where now $R_0=\sqrt{ (x-x_0)^2 + y^2 + z^2 }$, $R_1=\sqrt{ (x+x_0)^2 + y^2 + z^2 }$, $\rho=\sqrt{ y^2 + z^2 }$, and 
\begin{eqnarray}
\mathcal{K}_1 &=& \int_{-\zeta}^{\zeta}dz' e^{ \mathrm{i}\frac{ \omega z' }{v} \left(1 - \frac{ vn_1 z }{ c R_0 }  \right) }  = \frac{ 2\sin\left(\zeta \Xi_1\right) }{ \Xi_1 } \;, \label{Function K1} \\
\tilde{ \mathcal{K} }_1 &=& \int_{-\zeta}^{\zeta}dz' e^{ \mathrm{i}\frac{ \omega z' }{v} \left(1 - \frac{ vn_1 z }{ c R_1 }  \right) }  = \frac{ 2\sin\left(\zeta \tilde{ \Xi }_1\right) }{ \tilde{ \Xi }_1 } \;, \label{Function K1 tilde} \\
\mathcal{M} &=&  \int_{-\zeta}^{\zeta}dz' \exp \left[ \mathrm{i}z' \left( \frac{\omega}{v} - \varkappa \frac{z}{\rho} \right) \right] \;, \label{Function M1}
\end{eqnarray}
are the resulting integrals over $z'$ with 
\begin{equation} \label{Xi 1s}
\Xi_1= \frac{\omega }{v}\left( 1 - \frac{ vn_1 z }{ c R_0 } \right) \;,\; \tilde{ \Xi }_1=\frac{\omega }{v}\left( 1 - \frac{ vn_1 z }{ c R_1 } \right)\;.
\end{equation}
In Eq.~(\ref{E reflex}), $\mathrm{Res}$ is the residue of the indicated Fresnel coefficient at the simple pole $\varkappa$, which is the only physical root of the polynomial of degree 4 in $k_\parallel$ arising from the common denominator for all modified Fresnel coefficients and has units of wave number.

For the transmitted field, with $x_0\ll x$, we find
\begin{eqnarray}
&& \mathbf{E}_{2} (\mathbf{r};\omega) = \frac{ \mathrm{i}q\omega \mu_0 \mu_1(\omega)k_2 \mathcal{K}_2(z/r,\omega) }{ 4\pi k_1 r } e^{ \mathrm{i}k_2r } e^{ \mathrm{i} \xi x_0 } \nonumber\\
&& \times \left\{ T_{\mathrm{TM,TM}}^{12}(k_2\rho/r,\Delta_\Theta) \left(
\begin{array}{c}
  - xz/r^2 \\
   0 \\
   x^2/r^2 
\end{array}
\right) \right. \nonumber\\
&& \left. + T_{\mathrm{TE,TM}}^{12}(k_2\rho/r,\Delta_\Theta) \left(
\begin{array}{c}
  - y/r \\
   x/r \\
   0 
\end{array}
\right) \right\} \nonumber\\
&& +\frac{ q \omega \mu_0 \mu_1(\omega) }{ 4\pi k_1 } \sqrt{ \frac{ 2\pi\mathrm{i} \varkappa }{ \rho } }  e^{ \mathrm{i}\varkappa\rho + \mathrm{i}\sqrt{k_2^2 - \varkappa^2 }|x| } e^{ \mathrm{i}\sqrt{k_1^2 - \varkappa^2 }x_0 } \nonumber\\
&& \times \mathcal{M}(z/\rho,\omega) \left[ \frac{ \mathrm{Res}\left(T_{\mathrm{TM,TM}}^{12};\varkappa\right) }{ k_2 } \left(
\begin{array}{c}
   -\varkappa z/\rho  \\
  0 \\
  \sqrt{k_2^2 - \varkappa^2 }  
\end{array}
\right) \right. \nonumber\\
&& + \left. \mathrm{Res}\left(T_{\mathrm{TE,TM}}^{12};\varkappa\right) \left(
\begin{array}{c}
   -\varkappa y/ (\rho \sqrt{k_2^2 - \varkappa^2 } )   \\
   1 \\
   0 
\end{array}
\right) \right] ,\; \label{E trans}
\end{eqnarray}
where $r=\sqrt{ x^2 + y^2 + z^2 }$, $\xi=\sqrt{k_1^2 - k_2^2 (1 - x^2/r^2)} $, $\mathcal{M}$ is given by Eq.~(\ref{Function M1}) but must be now considered in the lower medium and 
\begin{equation}
\mathcal{K}_2 = \int_{-\zeta}^{\zeta}dz' e^{ \mathrm{i}\frac{ \omega z' }{v} \left(1 - \frac{ vn_2 z }{ c r }  \right) }  = \frac{ 2\sin\left(\zeta \Xi_2\right) }{ \Xi_2 } \;, \label{Function K2}
\end{equation}
is the corresponding integral over $z'$ with 
\begin{equation} \label{Xi 2}
\Xi_2= \frac{\omega }{v}\left( 1 - \frac{ vn_2 z }{ c r } \right) \;.
\end{equation}
%

At this stage it becomes necessary to discuss the physical meaning of the electric field given by Eqs.~(\ref{E reflex}) and (\ref{E trans}). First of all, we appreciate that two kinds of waves contribute to the electric field in both media: spherical and evanescent surface waves. As usual, the spherical-wave contribution decays with the inverse of the distance, where we identify two terms: one associated to the free-space contribution plus one related to the response of the interface proportional to the modified Fresnel coefficients $R_{\mathrm{TM,TM}}^{12}$ or $T_{\mathrm{TM,TM}}^{12}$ both being quadratic contributions of the topological parameter $\Delta_\Theta$. This response has a second contribution proportional to the crossed Fresnel coefficients $R_{\mathrm{TE,TM}}^{12}$ in medium 1 or $T_{\mathrm{TE,TM}}^{12}$ in medium 2, which leads to purely topological contributions because they are linear in the topological parameter $\Delta_\Theta$ and are absent in the standard case $\Delta_\Theta=0$. An important feature of this field is the VC cone arising from the integrals over $z'$, which lead to the functions $\mathcal{K}_1$ (\ref{Function K1}), $\tilde{ \mathcal{K} }_1$ (\ref{Function K1 tilde}) and $\mathcal{K}_2$ (\ref{Function K2}). From their corresponding arguments $\Xi_1$ (\ref{Xi 1s}), $\tilde{ \Xi }_1$ (\ref{Xi 1s}) and $\Xi_2$ (\ref{Xi 2}) of the sine function, we can define the Cherenkov angle that determines the VC cone aperture. Namely, we will have the following three Cherenkov angles:
\begin{eqnarray}
\cos\theta_1^{C}=\frac{z}{R}=\frac{c}{vn_1}&,& \cos\tilde{\theta}_1^{C}=\frac{z}{R_1}=\frac{c}{vn_1}\;, \label{Cherenkov reflex} \\ 
\cos\theta_2^{C}&=&\frac{z}{r}=\frac{c}{vn_2} \;, \label{Cherenkov trans}
\end{eqnarray}
which must lie in the interval $[0,\pi/2)$ otherwise there will not be VC radiation. Note that the Cherenkov angles of the free contribution and of the reflected one are equal in magnitude but different functionally because the reflected one arises from the electric field generated by the image charge. These angles differ from the transmissive one as found previously in Ref.~\cite{Bolotovskii}. This fact can be understood in simple terms when we recall the behavior of light in front of an interface, meaning that $\theta_1^{C}$ is the incidence angle of the VC radiation due to the free term of the electric field, $\tilde{\theta}_1^{C}$ is the reflected one both for the $R_{\mathrm{TM,TM}}^{12}$ and $R_{\mathrm{TE,TM}}^{12}$ terms, and the $\theta_2^{C}$ is the refracted one related to $T_{\mathrm{TM,TM}}^{12}$ and $T_{\mathrm{TE,TM}}^{12}$. As a consequence of the Cherenkov angles given by Eqs.~(\ref{Cherenkov reflex}) and (\ref{Cherenkov trans}), for this configuration a non-symmetric VC cone can arise \cite{Bolotovskii} and the reflective part or the transmissive part of the VC cone can be individually suppressed. The phases at both sides of the interface differ as may be expected, being $e^{ \mathrm{i}k_1 R_0}$ and $e^{ \mathrm{i}k_1 R_1 }$ due to the reflection response from the interface. On the other hand,  $e^{ \mathrm{i} \xi x_0 }$ accounts for the radiation transmitted through the interface, which could become a decaying exponential if $n_1<n_2$ meaning that the radiation would be dramatically attenuated in a similar way as happens to the dipolar radiation in front a single interface \cite{Novotny}. In contrast to the previously studied case where this phases difference was also found in the perpendicular motion towards the same interface but with same permittivities \cite{OJF-LFU-ORT}, here the possibility of a reversed VC radiation is ruled out because no phase allows negative angles. Physically speaking, this can be understood because the charge does not cross the interface at any point so the VC cone cannot be reflected backwards with respect to the charge's trajectory but only reflected forwards in the same direction of the particle's trajectory. Finally, we observe that the TE,TE polarization is no more present because it is of higher order than $r^{-2}$.

Apart from the spherical-wave contribution, we find surface waves in Eqs.~(\ref{E reflex}) and (\ref{E trans}). For the sake of discussion, we present their explicit form for the case of non magnetic media $\mu_1=\mu_2=1$. At order $\Delta_\Theta^2$, the reflected field is given by
\begin{eqnarray}
&&\mathbf{E}_{1}^{\;\mathrm{surface}} (\mathbf{r};\omega) = \frac{ q \omega \mu_0 }{ 4\pi n_1 } \sqrt{ \frac{ 2\pi\mathrm{i} \omega }{ \rho c } }  e^{ \mathrm{i}\varkappa\rho + \mathrm{i}\sqrt{k_1^2 - \varkappa^2 }(x+x_0) }  \nonumber\\
&& \times \mathcal{M}(z/\rho,\omega) \left[ \frac{ \mathcal{F}_1(\varepsilon_1,\varepsilon_2,\Delta_\Theta) }{ n_1 } \left( - \frac{z}{\rho}\mathbf{e}_{x} + \mathbf{e}_{z} \right) \right. \nonumber\\
&& + \left. \mathcal{F}_2(\varepsilon_1,\varepsilon_2,\Delta_\Theta) \left( - \frac{y}{\rho}\mathbf{e}_{x} + \mathbf{e}_{y} \right) 
\right]\;, \label{E reflex sw}
\end{eqnarray}
with $\mathcal{M}$ taking the next form: 
\begin{eqnarray}
\mathcal{M}  &=&  \int_{-\zeta}^{\zeta}dz' \exp\left\{  \mathrm{i}\frac{ \omega z' }{v} \left[ 1 - \frac{ vn_1 }{ c } \sin\tilde{\theta}_{B} \frac{ z }{ \rho } \right] \right\} \;, \nonumber\\
&=& \frac{ 2\sin\left(\zeta\Lambda\right) }{ \Lambda } \;,  \label{Function M simplified}
\end{eqnarray}
where
\begin{eqnarray}
\Lambda &=& \frac{\omega }{v} \left[ 1 - \frac{ vn_1 }{ c } \sin\tilde{\theta}_{B} \frac{ z }{ \rho }\right]\;,\\
\varkappa  &=& k_1 \sin\tilde{\theta}_{B} \;, \label{varkappa} \\
\sin\tilde{\theta}_{B} &=& \sqrt{\frac{\varepsilon_2 }{ \varepsilon_1 + \varepsilon_2 }} \left[ 1 + \frac{ \Delta_\Theta^2 \varepsilon_1 \varepsilon_2 }{ (\varepsilon_1-\varepsilon_2)^2 (\varepsilon_1+ \varepsilon_2) } \right] . \label{Brewster angle}
\end{eqnarray}
The transmitted field can be written in the following fashion:
\begin{eqnarray}
&&\mathbf{E}_{2}^{\;\mathrm{surface}} (\mathbf{r};\omega) = \frac{ q \omega \mu_0 }{ 4\pi n_1 } \sqrt{ \frac{ 2\pi\mathrm{i} \omega }{ \rho c } }  e^{ \mathrm{i}\varkappa\rho + \mathrm{i}\sqrt{k_2^2 - \varkappa^2 }|x| }  \nonumber\\
&& \times e^{ \mathrm{i}\sqrt{k_1^2 - \varkappa^2 }x_0 }\mathcal{M}(z/\rho,\omega) \nonumber\\
&& \times \left[ \frac{ \mathcal{F}_1(\varepsilon_1,\varepsilon_2,\Delta_\Theta) }{ n_2 } \left( -\frac{z}{\rho}\mathbf{e}_{x} + \mathbf{e}_{z} \right) \right. \nonumber\\
&& + \left. \mathcal{F}_2(\varepsilon_1,\varepsilon_2,\Delta_\Theta) \left( - \frac{y}{\rho}\mathbf{e}_{x} + \mathbf{e}_{y} \right) 
\right] \;, \label{E trans sw}
\end{eqnarray}
where $\mathcal{M}$ was just above defined in Eq.~(\ref{Function M simplified}). For Eqs.~(\ref{E reflex sw}) and (\ref{E trans sw}), $\mathcal{F}_j$ with $j=1,2$ are lengthy functions depending only on $\varepsilon_1,\varepsilon_2,\Delta_\Theta$ and their explicit expressions can be found in the Appendix \ref{C}. 

Let us now focus on the physics behind the surface-wave contribution. At first sight, we can split it into two kinds: cylindrical decaying as $\rho^{-1/2}$ and conical $\rho^{-3/2}$ \cite{Chew}, which differs from the lateral waves with a $\rho^{-2}$ dependence found for the transition radiation case in Ref.~\cite{OJF-SYB}. Here, the mathematical origin of these waves comes from the fact that all the relevant integrals over $k_\parallel$ of the electric field exhibit a simple pole at $\varkappa$. However, the cylindrical waves propagate along the plane parallel to the interface and can be physically interpreted as surface plasmon modes. Meanwhile, the conical ones decay rather faster than the previous ones and are the excited modes oscillating in the $\mathbf{e}_{x}$ direction perpendicular to the interface. Naturally, it is expected that for a charge traveling far away from the interface both contributions will not be excited. Furthermore, we see that the contributions modulated by $\mathcal{F}_1$ [Eq.~(\ref{curlyF1})] and $\mathcal{F}_2$ [Eq.~(\ref{curlyF2})] are of topological origin due to their quadratic dependence in the topological parameter $\Delta_\Theta$ meaning that they are absent in the standard case $\Delta_\Theta=0$. Another significant consequence of the topological parameter $\Delta_\Theta$ is the modification of the Brewster angle $\tilde{\theta}_{B}$ [Eq.~(\ref{Brewster angle})], as studied in the work \cite{Chang-Yang}, which dictates the propagation of both surface waves by means of $\varkappa$ [Eq.~(\ref{varkappa})]. Interestingly, in order to exist, we observe that the surface waves must satisfy the condition
\begin{equation}\label{VC condition sw}
\frac{ vn_1 }{ c } \sin\tilde{\theta}_{B} \frac{ z }{ \sqrt{ y^2 + z^2} }=1,
\end{equation}
which comes directly from the function $\mathcal{M}$ [Eq.~(\ref{Function M simplified})]. Equation (\ref{VC condition sw}) can be regarded as a different VC condition where the modified Brewster angle [Eq.~(\ref{Brewster angle})] has a key role.  For the case $\varkappa>k_{1,2}$ we found that the conical and the cylindrical waves can be evanescent because the corresponding squared root becomes imaginary providing an exponential decay of both surface waves. Otherwise, if $\varkappa<k_{1,2}$, they will only decay algebraically.

\begin{figure*}[ht]
\centering
\subfloat[]{
\label{g:1}
\includegraphics[width=0.3\textwidth]{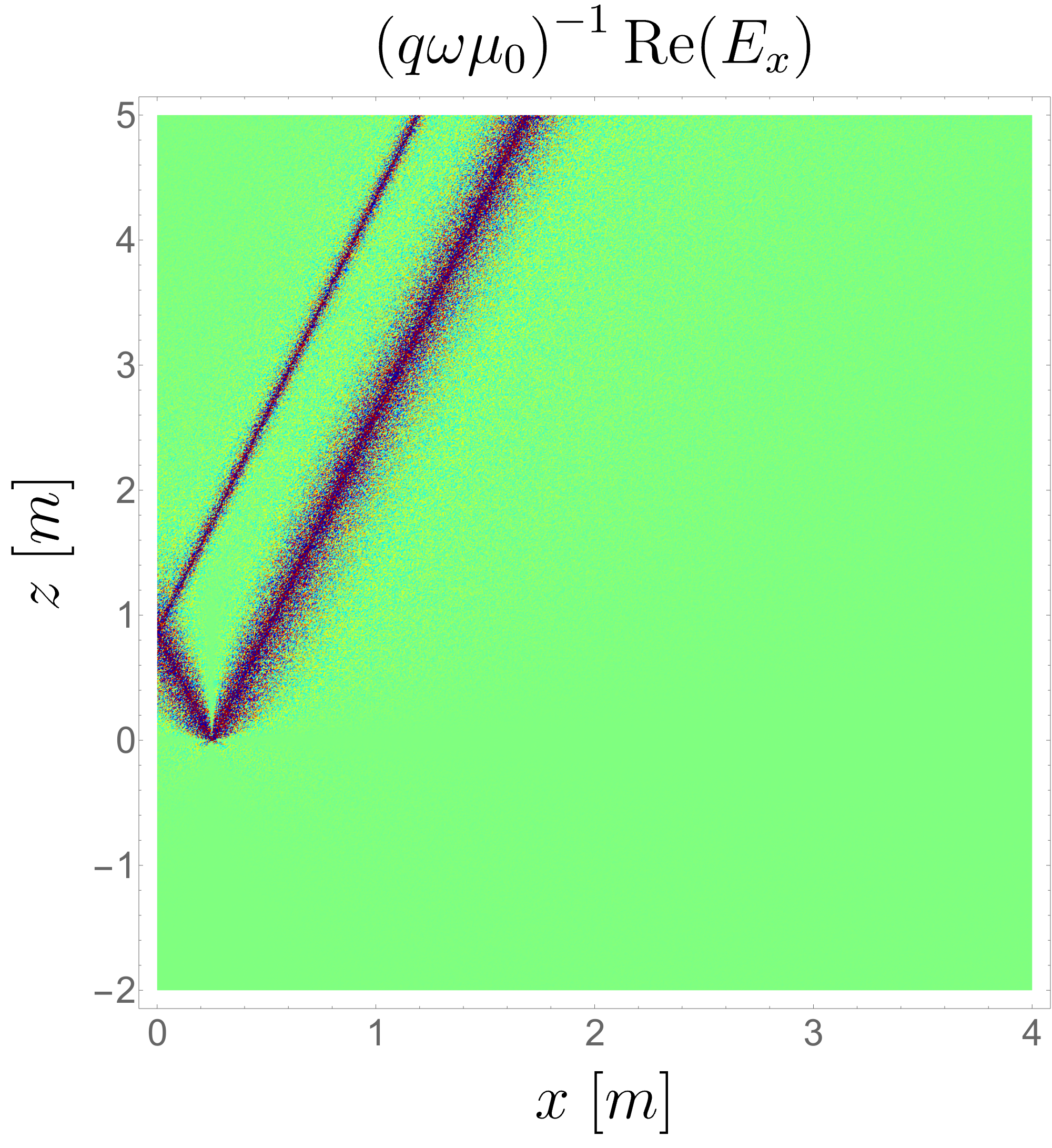}}
\subfloat[]{
\label{g:2}
\includegraphics[width=0.3\textwidth]{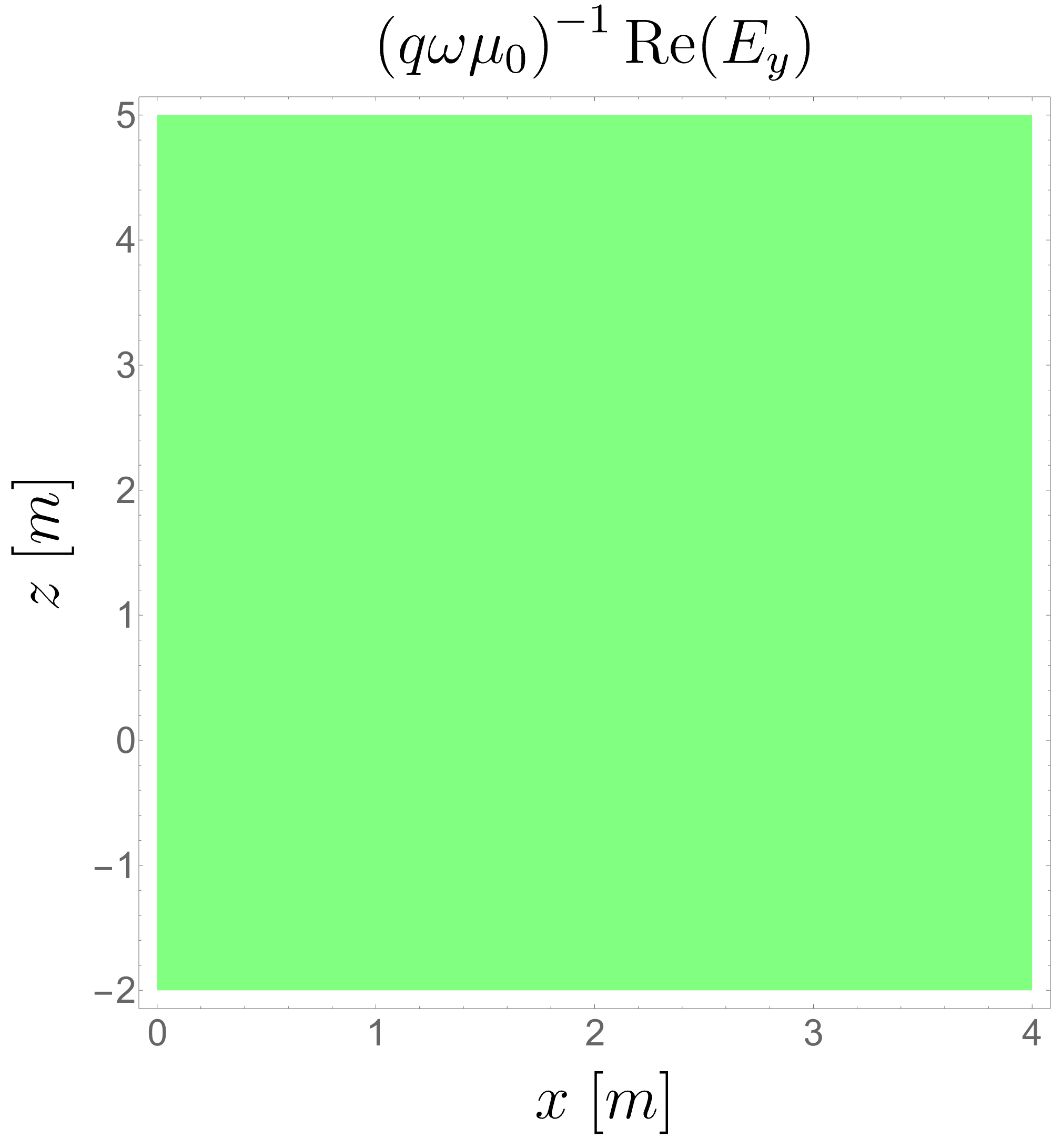}}
\subfloat[]{
\label{g:3}
\includegraphics[width=0.3\textwidth]{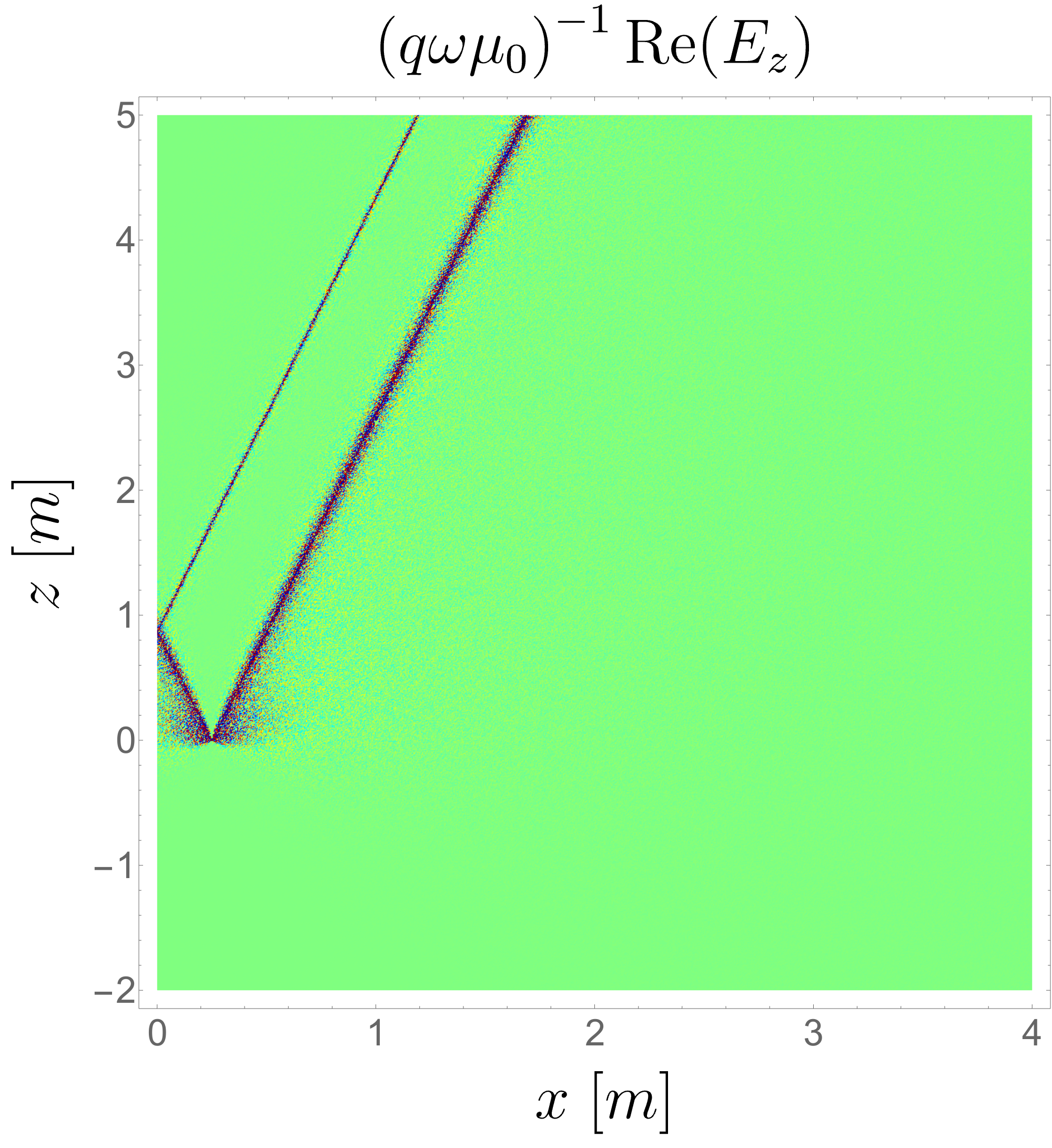}}

\subfloat[]{
\label{g:4}
\includegraphics[width=0.3\textwidth]{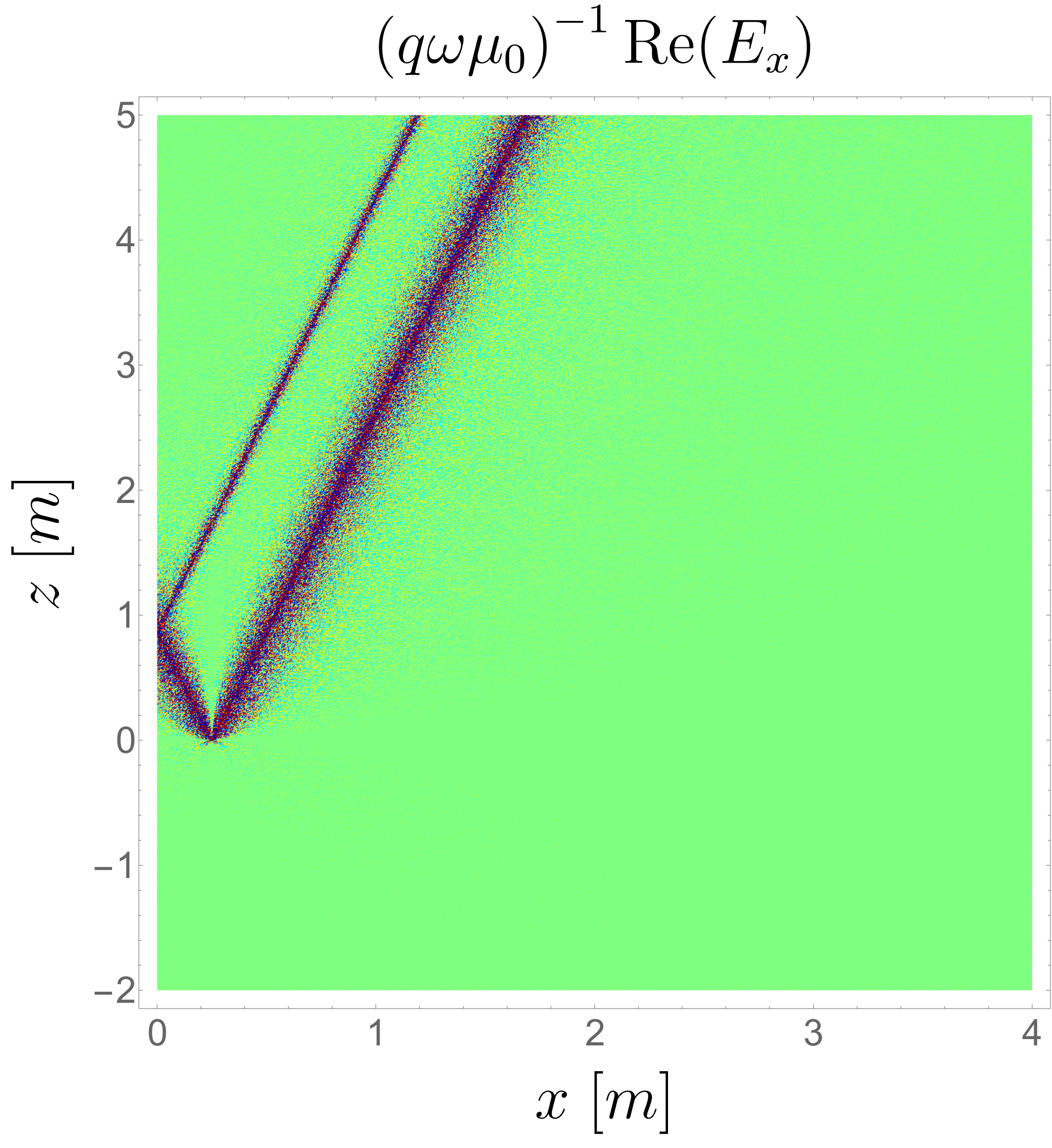}}
\subfloat[]{
\label{g:5}
\includegraphics[width=0.3\textwidth]{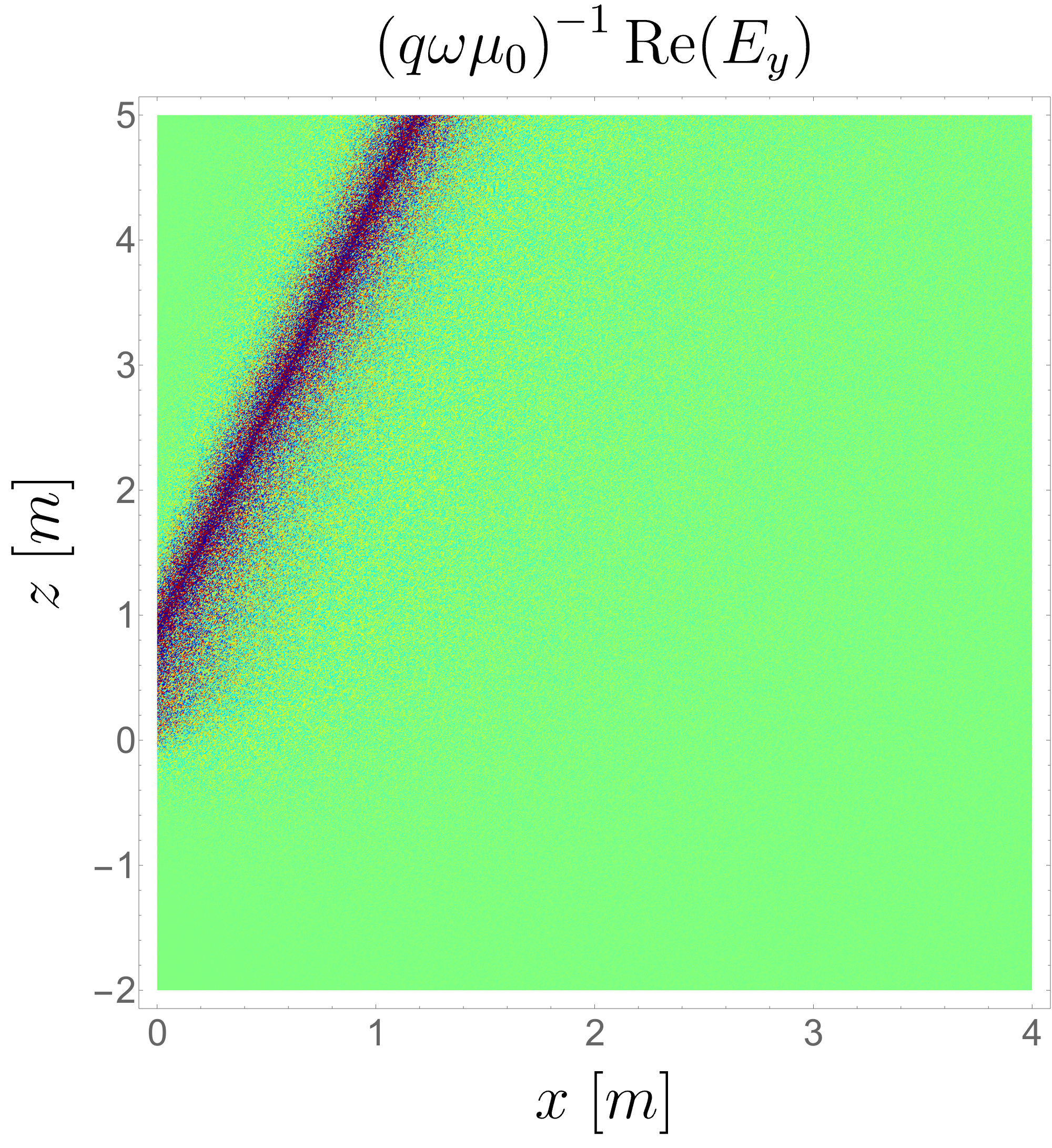}}
\subfloat[]{
\label{g:6}
\includegraphics[width=0.3\textwidth]{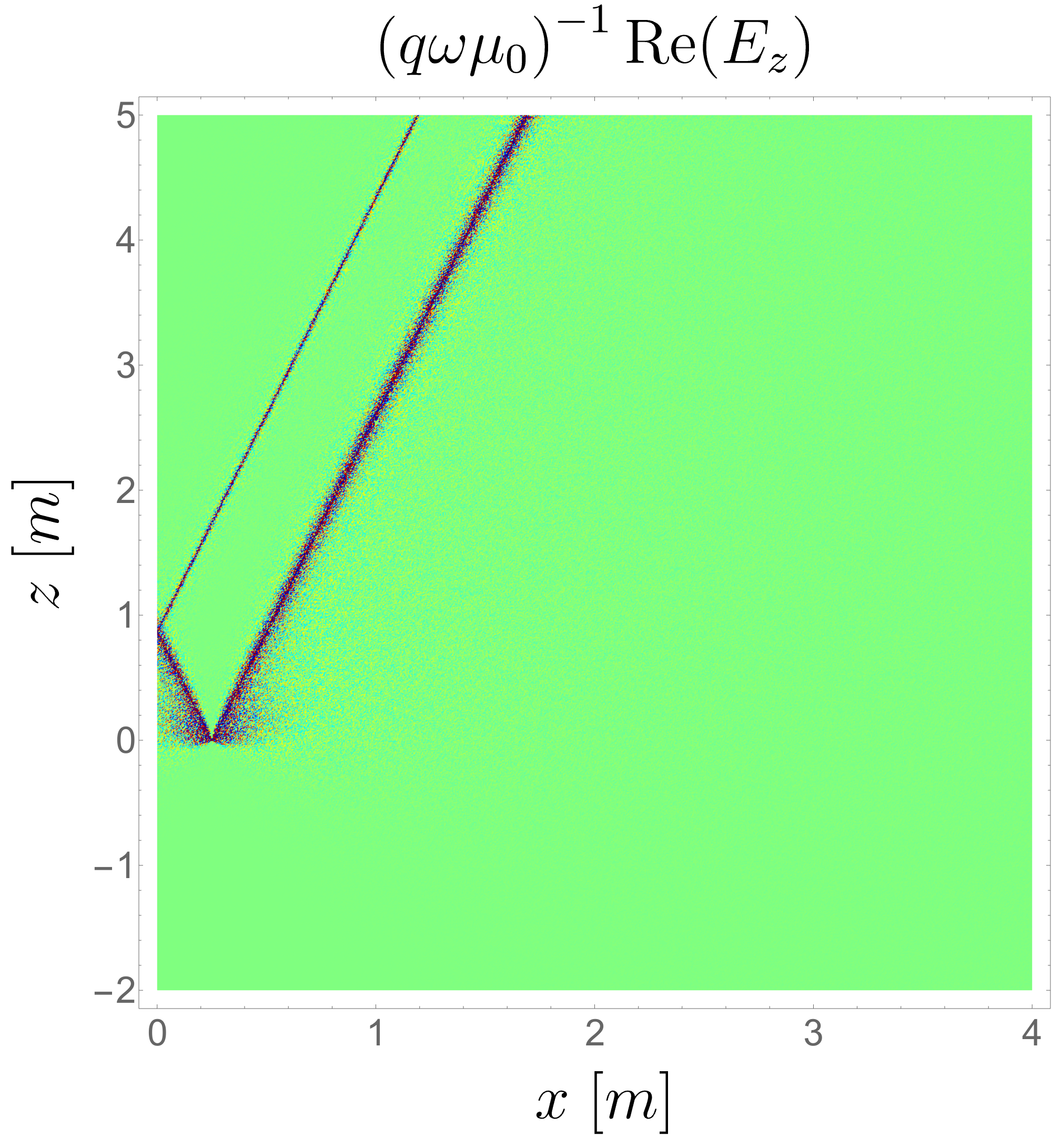}}

\caption{ The electric field pattern (real part) in the positive $x-z$ plane (reflection region) for a charge particle with single frequency $\omega= 6.0450\times10^{14}\,\, \mathrm{s}^{-1}$, constant velocity $v=0.95\,c$ parallel to the $x$ axis at a height $x_0=0.25\,\mathrm{m}$ from an interface constituted by standard dielectrics with $\varepsilon_1=1.2$, $\mu_1=1$ at the upper layer and another one with $\varepsilon_2=4$, $\mu_2=1$ at the lower one as depicted in plots (a)--(c). The plots (d)--(f) show again the electric field for an interface conformed by the same upper medium but now with the topological insulator TlBiSe$_2$ with $\varepsilon_2=4$, $\mu_2=1$, and $\Delta_\Theta=11\alpha$ a the lower one. The electric field is dimensionless. The charge moves from the bottom to the top. }  
%
\label{Electric Field Patterns 1}
\end{figure*}

\begin{figure*}[ht]
\centering
\subfloat[]{
\label{g:7}
\includegraphics[width=0.3\textwidth]{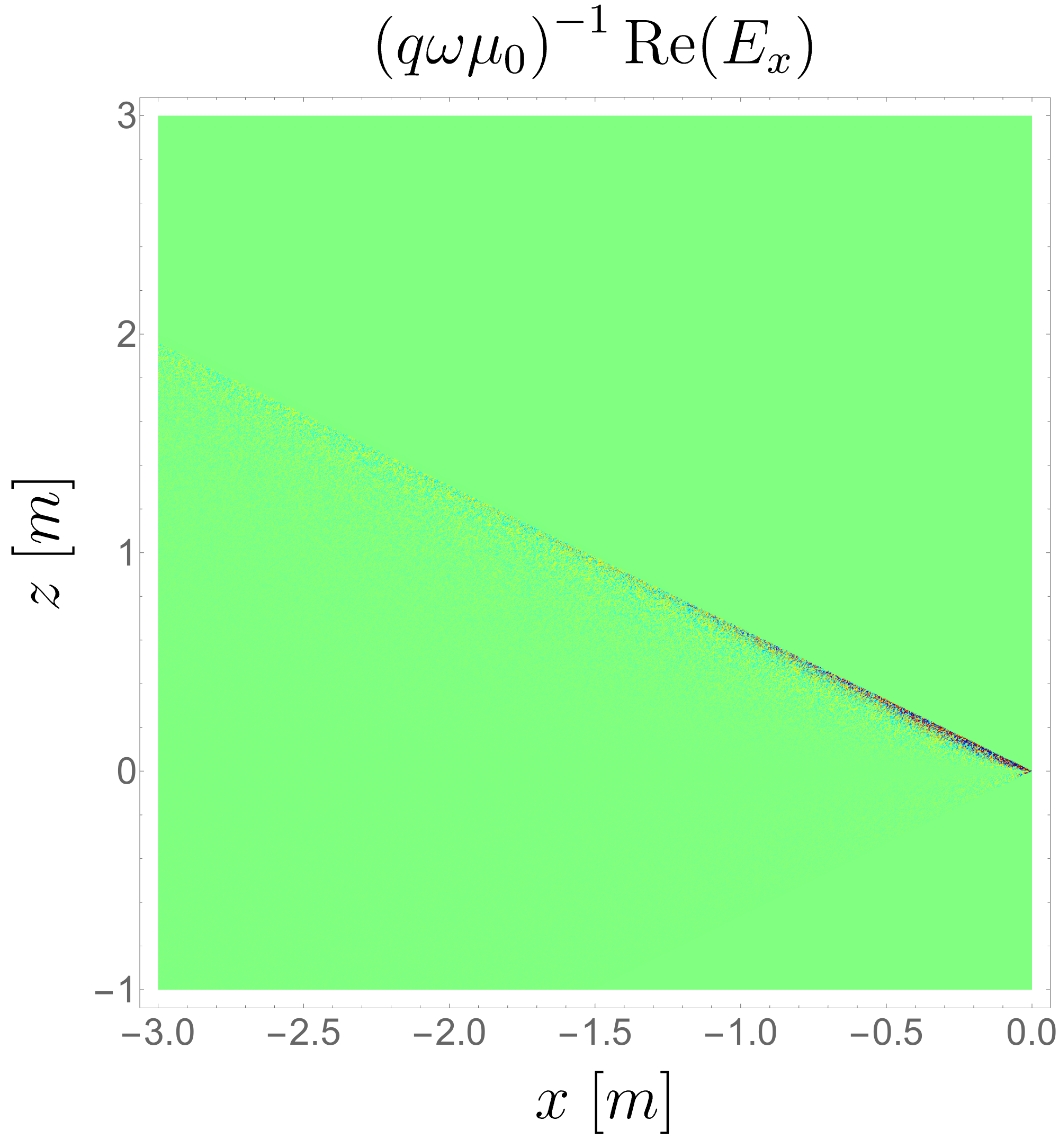}}
\subfloat[]{
\label{g:8}
\includegraphics[width=0.3\textwidth]{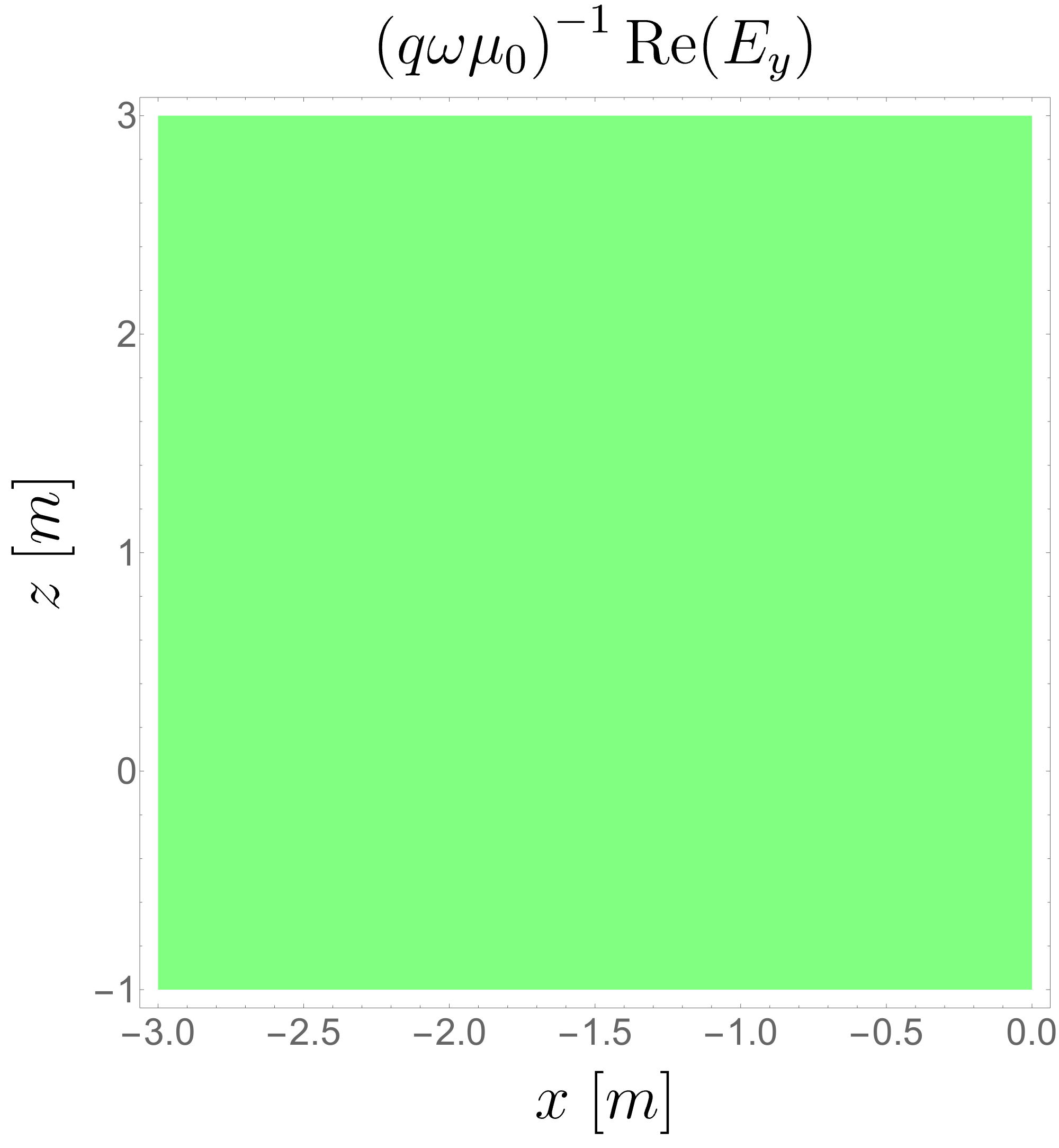}}
\subfloat[]{
\label{g:9}
\includegraphics[width=0.3\textwidth]{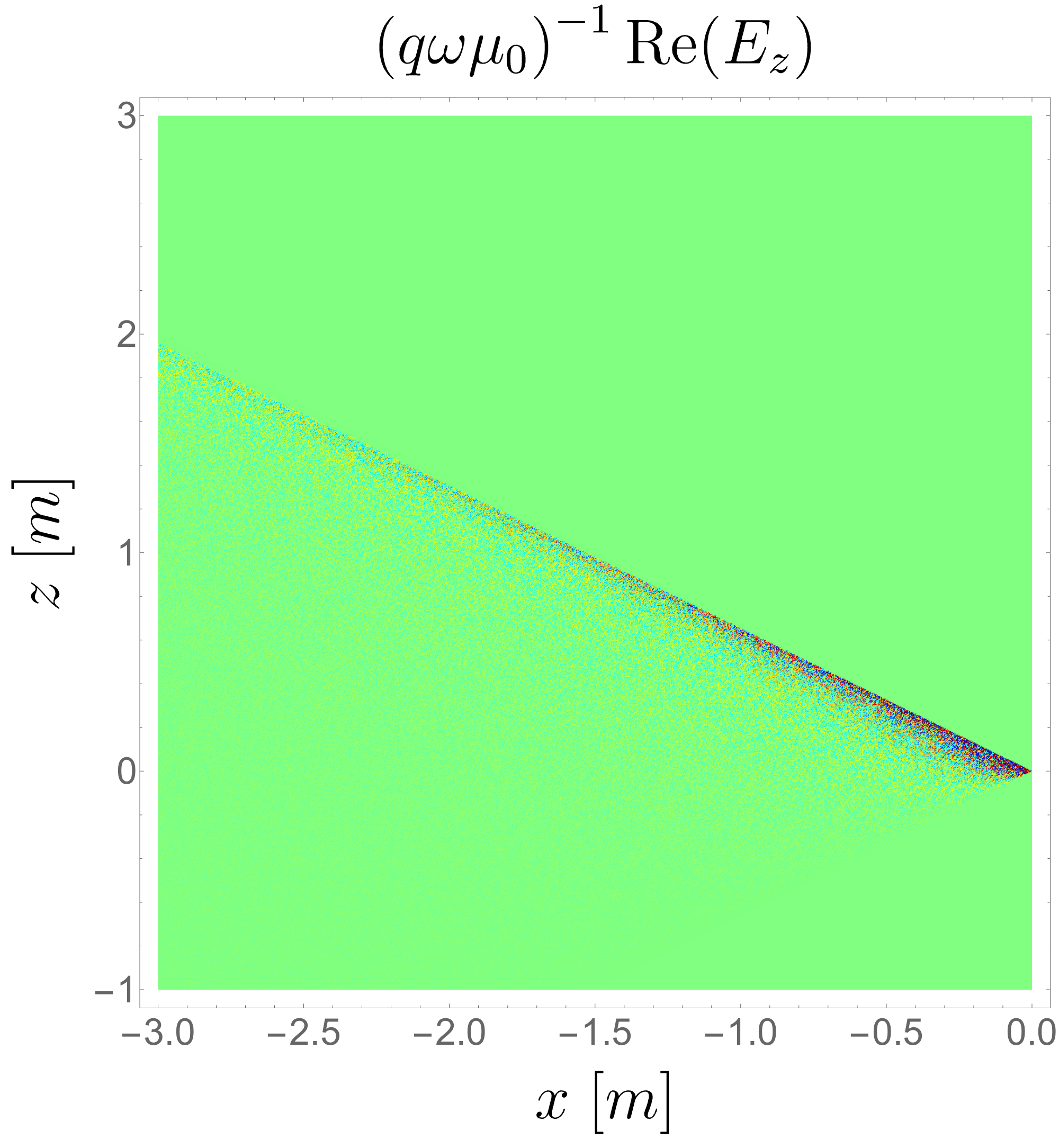}}

\subfloat[]{
\label{g:10}
\includegraphics[width=0.3\textwidth]{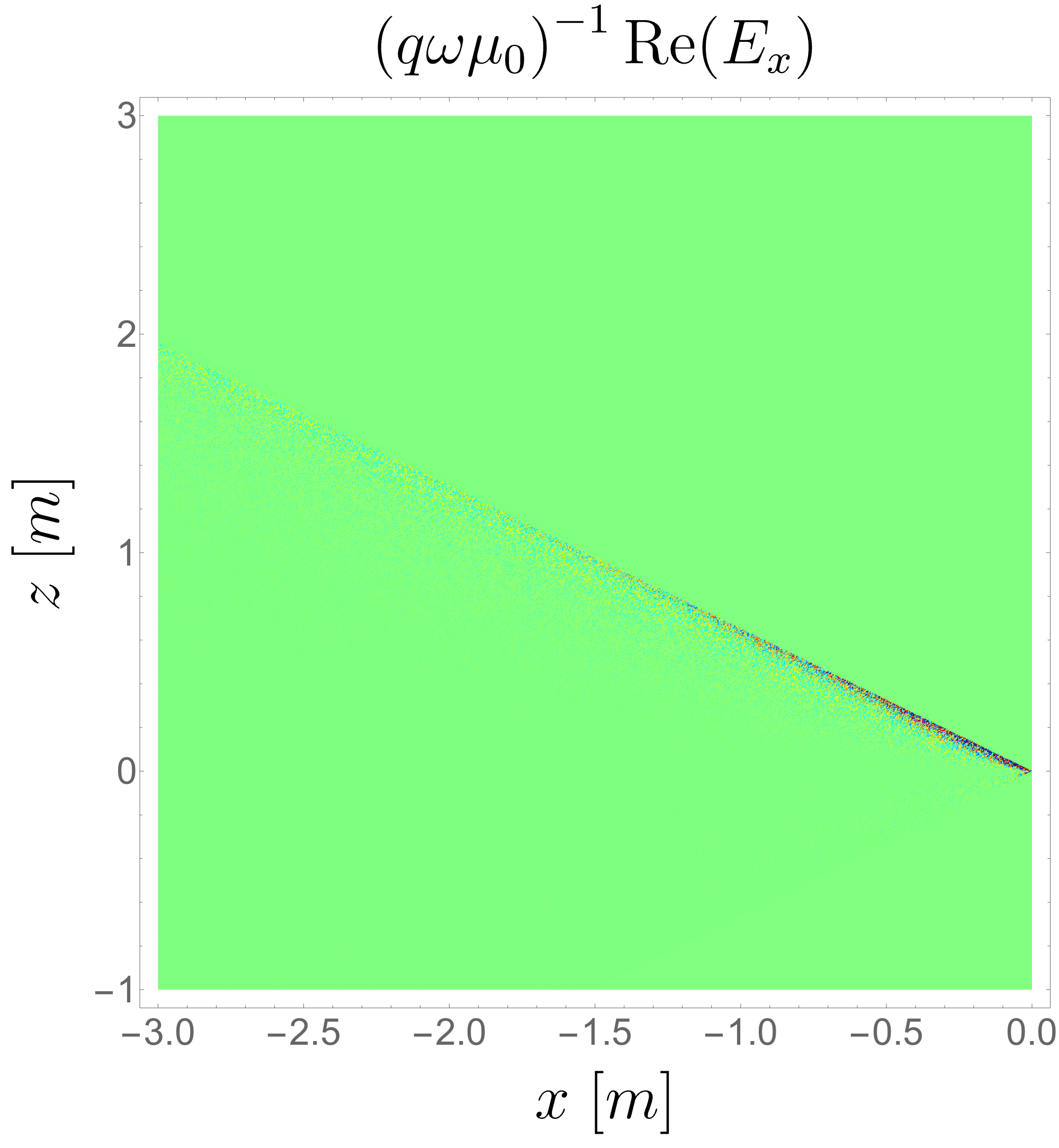}}
\subfloat[]{
\label{g:11}
\includegraphics[width=0.3\textwidth]{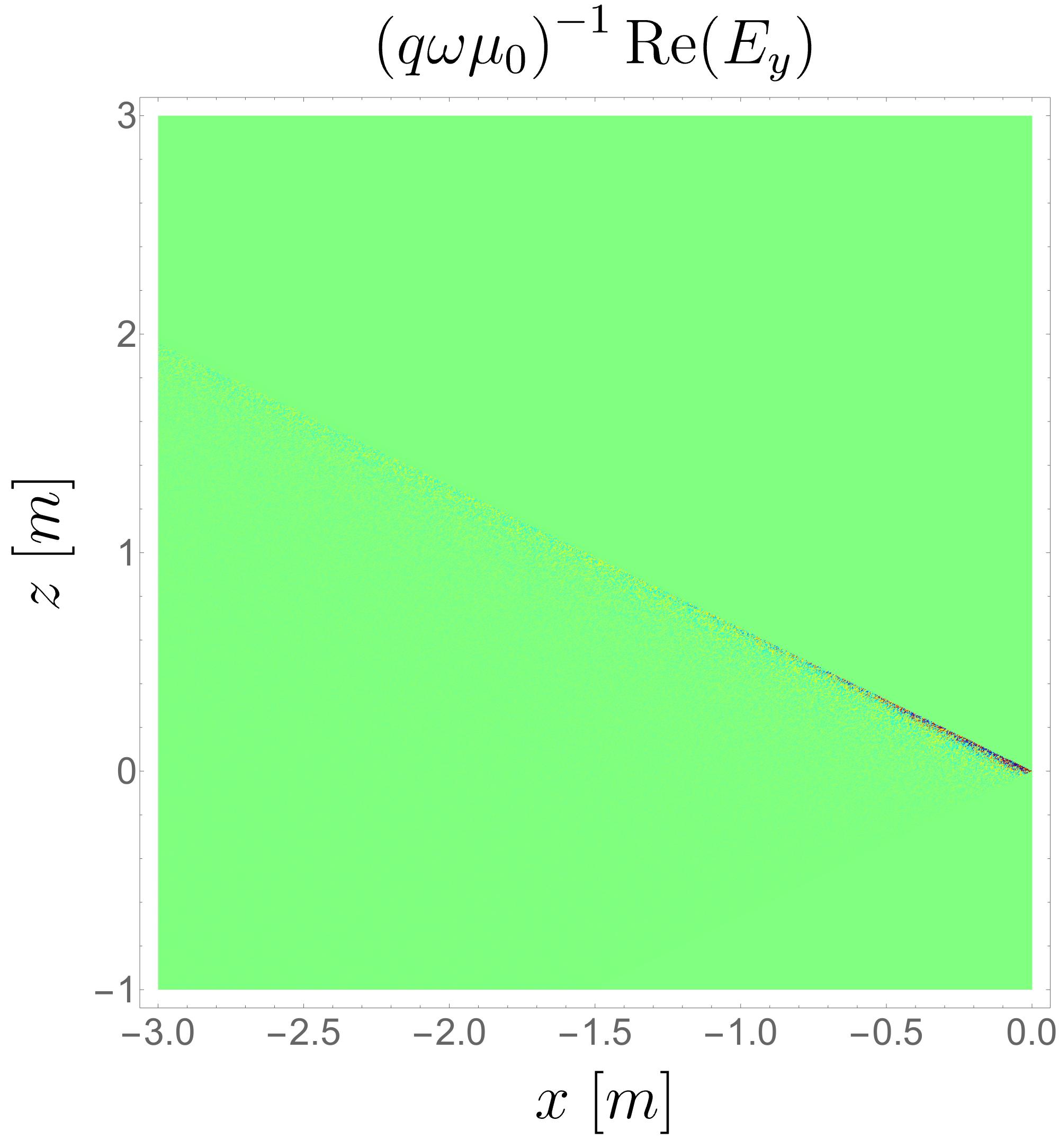}}
\subfloat[]{
\label{g:12}
\includegraphics[width=0.3\textwidth]{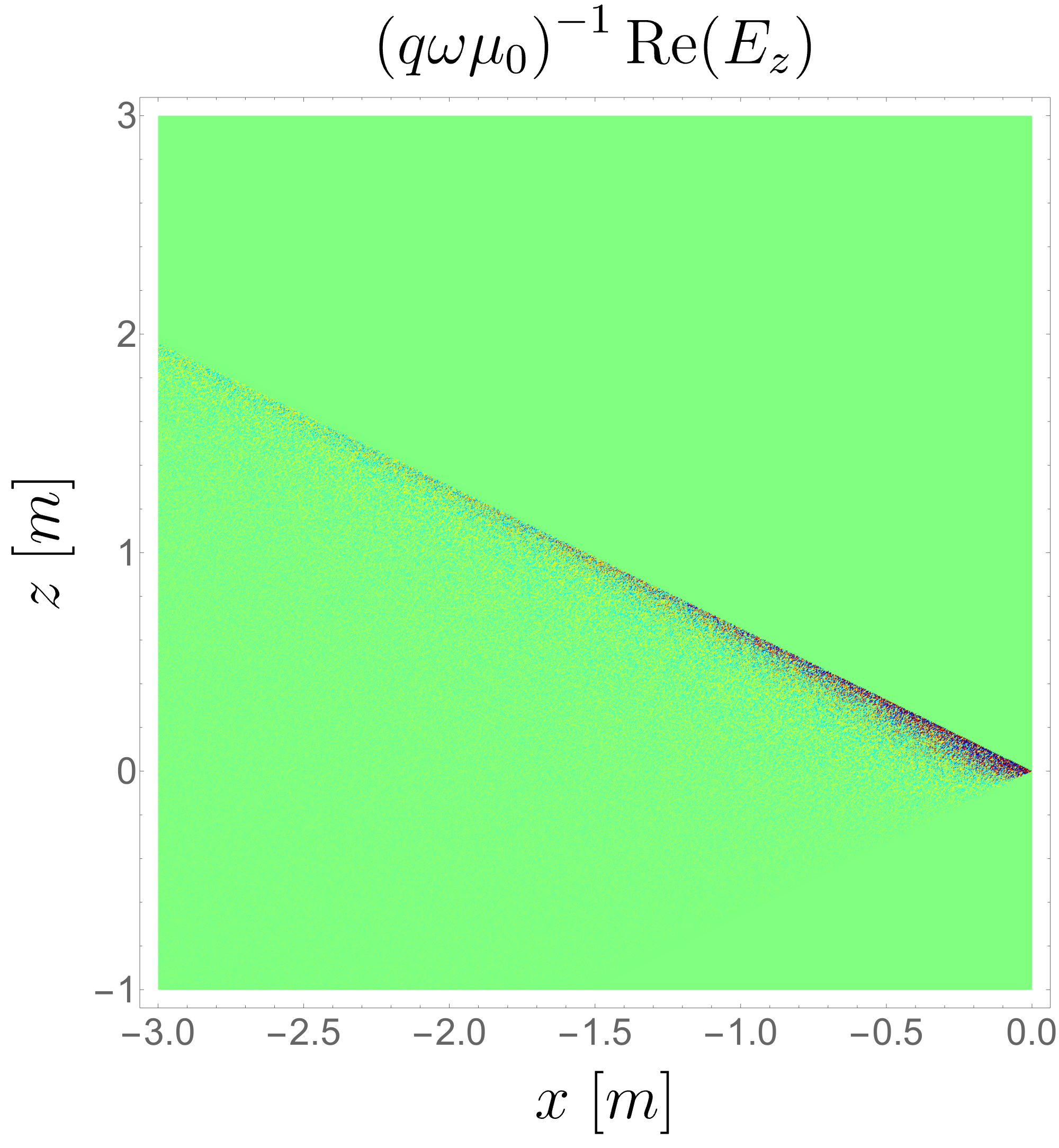}}

\caption{ The electric field pattern (real part) in the negative $x-z$ plane (transmission region) for a charge particle with single frequency $\omega= 6.0450\times10^{14}\,\, \mathrm{s}^{-1}$, constant velocity $v=0.9\,c$ parallel to the $x$ axis at a height $x_0=1\times10^{-4}\,\mathrm{m}$ from an interface constituted by standard dielectrics with $\varepsilon_1=1.2$, $\mu_1=1$ at the upper layer and another one with $\varepsilon_2=4$, $\mu_2=1$ at the lower one as depicted in plots (a)--(c). The plots (d)--(f) show again the electric field for an interface conformed by the same upper medium but now with the topological insulator TlBiSe$_2$ with $\varepsilon_2=4$, $\mu_2=1$, and $\Delta_\Theta=11\alpha$ a the lower one. The electric field is dimensionless. The charge moves from the bottom to the top. }  
%
\label{Electric Field Patterns 2}
\end{figure*}

To conclude this section, and for illustrative purposes, we show the radiation part of electric field at both sides of the interface two in Figs.~\ref{Electric Field Patterns 1} and \ref{Electric Field Patterns 2}.  The first column of Figs.~\ref{Electric Field Patterns 1} and \ref{Electric Field Patterns 2} displays the $x$ component of the electric field, and the second and the third columns feature the $y$ and $z$ components, respectively. The first row Figs.~\ref{Electric Field Patterns 1} and  \ref{Electric Field Patterns 2} exhibits the electric field components for a bi-layer configuration constituted by two standard dielectrics with $\varepsilon_1=1.2$, $\mu_1=1$ and $\Delta_\Theta=0$ at the upper layer and another one with $\varepsilon_2=4$, $\mu_2=1$ and $\Delta_\Theta=0$ at the lower one. The second row of these figures displays again the electric field components but now for an interface conformed by the same standard dielectric with $\varepsilon_1=1.2$, $\mu_1=1$ and $\Delta_\Theta=0$ at the upper layer and the topological insulator TlBiSe$_2$ with $\varepsilon_2=4$, $\mu_2=1$ and $\Delta_\Theta=11\alpha$ at the lower one \cite{TlBiSe2}. The charge moves with a velocity $v=0.95\,c$ from the bottom to the top at a distance $x_0=0.25\,\mathrm{m}$ to the interface located at $x=0$ in Fig.~\ref{Electric Field Patterns 1} and in Fig.~\ref{Electric Field Patterns 2} with a velocity $v=0.9\,c$ again from the bottom to the top at a distance $x_0=1\times10^{-4}\,\mathrm{m}$ to the interface in the reflection region. The exponential attenuation of the electric field in Fig.~\ref{Electric Field Patterns 2} is quite clear at the transmission region showing a notable reduction of the radiation lobes in comparison with the ones at the reflection region of Fig.~\ref{Electric Field Patterns 1}.

Finally, we give two comments regarding the Figs.~\ref{Electric Field Patterns 1} and \ref{Electric Field Patterns 2}. First, we remark that the typical VC cone always conserves its form without any modification. And finally, we highlight that in comparison with the standard electrodynamics case ($\Delta_\Theta=0$) the component $E_y$ is different from zero and two orders of magnitude smaller than the other two components, which will contribute to enhance the effect as Figs.~\ref{g:2}, \ref{g:5}, \ref{g:8}, and \ref{g:11}  clearly show.

\section{Angular Distribution of Energy}\label{ANGULAR}

In this section, we will compute the angular distribution of radiation in the upper and lower hemispheres focusing on the spherical wave contribution. To this end, we will henceforth assume that $x_0\ll x$, therefore the inverse distances $R_0$, $R_1\simeq r$ and the exponents $R_0\simeq r -x_0x/r$, $R_1\simeq r +x_0x/r$ allowing us to rewrite the electric fields of Eqs.~(\ref{E reflex}) and (\ref{E trans}) as follows 
\begin{eqnarray}
&& \mathbf{E}_{1} (\mathbf{r};\omega) = \frac{ \mathrm{i}q\omega \mu_0 \mu_1(\omega) \mathcal{K}_1(\theta,\omega)  }{ 4\pi r } e^{ \mathrm{i}k_1 r }  \nonumber\\
&& \times  \left\{ e^{ - \mathrm{i}k_1 x_0x/r } \left(\mathbf{e}_{z}-\cos\theta\mathbf{e}_{r,1} \right) \right. \nonumber\\
&&+ e^{ \mathrm{i}k_1 x_0x/r } \sin\theta \left[ \cos\phi_1 R_{\mathrm{TM,TM}}^{12}(\theta,\phi_1,\Delta_\Theta) \left( \mathbf{e}_{z}\times \mathbf{e}_{r,1} \right) \right. \nonumber\\
&& \left.\left. - R_{\mathrm{TE,TM}}^{12}(\theta,\phi_1,\Delta_\Theta)  \mathbf{e}_{\phi1}  \right] \right\}, \label{E rad reflex}
\end{eqnarray}
\begin{eqnarray}
&& \mathbf{E}_{2} (\mathbf{r};\omega) = \frac{ \mathrm{i}q\omega \mu_0 \mu_1(\omega)k_2 \mathcal{K}_2(\theta,\omega) }{ 4\pi k_1 r } e^{ \mathrm{i} ( k_2r + \xi x_0) } \nonumber\\
&& \times \sin\theta\left[ \cos\phi_2 T_{\mathrm{TM,TM}}^{12}(\theta,\phi_2,\Delta_\Theta) \left( \mathbf{e}_{z}\times \mathbf{e}_{r,2} \right) \right. \nonumber\\
&& \left. + T_{\mathrm{TE,TM}}^{12}(\theta,\phi_2,\Delta_\Theta)   \mathbf{e}_{\phi2} \right] , \label{E rad trans}
\end{eqnarray}
where $\mathbf{e}_{r1,2}=(\sin\theta\cos\phi_{1,2},\sin\theta\sin\phi_{1,2},\cos\theta)$ and $\mathbf{e}_{\phi1,2}=(-\sin\phi_{1,2},\cos\phi_{1,2},0)$. Here $\theta\in[0,\pi]$ denote the zenith angle of the observer at both sides of the interface. The azimuthal angle $\phi_{1,2}$ is defined for the upper hemisphere in the interval $I_{\mathrm{UH}}=[0,\pi/2]\cup[3\pi/2,2\pi]$ and $I_{\mathrm{LH}}=[\pi/2,3\pi/2]$ for the lower hemisphere. By means of these basis we appreciate that the free space contribution oscillates throughout the non-intuitive direction defined by $\mathbf{e}_{z}-\cos\theta\mathbf{e}_{r,1}$. Then, the contributions related to $R_{\mathrm{TM,TM}}^{12}$ and $T_{\mathrm{TM,TM}}^{12}$ oscillate in the direction $\mathbf{e}_{z}\times \mathbf{e}_{r1,2}$ outside the interface located at $x=0$. And, the contributions proportional to $R_{\mathrm{TE,TM}}^{12}$ and $T_{\mathrm{TE,TM}}^{12}$ oscillate perpendicular to the interface throughout the $\mathbf{e}_{\phi1,2}$ direction.

We notice that the three vectors $\mathbf{E}_{1,2}$, $\mathbf{B}_{1,2}$, and $\mathbf{e}_{r1,2}$ define a right-handed triad resulting in the Poynting vector for the material medium 1 and 2 \cite{OJF-SYB}. Then, the angular distribution of the radiated energy per unit frequency in the interval $z\in(-\zeta,\zeta)$ over the solid angle can be defined through the electric field $\mathbf{E}_{1,2}$ in the following fashion \cite{Jackson,Schwinger}:
\begin{equation} \label{Angular distribution}
\frac{ d^2 \mathcal{E}_j }{ d\omega d\Omega_j } = \sqrt{ \frac{ \varepsilon_0\varepsilon_j }{ \mu_0 \mu_j} } \frac{ r^2 }{ \pi } \mathbf{E}_j^*(\mathbf{r};\omega) \cdot \mathbf{E}_j(\mathbf{r};\omega) \;, \\
\end{equation}
with $j=1,2$ and $d\Omega_j=\sin\theta d\theta d\phi_j$ being the solid angle at the corresponding medium. 

Before computing the two angular distributions, we discuss an important tool that will simplify their calculations. In calculating the function $\mathcal{K}_1$  [Eq.~(\ref{Function K1})] and $\mathcal{K}_2$ [Eq.~(\ref{Function K2})], if we consider the limit $\zeta \gg v/\omega$ meaning effectively $\zeta\rightarrow \infty$, we encounter expressions like $\sin(\zeta aN)/(a N)$ which  behave as $\pi \delta(a N)$ \cite{Lighthill}. We will take advantage of this delta-like behavior by setting all terms involving $\cos\theta$ equal to $c/vn_j$. 

\subsection{Upper hemisphere}\label{ANGULAR UH}
After applying Eq.~(\ref{Angular distribution}) to the spherical contribution of the electric field (\ref{E rad reflex}), we find:
\begin{equation} \label{Angular distribution UH}
\frac{ d^2 \mathcal{E}_1 }{ d\omega d\Omega_1 } = \sum_{j=1}^{5} \frac{ d^2 \mathcal{E}_1^{ (j) } }{ d\omega d\Omega_1 } \;,
\end{equation}
with leading-order terms in $1-1/\beta^2n_1^2$ being
\begin{equation}\label{ADUH 1}
\frac{ d^2 \mathcal{E}_1^{(1)} }{ d\omega d\Omega_1 } = \sqrt{ \frac{ \varepsilon_0 \varepsilon_1 }{ \mu_0 \mu_1 } } \frac{ q^2 \omega^2 \mu_0^2 \mu^2_1 \mathcal{K}_1^2(\theta,\omega) }{ 16\pi^3 } \left( 1 - \frac{ 1 }{ \beta^2n_1^2 } \right) \;,
\end{equation}
\begin{eqnarray}
\frac{ d^2 \mathcal{E}_1^{(2)} }{ d\omega d\Omega_1 } &=& - 2 \frac{ d^2 \mathcal{E}_1^{(1)} }{ d\omega d\Omega_1 }\mathrm{Re}\left[ R_{\mathrm{TM,TM}}^{12}(\theta_1^{C},\phi_1,\Delta_\Theta) \right] \cos^2\phi_1 \nonumber \\
&& \times \cos\left( 2k_1 x_0 \sin\theta_1^{C} \cos\phi_1 \right)\;, \label{ADUH 2}
\end{eqnarray}
\begin{eqnarray}
&&\frac{ d^2 \mathcal{E}_1^{(3)} }{ d\omega d\Omega_1 } = \frac{2}{ \beta n_1 } \frac{ d^2 \mathcal{E}_1^{(1)} }{ d\omega d\Omega_1 } \cos\phi_1\sin\phi_1 \nonumber \\
&& \times \mathrm{Re}\left[ R_{\mathrm{TM,TM}}^{12}(\theta_1^{C},\phi_1,\Delta_\Theta) R_{\mathrm{TE,TM}}^{12\, *}(\theta_1^{C},\phi_1,\Delta_\Theta) \right], \quad\quad \label{ADUH 3}
\end{eqnarray}
\begin{equation}
\frac{ d^2 \mathcal{E}_1^{(4)} }{ d\omega d\Omega_1 } = \frac{ d^2 \mathcal{E}_1^{(1)} }{ d\omega d\Omega_1 } \big| R_{\mathrm{TM,TM}}^{12}(\theta_1^{C},\phi_1,\Delta_\Theta) \big|^2 \cos^2\phi_1 \;, \label{ADUH 4}
\end{equation}
\begin{equation}
\frac{ d^2 \mathcal{E}_1^{(5)} }{ d\omega d\Omega_1 } =  \frac{ d^2 \mathcal{E}_1^{(1)} }{ d\omega d\Omega_1 } \big| R_{\mathrm{TE,TM}}^{12}(\theta_1^{C},\phi_1,\Delta_\Theta) \big|^2 \;, \label{ADUH 5}
\end{equation}
where $\beta=v/c$, $\sin\theta_1^{C}=\sqrt{1 -  1/(\beta^2n_1^2) }$, and we assumed frequency-independent values for $\varepsilon_1, \mu_1, \varepsilon_2$, and $\mu_2$ in order to apply Eq.~(\ref{Angular distribution}), as well as in the rest of this paper. Here the $\delta$-like behavior of the function $\mathcal{K}_1$ described above was used implying an evaluation of the involved Fresnel coefficients at the Cherenkov angle $\theta_1^{C}$ [Eq.~(\ref{Cherenkov reflex})]. Also due to this application the characteristic factor $1-1/\beta^2n_1^2$ of the VC radiation appears as a global factor modulating the total angular distribution (\ref{Angular distribution UH}), which will always be positive. 

The important limit $\varepsilon_1=\varepsilon_2$, $\mu_1=\mu_2$, and $\Delta_\Theta\neq0$ describing a pure $\Delta_\Theta$ interface derived directly from Eqs.~(\ref{ADUH 1})--(\ref{ADUH 5}) reads as
\begin{equation}\label{ADUH topo}
\begin{aligned}
&\frac{ d^2 \mathcal{E}_1 }{ d\omega d\Omega_1 } = \frac{ d^2 \mathcal{E}_1^{(1)} }{ d\omega d\Omega_1 } \left[ 1 - \left( R_{\mathrm{TM,TM}}^{11} \right)^2 \cos^2\phi_1  \right. \\
&- 2  R_{\mathrm{TM,TM}}^{11} \cos^2\phi_1 \cos\left( 2k_1 x_0 \sin\theta_1^{C} \cos\phi_1 \right) \\
& \left. + \left( R_{\mathrm{TE,TM}}^{11} \right)^2  + \frac{2}{ \beta n_1 } R_{\mathrm{TM,TM}}^{11} R_{\mathrm{TE,TM}}^{11} \cos\phi_1\sin\phi_1 \right] ,
\end{aligned}
\end{equation}
where we remark that the Fresnel coefficients are now real constants determined by $\varepsilon_1,\mu_1$ and $\Delta_\Theta$.
\begin{figure*}[ht]
\centering
\subfloat[]{
\label{g:13}
\includegraphics[width=0.5\textwidth]{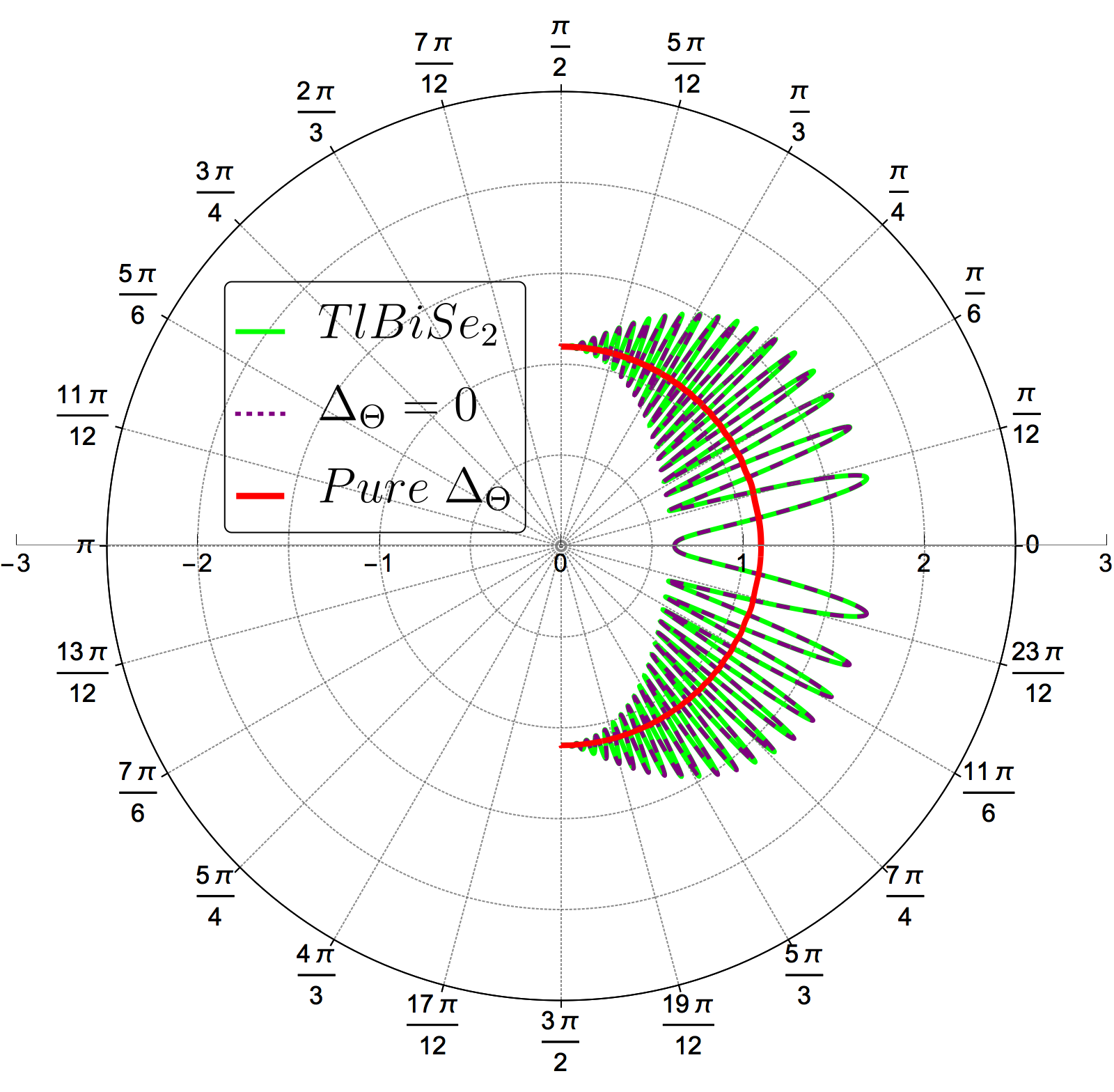}}
\subfloat[]{
\label{g:14}
\includegraphics[width=0.5\textwidth]{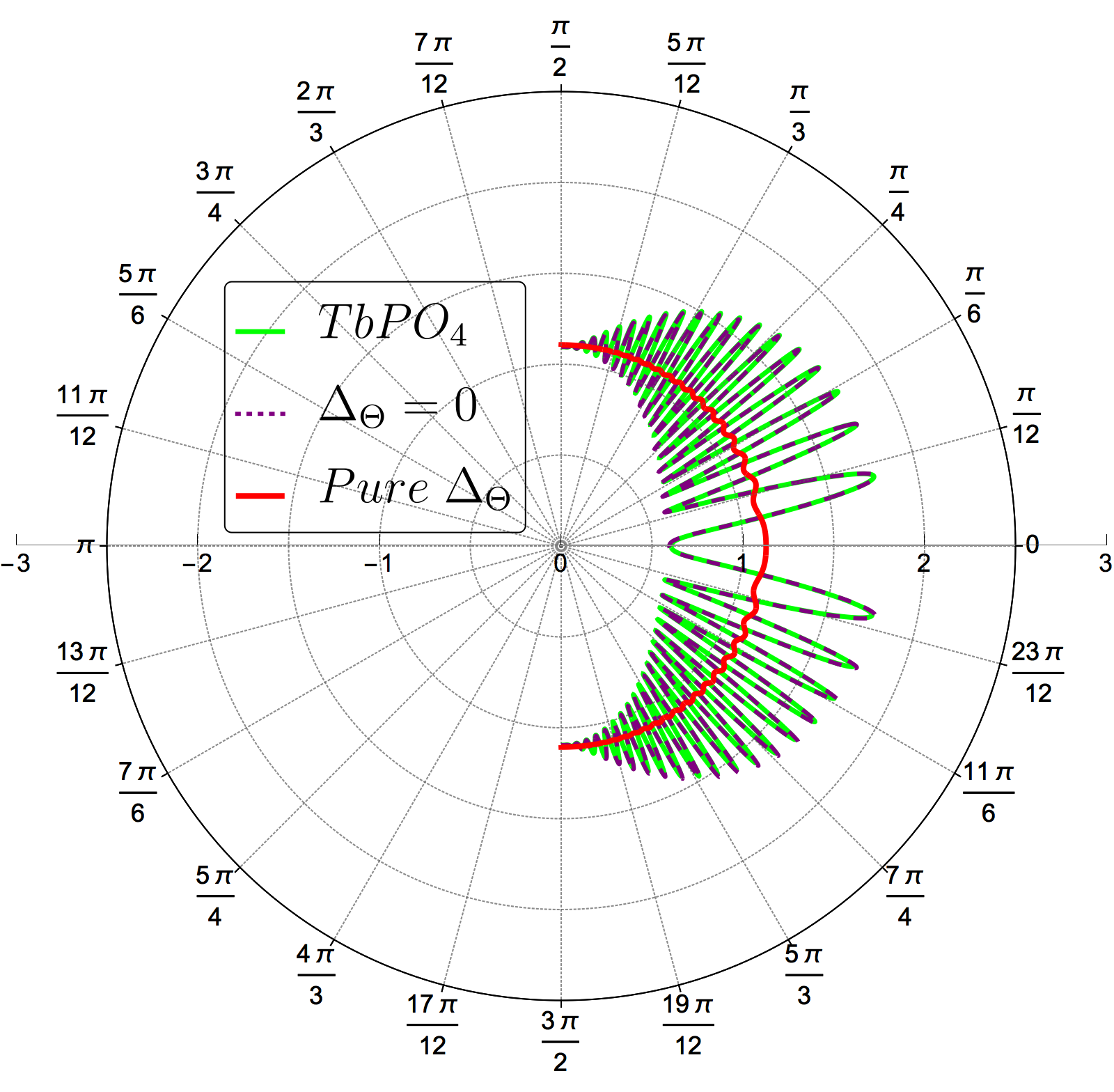}}

\caption{ Polar plots of the angular distribution of parallel VC radiation along the VC cone in the reflection zone generated by a particle with $v=0.95\,c$ parallel to the $z$ axis at a height $x_0=1\times10^{-4}\,\mathrm{m}$ from the interface and with single frequency $\omega= 6.0450\times10^{14}\,\, \mathrm{s}^{-1}$. A comparison between three different scenarios is considered. The purple line corresponds to a dielectric--dielectric interface ($\Delta_\Theta=0$) with different permittivities $\varepsilon_1\neq\varepsilon_2$. The green solid line represents the dielectric--TI interface ($\Delta_\Theta\neq0$). And, the red dotdashed line stands for the pure $\Delta_\Theta$ case when $\varepsilon_1=\varepsilon_2$ and keeping the parameter $\Delta_\Theta\neq0$. The polar plots (a) compare these scenarios when the topological insulator TlBiSe$_2$ is at the lower layer. Meanwhile the polar plots (b) compare these scenarios but with the magnetoelectric TbPO$_4$ located at the lower layer. Both materials are non magnetic ones. The radial axis indicates the dimensionless angular distribution in the respective direction. The charge moves along a vertical line that crosses the origin outside the sheet of paper. }
\label{PLOTS 1}
\end{figure*}

Some comments regarding this angular distribution are now in order. The expression (\ref{Angular distribution UH}), as can be contrasted with its addends (\ref{ADUH 1})--(\ref{ADUH 5}), has a linear contribution in the topological parameter $\Delta_\Theta$, depends explicitly on the particle's distance $x_0$ to the interface, and is neither an even function of the angle $\phi_1$ nor an odd one, the last fact being a consequence of the azimuthal symmetry breaking of the problem (recall Fig.~\ref{REGIONS}). The first term (\ref{ADUH 1}) is the free term corresponding to the standard VC radiation. The second term (\ref{ADUH 2}) is an interference term due to the change of phase gained from the electromagnetic field after reflection by the interface, whose interference effects can be disregarded if $\cos\left( 2k_1 x_0 \sin\theta_1^{C} \cos\phi_1 \right)\simeq 1$. The third term (\ref{ADUH 3}) arises from the interference between the TM,TM and the TM,TE polarizations and is linear in the topological parameter $\Delta_\Theta$. This is evidence of the topological magnetoelectric effect which is absent in the standard case ($\Delta_\Theta=0$). Nevertheless, if we apply the ideas of fields generated by dynamical images of the work \cite{OJF-LFU-ORT}, extensions of the original ones developed in Refs.~\cite{Qi Science,Wilczek}, this term is the result of the interference between the electric field of an image magnetic monopole of strength $g_1=qR_{\mathrm{TE,TM}}^{12}\sim -q\Delta_\Theta$ 
and the one generated by an image electric charge with magnitude $q'=qR_{\mathrm{TM,TM}}^{12}\sim q(\varepsilon_2-\varepsilon_1+\Delta_\Theta^2)$
. Then, the fourth term (\ref{ADUH 4}) describes the interaction of the image electric charge with the interface and analogously the last term (\ref{ADUH 5}) is related to the response of the interface to the moving image magnetic monopole.

In addition, we observe that although the parallel VC radiation will be always emitted in the typical VC cone (recall Figs.~\ref{Electric Field Patterns 1}), it will not be emitted uniformly because the functions involved in the terms (\ref{ADUH 2})-(\ref{ADUH 5}) depend explicitly on the azimuthal angle $\phi_1$, i.e. the parallel VC radiation is not uniform over different generatrices of the VC cone \cite{Bolotovskii}. Such non uniformity is shown in the polar plots of Figs.~\ref{g:13} and \ref{g:14}, where we plot the angular distribution (\ref{Angular distribution UH}) divided by $\mathcal{K}_1^2(\theta,\omega)\left(1-c^2/(v^2n_1^2)\right)\sqrt{\varepsilon_0}q^2\omega^2\mu_0^2/(16\pi^3\sqrt{\mu_0}) $ to appreciate the angular behavior, the interface lies on the line $(\pi/2-3\pi/2)$ and the charge moves along a vertical line that crosses the origin outside the sheet of paper with $v=0.95\,c$ at a height $x_0=1\times10^{-4}\,\mathrm{m}$ from the interface. Fig.~\ref{g:13} illustrates this by comparing three different interfaces: dielectric--dielectric ($\Delta_\Theta=0$), dielectric--TI and pure $\Delta_\Theta$ ($\varepsilon_1=\varepsilon_2$ and $\Delta_\Theta\neq0$). First, the dielectric--dielectric case having a dielectric with $\varepsilon_1=1.2$, $\mu_1=1$ at the upper layer and another dielectric with $\varepsilon_2=4$, $\mu_2=1$ at the lower layer corresponds to the dashed purple line. Second, the green solid line represents the dielectric--TI case with the same upper medium but the topological insulator TlBiSe$_2$ with $\varepsilon_2=4$, $\mu_2=1$, and $\Delta_\Theta=11\alpha$ at the lower layer \cite{TlBiSe2}. And, the red dotted-dashed line stands for the pure $\Delta_\Theta$ case given by Eq.~(\ref{ADUH topo}) with the upper and lower media having $\varepsilon_1=\varepsilon_2=1.2$, $\mu_1=\mu_2=1$, and $\Delta_\Theta=11\alpha$, which is smaller than the purple and green ones. In Fig.~\ref{g:13} we notice a big difference between the pure $\Delta_\Theta$ case and the other two, where the former one presents an almost uniform distribution of the radiation because the interference effects are suppressed due to the small value of $\Delta_\Theta$ in contrast with the other two angular distributions. This behavior reinforces the idea that the topological parameter mimics a permittivity as analyzed in Refs.~\cite{Crosse-Fuchs-Buhmann,OJF-SYB,OJF-LFU} for other kinds of radiation. 

In Fig.~\ref{g:14} we compare the same three cases but now for a material with a stronger topological parameter. Figure \ref{g:14} has the same specifications of Fig.~\ref{g:13}. Here, the dielectric--dielectric case (dashed purple line) has a standard dielectric medium with $\varepsilon_1=1.2$, $\mu_1=1$ at the upper layer and another standard dielectric with $\varepsilon_2=3.4969$, $\mu_2=1$ at the lower layer. Then, the dielectric--TI case (green solid line) is constituted by the same upper medium and the magnetoelectric TbPO$_4$ with $\varepsilon_2=3.4969$, $\mu_2=1$, and $\Delta_\Theta=0.22$ at the lower layer \cite{TbPO4,Rivera}. Lastly, the pure $\Delta_\Theta$ case (red dotted-dashed line) given by Eq.~(\ref{ADUH topo}) is presented here for $\varepsilon_1=\varepsilon_2=1.2$, $\mu_1=\mu_2=1$, and $\Delta_\Theta=0.22$ being smaller greater than the purple and the green lines. Interestingly, we appreciate an unexpected asymmetry with respect to the polar axis between the angular distribution of the dielectric--dielectric case and the one corresponding to the magnetoelectric TbPO$_4$, because the latter one results to be slightly bigger than the former one in the interval $\phi_1\in[0,\pi/2]$ and vice versa in the interval $\phi_1\in[3\pi/2,2\pi]$. This asymmetry is also present in Fig.~\ref{g:13} but is difficult to appreciate there and is absent in the dielectric--dielectric case. We can trace back its origin to the interference term given by Eq.~(\ref{ADUH 3}) which is sensible to a sign change in $\Delta_\Theta$ as certain strong 3D TIs present. We observe that this asymmetry can be sharper here when the interference pattern coming from the term (\ref{ADUH 2}) is suppressed, i.e., when $\cos\left( 2k_1 x_0 \sin\theta_1^{C} \cos\phi_1 \right)\simeq 1$. Regarding the pure $\Delta_\Theta$ case, we predominantly observe an inhomogeneous distribution whose interference pattern becomes sharp when $\phi_1\rightarrow0$ due to the higher value of the topological parameter $\Delta_\Theta$.

\subsection{Lower hemisphere}\label{ANGULAR LH}
Here we apply Eq.~(\ref{Angular distribution}) to the spherical contribution of the electric field (\ref{E rad trans}) and after assuming that $n_1<n_2$, we find:
\begin{equation} \label{Angular distribution LH}
\frac{ d^2 \mathcal{E}_2 }{ d\omega d\Omega_2 } = \sum_{j=1}^{3} \frac{ d^2 \mathcal{E}_2^{(j)} }{ d\omega d\Omega_2 }
\end{equation}
having the following leading-order terms in $1-1/\beta^2n_2^2$: 
\begin{eqnarray}
&&\frac{ d^2 \mathcal{E}_2^{(1)} }{ d\omega d\Omega_2 } = 2 \sqrt{ \frac{ \varepsilon_0 \varepsilon_2 }{ \mu_0 \mu_2 } } \frac{ q^2 \omega^2 \mu_0^2 \mu^2_1 k_2^2 \mathcal{K}_2^2(\theta,\omega) }{ 16 \pi^3 k_1^2 \beta n_2  } \left( 1 - \frac{ 1 }{ \beta^2n_2^2 } \right)  \nonumber\\
&& \times \mathrm{Re}\left[ T_{\mathrm{TM,TM}}^{12}(\theta_2^{C},\phi_2,\Delta_\Theta) T_{\mathrm{TE,TM}}^{12\,*}(\theta_2^{C},\phi_2, \Delta_\Theta) \right] \nonumber \\
&& \times e^{-2 \Phi x_0 } \cos\phi_2\sin\phi_2 \;, \label{ADLH 1}
\end{eqnarray}
\begin{eqnarray}
\frac{ d^2 \mathcal{E}_2^{(2)} }{ d\omega d\Omega_2 } &=& \sqrt{ \frac{ \varepsilon_0 \varepsilon_2 }{ \mu_0 \mu_2 } } \frac{ q^2 \omega^2 \mu_0^2 \mu^2_1 k_2^2 \mathcal{K}_2^2(\theta,\omega) }{ 16 \pi^3 k_1^2 } \left( 1 - \frac{ 1 }{ \beta^2n_2^2 } \right)  \nonumber\\
&& \times \big| T_{\mathrm{TM,TM}}^{12}(\theta_2^{C},\phi_2, \Delta_\Theta) \big|^2 \cos^2\phi_2 \,  e^{-2 \Phi x_0 } , \;\;\; \quad
\label{ADLH 2}
\end{eqnarray}
\begin{eqnarray}
\frac{ d^2 \mathcal{E}_2^{(3)} }{ d\omega d\Omega_2 } &=& \sqrt{ \frac{ \varepsilon_0 \varepsilon_2 }{ \mu_0 \mu_2 } } \frac{ q^2 \omega^2 \mu_0^2 \mu^2_1 k_2^2 \mathcal{K}_2^2(\theta,\omega) }{ 16 \pi^3 k_1^2 } \left( 1 - \frac{ 1 }{ \beta^2n_2^2 } \right)  \nonumber\\
&& \times  \big| T_{\mathrm{TE,TM}}^{12}(\theta_2^C,\phi_2,\Delta_\Theta) \big|^2 e^{-2 \Phi x_0 } \;, \label{ADLH 3}
\end{eqnarray}
where we defined
\begin{equation}\label{Phi}
\Phi=\frac{\omega}{v}\sqrt{ 1 - \beta^2n_1^2  + \sin^2\phi_2 \left( \beta^2n_2^2  - 1 \right)  } \;.
\end{equation}
To obtain Eqs.(\ref{ADLH 1})--(\ref{ADLH 3}) we employed the same mathematical procedure and assumptions of Sec. \ref{ANGULAR UH}. Consequently, the characteristic factor $1-1/\beta^2n_2^2$ of the VC radiation appears as a global factor modulating the total angular distribution (\ref{Angular distribution LH}), which will always be positive. Importantly, results the global factor $e^{-2 \Phi x_0 }$ coming directly from the factor $e^{ \mathrm{i} \xi x_0 }$ of the electric field (\ref{E rad trans}), which is a great difference in comparison with its upper hemisphere analog and becomes a real decaying exponential due to the assumption $n_1<n_2$, otherwise, it simply cancels out with its conjugate in Eq.~(\ref{Angular distribution}). Thus,  parallel VC radiation is highly suppressed in this region for distances far away of the interface as the corresponding electric field plot shows in Fig.~\ref{Electric Field Patterns 2}. We will address this issue in the next section regarding the radiated energy.

The relevant limit $\varepsilon_1=\varepsilon_2$, $\mu_1=\mu_2$ and $\Delta_\Theta\neq0$ describing a pure $\Delta_\Theta$ interface derived directly from Eqs.~(\ref{ADLH 1})--(\ref{ADLH 3}) reads as
\begin{equation}\label{ADLH topo}
\begin{aligned}
 \frac{ d^2 \mathcal{E}_2 }{ d\omega d\Omega_2 } &= \sqrt{ \frac{ \varepsilon_0 \varepsilon_2 }{ \mu_0 \mu_2 } } \frac{ q^2 \omega^2 \mu_0^2 \mu^2_2 \mathcal{K}_2^2(\theta,\omega) }{ 16 \pi^3 } \left( 1 - \frac{ 1 }{ \beta^2n_2^2 } \right) \\
& \times \left[ \left( T_{\mathrm{TM,TM}}^{22} \right)^2 \cos^2\phi_2 + \left( T_{\mathrm{TE,TM}}^{22} \right)^2 \right. \\
& \left. + \frac{2}{ \beta n_2 } T_{\mathrm{TM,TM}}^{22} T_{\mathrm{TE,TM}}^{22} \cos\phi_2\sin\phi_2 \right] \;,
\end{aligned}
\end{equation}
where we emphasize that the Fresnel coefficients are real numbers and the absence of the exponential factor.

%
%
%
\begin{figure*}[ht]
\centering
\subfloat[]{
\label{g:15}
\includegraphics[width=0.5\textwidth]{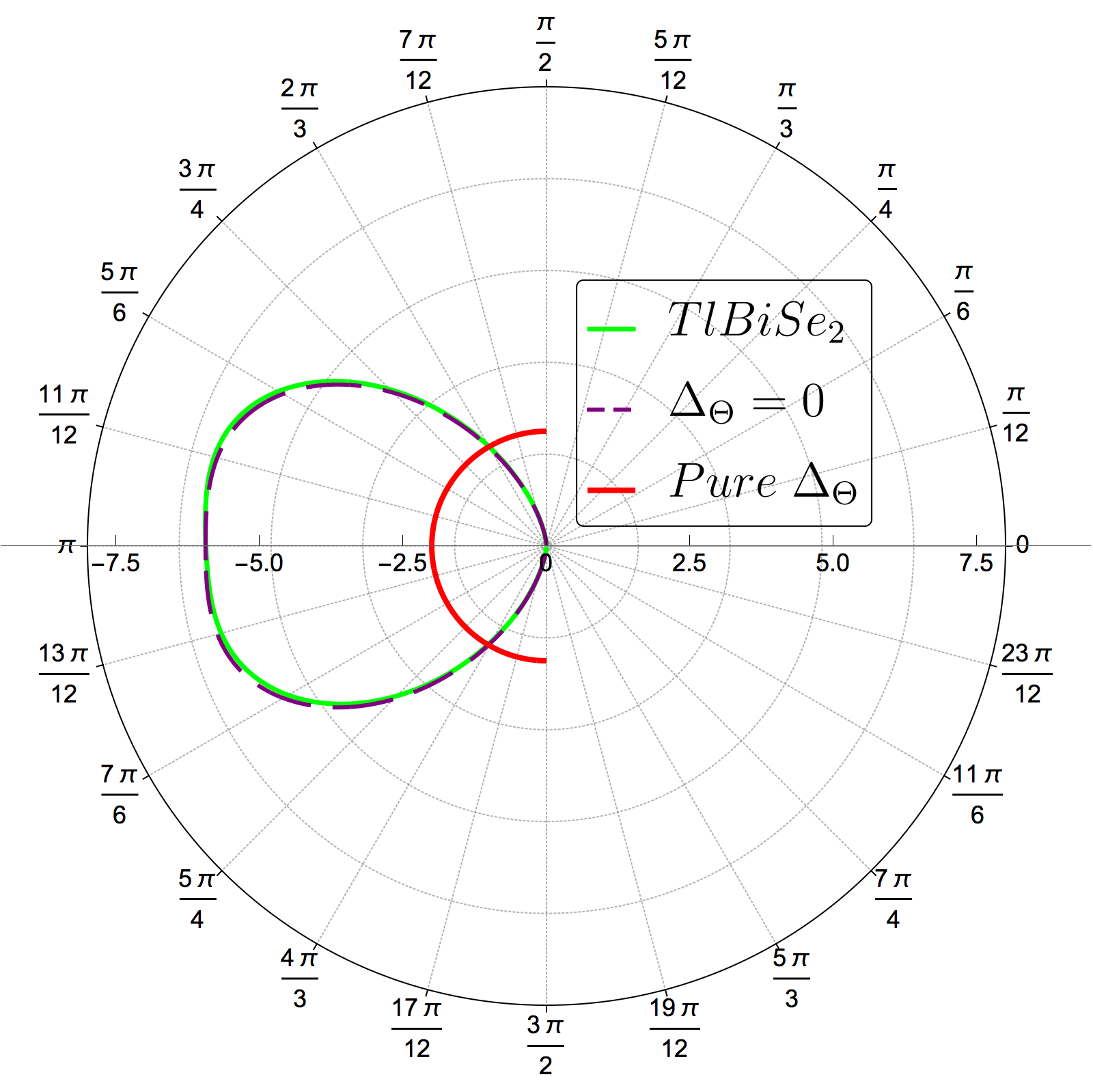}}
\subfloat[]{
\label{g:16}
\includegraphics[width=0.5\textwidth]{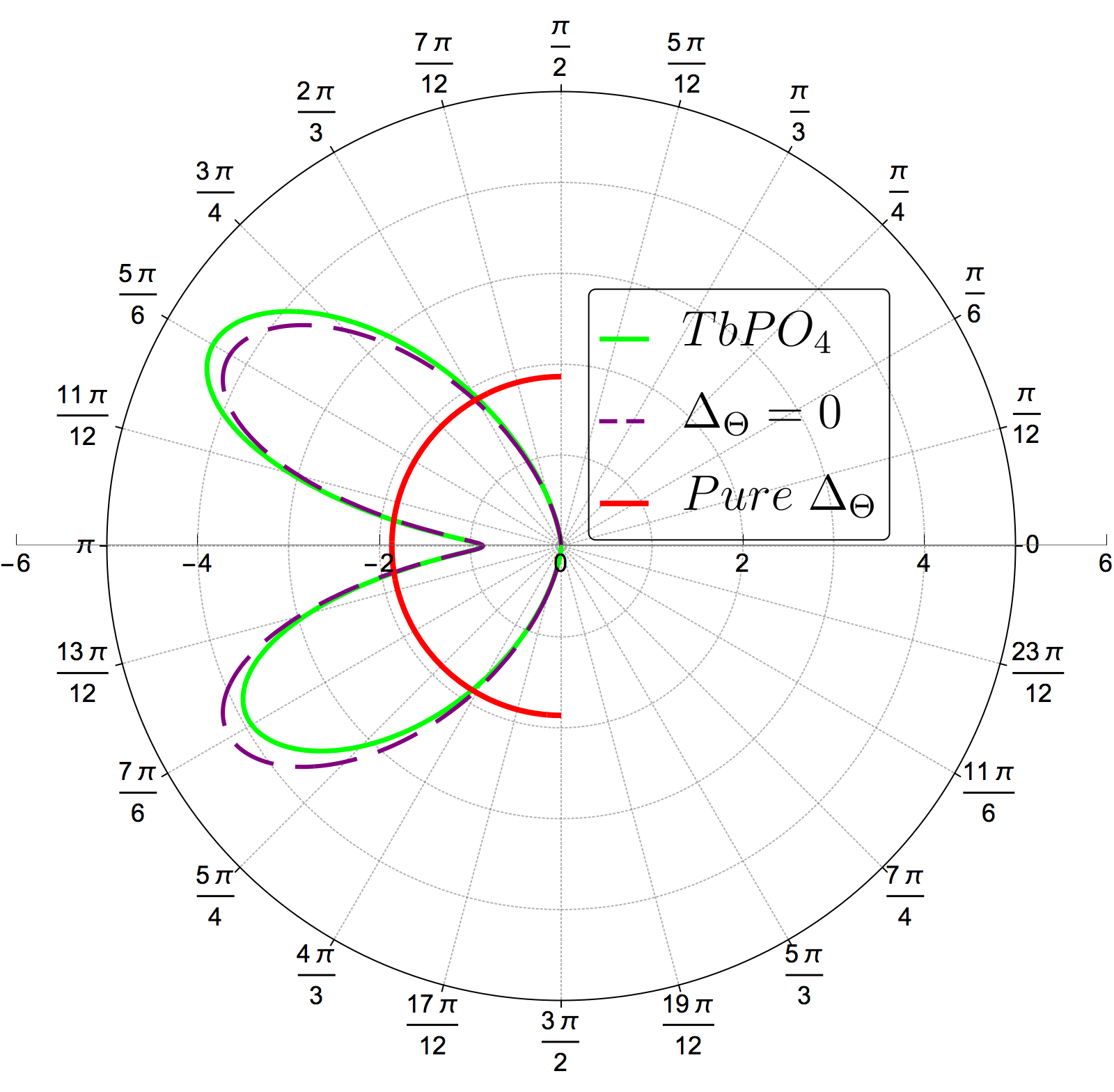}}

\caption{ Polar plots of the angular distribution of parallel VC radiation along the VC cone in the transmission zone generated by a particle with single frequency $\omega= 6.0450\times10^{14}\,\, \mathrm{s}^{-1}$. A comparison between three different scenarios is considered. The purple line corresponds to a dielectric--dielectric interface ($\Delta_\Theta=0$) with different permittivities $\varepsilon_1\neq\varepsilon_2$. The green solid line represents the dielectric--TI interface ($\Delta_\Theta\neq0$). The red dotted-dashed line stands for the pure $\Delta_\Theta$ case when $\varepsilon_1=\varepsilon_2$ and keeping the parameter ($\Delta_\Theta\neq0$). The polar plots (a) compare these scenarios when the topological insulator TlBiSe$_2$ is at the lower layer with a charge velocity $v=0.75\,c$. Meanwhile the polar plots (b) compare these scenarios but with the magnetoelectric TbPO$_4$ located at the lower layer and with a charge velocity $v=0.9\,c$. Both materials are non magnetic ones. The radial axis indicates the dimensionless angular distribution in the respective direction. The charge moves along a vertical line that crosses the origin outside the sheet of paper. }
\label{PLOTS 2}
\end{figure*}

Notwithstanding that the angular distribution (\ref{Angular distribution LH}) has almost the same characteristics as its upper hemisphere counterpart, we appreciate the absence of the free term, since the particle moves at the other side of the interface. By adapting the same arguments of the previous section, we can understand the three terms (\ref{ADLH 1})-(\ref{ADLH 3}) via the images interpretation but now the corresponding image magnetic monopole has the strength $g_2=qT_{\mathrm{TE,TM}}^{12}\sim q\Delta_\Theta$ 
and the magnitude of the image electric charge is $q''=qT_{\mathrm{TM,TM}}^{12}\sim q\varepsilon_1 n_2/n_1$. 

As in the reflective region, despite the parallel VC radiation always being emitted in the typical VC cone (recall Figs.~\ref{Electric Field Patterns 2}), the non uniform radiation emission over different generatrices of the VC cone is present because the functions involved in the terms (\ref{ADLH 1})--(\ref{ADLH 3}) depend explicitly on the azimuthal angle $\phi_2$  \cite{Bolotovskii}. We show this non-uniformity in the polar plots of Figs.~\ref{g:15} and \ref{g:16}, where we plot the angular distribution (\ref{Angular distribution LH}) divided by $\mathcal{K}_2^2(\theta,\omega)e^{-2 \Phi x_0 }\left(1-c^2/(v^2n_2^2)\right)\sqrt{\varepsilon_0}q^2\omega^2\mu_0^2/(16\pi^3\sqrt{\mu_0}) $ to appreciate the angular behavior. Figs.~\ref{g:15} and \ref{g:16} have the same specifications of Figs.~\ref{g:13} and \ref{g:14} and compare the same three different interfaces: dielectric--dielectric ($\Delta_\Theta=0$), dielectric--TI and  pure $\Delta_\Theta$ ($\varepsilon_1=\varepsilon_2$ and $\Delta_\Theta\neq0$), but they display two different velocities $v=0.75\,c$ and $v=0.9\,c$ respectively. The pure $\Delta_\Theta$ case (red dotted-dashed line) of Eq.~(\ref{ADLH topo}) results to be clearly smaller than the purple and green ones being the opposite situation of Figs.~\ref{g:13} and \ref{g:14}. Here we notice that the pure $\Delta_\Theta$ case presents an uniform distribution of the radiation expected in a homogeneous medium in contrast with the other two angular distributions.

Finally, we highlight the enhancement of the asymmetry with respect to the polar axis between the angular distribution of the standard case and the one corresponding to both the topological insulator TlBiSe$_2$ and the magnetoelectric TbPO$_4$ in comparison to what is observed at the reflection region in Fig.~\ref{PLOTS 1}. Consistently, this asymmetry is due again to the interference term given by Eq.~(\ref{ADLH 1}) and is of course sensible to a sign change in $\Delta_\Theta$ too.

\section{Radiated Energy} \label{ENERGY}
In this section we calculate the energy per unit frequency radiated by the charge on its path from $-\zeta$ to $+\zeta$ at both sides of the interface. To this end, we must integrate over the solid angle $\Omega_j$ to obtain $d\mathcal{E}_j/d\omega$.

\subsection{Upper hemisphere}\label{ENERGYUH}
By integrating the angular distribution of energy (\ref{Angular distribution UH}) over the solid angle $\Omega_1$ in the upper hemisphere, we have
\begin{equation}\label{TREUH}
\frac{ d \mathcal{E}_1 }{ d\omega } = \sum_{j=1}^{5} \int_{\Omega_{1}} \frac{ d^2 \mathcal{E}_1^{(j)} }{ d\omega d\Omega_1 } \;,
\end{equation}
which has the following five at order $1-1/\beta^2n_1^2$:  
\begin{equation}\label{TREUH 1}
\frac{ d \mathcal{E}_1^{(1)} }{ d\omega} = \frac{ q^2 \omega  \mu_1 L }{ 8\pi \varepsilon_0 c^2 } \left( 1 - \frac{ 1 }{ \beta^2n_1^2 } \right) \;,
\end{equation}
\begin{eqnarray}
\frac{ d \mathcal{E}_1^{(2)} }{ d\omega } &=& - \frac{2}{\pi} \frac{ d \mathcal{E}_1^{(1)} }{ d\omega } \int_{I_{\mathrm{UH}}} d\phi_1 \cos\left( 2k_1 x_0 \sin\theta_1^{C} \cos\phi_1 \right)  \nonumber\\
&& \times \mathrm{Re}\left[ R_{\mathrm{TM,TM}}^{12}(\theta_1^{C},\phi_1,\Delta_\Theta) \right] \cos^2\phi_1 \;, \label{TREUH 2}
\end{eqnarray}
\begin{equation}\label{TREUH 3} 
\frac{ d \mathcal{E}_1^{(3)} }{ d\omega } = 0 \;,
\end{equation}
\begin{equation}\label{TREUH 4}
\frac{ d \mathcal{E}_1^{(4)} }{ d\omega } = \frac{ 1 }{ \pi } \frac{ d \mathcal{E}_1^{(1)} }{ d\omega } \int_{I_{\mathrm{\mathrm{UH}}}} d\phi_1  \cos^2\phi_1 \big| R_{\mathrm{TM,TM}}^{12}(\theta_1^{C},\phi_1,\Delta_\Theta) \big|^2 ,
\end{equation}
\begin{equation}\label{TREUH 5}
\frac{ d \mathcal{E}_1^{(5)} }{ d\omega } = \frac{ 1 }{ \pi } \frac{ d \mathcal{E}_1^{(1)} }{ d\omega } \int_{I_{\mathrm{UH}}} d\phi_1 \big| R_{\mathrm{TE,TM}}^{12}(\theta_1^{C},\phi_1,\Delta_\Theta) \big|^2 \;,
\end{equation}
where we recalled that $I_{\mathrm{UH}}=[0,\pi/2]\cup[3\pi/2,2\pi]$ and introduced the total length $L=2\zeta$ traveled by the particle, thus obtaining that Eq.~(\ref{TREUH 1}) is a half of the standard result \cite{Panofsky} because we are restricted to the upper hemisphere. For full details, please see Appendix \ref{D}.

Now, we will carry out the integrals over $\phi_1$ for the interesting case when the parallel VC radiation presents total internal reflection at the upper medium, which is of practical interest for waveguides. To this end, we choose a TI-vacuum interface with a nonmagnetical strong 3D TI with $\mu_1=1$ as upper medium. As discussed in Sec. \ref{CHARGE}, this will mean that all the radiation is reflected and the VC radiation is absent in vacuum (the transmitted region $x<0$). However, the integral of the second term of the angular distribution given by Eq.~(\ref{ADUH 2}) is very difficult to solve analytically but this difficulty can be overcome by assuming $\cos\left( 2k_1 x_0 \sin\theta_1^{C} \cos\phi_1 \right)\simeq 1$. After transforming the interval of integration $I_{\mathrm{UH}}$ to $[-\pi/2,\pi/2]$ the resulting integrals at order $\Delta_\Theta^2$ are 
%
\begin{eqnarray}
\frac{ d \mathcal{E}_1^{(2)} }{ d\omega } &=& \frac{ q^2 \omega L }{ 8\pi \varepsilon_0 c^2 } \frac{ 1 }{ (\varepsilon_1 + 1) } \left\{ \left( 1 - \frac{ 1 }{ \beta^2n_1^2 } \right)  \mathcal{R}_1(\varepsilon_1, \Delta_\Theta) \right. \nonumber\\
&& +\frac{ 2 \varepsilon_1 \mathcal{R}_2(\varepsilon_1, \Delta_\Theta) }{ \varepsilon_1 + 1 } \left[ 1 - \frac{ \sqrt{\varepsilon_1(\varepsilon_1 -1)} }{ \Gamma(\varepsilon_1, \beta, \Delta_\Theta) } \right] \nonumber\\
&& \left. + \frac{ \varepsilon_1(\varepsilon_1 -1) }{ \varepsilon_1 - 1 - 2\Delta_\Theta^2 } \left[ \frac{ \sqrt{\varepsilon_1(\varepsilon_1 -1)} }{ \Gamma(\varepsilon_1, \beta, \Delta_\Theta) } -1 \right] \right\} \;, \label{TREUH final 2}
\end{eqnarray}
\begin{eqnarray}
\frac{ d \mathcal{E}_1^{(4)} }{ d\omega } &=& \frac{ q^2 \omega L }{ 16 \pi \varepsilon_0 c^2 } \frac{ 1 }{ (\varepsilon_1 + 1) } \nonumber\\
&& \times \left\{ \left( 1 - \frac{ 1 }{ \beta^2n_1^2 } \right)  \left(\varepsilon_1 - 1\right) \mathcal{R}_3(\varepsilon_1, \Delta_\Theta) \right. \nonumber\\
&& +\frac{ 2 \left(\varepsilon_1-1\right) \mathcal{R}_4(\varepsilon_1, \Delta_\Theta) }{ \varepsilon_1 + 1 } \left[ 1 - \frac{ \sqrt{\varepsilon_1(\varepsilon_1 -1)} }{ \Gamma(\varepsilon_1, \beta, \Delta_\Theta) } \right]  \nonumber\\
&& + \left. \frac{ 2\varepsilon_1 (\varepsilon_1 - 1) }{ (\varepsilon_1 - 1 - 2\Delta_\Theta^2) } \left[ \frac{ \sqrt{\varepsilon_1 (\varepsilon_1 - 1)} }{ \Gamma(\varepsilon_1, \beta, \Delta_\Theta) } -1 \right] \right\} , \;\;\quad \label{TREUH final 4}
\end{eqnarray}

\begin{eqnarray}
\frac{ d \mathcal{E}_1^{(5)} }{ d\omega } &=& \frac{ q^2 \omega L }{ 2 \pi \varepsilon_0 c^2 } \left( 1 - \frac{ 1 }{ \beta^2n_1^2 } \right) \frac{ \Delta_\Theta^2 }{ (\varepsilon_1 + 1)(\varepsilon_1 - 1 - 2\Delta_\Theta^2) } \nonumber\\
&& \times \left[ \frac{ \sqrt{\varepsilon_1(\varepsilon_1 -1)} }{ \Gamma(\varepsilon_1, \beta, \Delta_\Theta) } -1 \right], \label{TREUH final 5}
\end{eqnarray}
where we have defined
\begin{eqnarray}
&& \mathcal{R}_1(\varepsilon_1, \Delta_\Theta) = \frac{ \varepsilon_1^2 + 1 + 2\Delta_\Theta^2 }{ \varepsilon_1 - 1 - 2\Delta_\Theta^2 }\;, \\
&& \mathcal{R}_2(\varepsilon_1, \Delta_\Theta) = \frac{ \varepsilon_1^2 + 1 + 2\Delta_\Theta^2 }{ \varepsilon_1 - 1 - 4\Delta_\Theta^2 }\;, \\
&& \mathcal{R}_3(\varepsilon_1, \Delta_\Theta) = \frac{ \varepsilon_1 + 1 + 2\Delta_\Theta^2 }{ \varepsilon_1 - 1 - 2\Delta_\Theta^2 }\;, \\
&& \mathcal{R}_4(\varepsilon_1, \Delta_\Theta) = \frac{ \varepsilon_1 + 1 + 2\Delta_\Theta^2 }{ \varepsilon_1 - 1 - 4\Delta_\Theta^2 }\;, 
\end{eqnarray}
%
\begin{eqnarray}
&&\Gamma(\varepsilon_1, \beta, \Delta_\Theta) = \nonumber\\
&& \sqrt{ \varepsilon_1 (\varepsilon_1 - 1) - (\varepsilon_1 + 1)\left( 1 - \frac{ 1 }{ \beta^2n_1^2 } \right)(\varepsilon_1 - 1 - 2\Delta_\Theta^2) }, \nonumber\\ \label{Gamma}
\end{eqnarray}
with $\Gamma(\varepsilon_1, \beta, \Delta_\Theta)$ taken as a real function. In this way, the radiated energy at the upper hemisphere (\ref{TREUH}) will consist of the sum of (\ref{TREUH 1}) with $\mu_1=1$ and Eqs.~(\ref{TREUH final 2})--(\ref{TREUH final 5}), which will be physically acceptable only when positive. We observe that due to the vanishing interference term (\ref{TREUH 3}), the radiated energy is not sensitive to sign changes on $\Delta_\Theta$.

\subsection{Lower hemisphere}
Analogously, through the method described immediately above, the energy per unit frequency radiated by the charge in the lower hemisphere results
\begin{equation}\label{TRELH}
\frac{ d \mathcal{E}_2 }{ d\omega } = \sum_{j=1}^{3} \int_{\Omega_{2}} \frac{ d^2 \mathcal{E}_2^{(j)} }{ d\omega d\Omega_2 } \;,
\end{equation}
where the adding terms at order $1-1/\beta^2n_2^2$ are the following 
\begin{eqnarray}
\frac{ d \mathcal{E}_2^{(1)} }{ d\omega } &=& 0 ,\; \label{TRELH 1} 
%
%
%
%
\end{eqnarray}
\begin{eqnarray}
\frac{ d \mathcal{E}_2^{(2)} }{ d\omega } &=& \frac{ q^2 \omega  \mu_1 \varepsilon_2  L }{ 8\pi^2 \varepsilon_0 \varepsilon_1 c^2 } \left( 1 - \frac{ 1 }{ \beta^2n_2^2 } \right) \int_{\pi/2}^{3\pi/2} d\phi_2  e^{-2 \Phi x_0 } \nonumber\\ 
&& \times \big| T_{\mathrm{TM,TM}}^{12}(\theta_2^{C},\phi_2, \Delta_\Theta) \big|^2 \cos^2\phi_2 \;, \label{TRELH 2}
\end{eqnarray}
\begin{eqnarray}
\frac{ d \mathcal{E}_2^{(3)} }{ d\omega } &=& \frac{ q^2 \omega  \mu_1 \varepsilon_2  L }{ 8\pi^2 \varepsilon_0 \varepsilon_1 c^2 } \left( 1 - \frac{ 1 }{ \beta^2n_2^2 } \right) \nonumber\\ 
&& \times \int_{\pi/2}^{3\pi/2} d\phi_2  e^{-2 \Phi x_0 } \big| T_{\mathrm{TE,TM}}^{12}(\theta_2^C,\phi_2,\Delta_\Theta) \big|^2 .\nonumber\\\label{TRELH 3}
\end{eqnarray}

The first integral corresponding to the interference term is zero because the integrand is an odd function and after a change of variable one may shift the interval of integration to $[-\pi/2,\pi/2]$. The integrals can be carried out if we consider only the motion of a particle moving very close to the interface. So, from the exponential factor and Eq.~(\ref{Phi}) we find that the cutoff frequency at which the total radiated energy at the angle $\phi_2$ begins to be sharply suppressed is given by 
\begin{equation}\label{omega cut}
\omega_{\mathrm{cut}} = \frac{ v }{ 2x_0 \sqrt{ 1 - n_1^2\beta^2 + \sin^2\phi_2\left( n_2^2\beta^2 - 1 \right) } } \;,
\end{equation}
which is exactly the same cutoff frequency reported by Bolotovskii \cite{Bolotovskii} but with $n_1\neq1$. This coincidence and the absence of a dependency on the topological parameter $\Delta_\Theta$ can be understood by recalling that $\omega_{\mathrm{cut}}$ arises from $\Phi$ [Eq.~(\ref{Phi})], which in turn originates from the exponent $\xi$ of the electric field [Eq.~(\ref{E rad trans})]. Since the latter is characterized only by the refractive indices defined through bulk properties, therefore, $\omega_{\mathrm{cut}}$ must be unaffected by the topological parameter $\Delta_\Theta$.

At this point, we will focus on the situation when the particle moves in vacuum parallel to a nonmagnetical strong three-dimensional TI interface, which is of practical interest for particle detectors. In this way, the VC radiation in vacuum (the reflected region $x>0$) will be automatically ruled out and all the possible radiation will be transmitted to the TI. So, by assuming the condition (\ref{omega cut}) over the field's frequency, we can set the exponential factor equal to one and work out the integral over $\phi_2$ in the two remaining expressions (\ref{TRELH 2}) and (\ref{TRELH 3}) obtaining at order $\Delta_\Theta^2$: 
\begin{equation}\label{TRELH final 2}
\begin{aligned}
&\frac{ d \mathcal{E}_2^{(2)} }{ d\omega } = \frac{ q^2 \omega  \varepsilon_2  L }{ 2\pi \varepsilon_0 c^2 } \left\{ -\frac{ 1 }{ 2(\varepsilon_2 + 1) } \right. \\
& + \frac{ 1 - \beta^{-2} }{ (\varepsilon_2 + 1)(\varepsilon_2 - 1 - 2\Delta_\Theta^2) } \left[ 1 - \frac{ \sqrt{\varepsilon_2(\varepsilon_2 -1)} }{ \Gamma(\varepsilon_2, \beta, \Delta_\Theta) } \right] \\
& + \left. \frac{ \beta^2 \varepsilon_2^3 \Gamma(\varepsilon_2, \beta, \Delta_\Theta) }{ (\varepsilon_2 + 1)(\varepsilon_2 - 1 - 2\Delta_\Theta^2) } 
%
\left[ \frac{ \varepsilon_2 - 1 }{ \Gamma(\varepsilon_2, \beta, \Delta_\Theta) } - \frac{ \sqrt{\varepsilon_2 -1} }{2\varepsilon_2 \sqrt{\varepsilon_2} }  \right]  \right\},
\end{aligned}
\end{equation}
%
%
%
\begin{eqnarray}
\frac{ d \mathcal{E}_2^{(3)} }{ d\omega } &=& \frac{ q^2 \omega  \varepsilon_2  L }{ 2\pi \varepsilon_0 c^2 } \left( 1 - \frac{ 1 }{ \beta^2\varepsilon_2 } \right) \frac{ \Delta_\Theta^2 }{ (\varepsilon_2 + 1)(\varepsilon_2 - 1 - 2\Delta_\Theta^2) } \nonumber\\
&& \times \left[ \frac{ \sqrt{\varepsilon_2(\varepsilon_2 -1)} }{ \Gamma(\varepsilon_2, \beta, \Delta_\Theta) } -1 \right], \label{TRELH final 3}
\end{eqnarray}
where $\Gamma(\varepsilon_2, \beta, \Delta_\Theta)$ was defined on Eq.~(\ref{Gamma}) and also the interval of integration was shifted to $[-\pi/2,\pi/2]$. Again $\Gamma(\varepsilon_2, \beta, \Delta_\Theta)$ must be real in expressions (\ref{TRELH final 2}) and (\ref{TRELH final 3}). By adding these terms, the radiated energy at the lower hemisphere (\ref{TRELH}) can be found, which will be physically consistent when positive. And, as found in the radiated energy at the upper hemisphere, this radiated energy is insensible to changes on the topological parameter $\Delta_\Theta$ because of the vanishing interference term (\ref{TRELH 1}).

Inspecting Eqs.~(\ref{TREUH final 2})--(\ref{TREUH final 5}) for the radiated energy in the upper hemisphere and the corresponding ones for the lower hemisphere given by Eqs.~(\ref{TRELH final 2}) and (\ref{TRELH final 3}), we appreciate an intricate algebraic dependency on the velocity as  Bolotovskii's single result \cite{Bolotovskii}. Despite the similarities, they do not coincide in the dielectric--dielectric case ($\Delta_\Theta=0$) because Bolotovskii's expression is valid in all space resulting in it being impossible to distinguish correctly between the reflection and transmissive contributions of the electric field. In contrast, ours is naturally split into reflection and transmissive contributions due to the Green's function method. The final comment of this section is to stress that this algebraic dependency on the velocity is completely different in comparison with the logarithmic one that we obtained also by means of the Green's function method for transition radiation (perpendicular motion towards the interface) near this kind of materials \cite{OJF-SYB}.

\subsection{Pure $\Delta_\Theta$ interface}\label{ENERGY Pure Delta}
It is worth to study the radiated energy for the pure $\Delta_\Theta$ case, i.e., an interface with equal permittivities $\varepsilon_1=\varepsilon_2$, same permeabilities $\mu_1=\mu_2$ but $\Delta_\Theta\neq 0$. Consequently, the involved reflection and transmission coefficients no longer depend on the angles and become constants. Thus, after solving the integrals (\ref{TREUH 2})--(\ref{TREUH 5}) including interference effects, the radiated energy at the upper hemisphere (\ref{TREUH}) results in
\begin{eqnarray}
&&\frac{ d \mathcal{E}_1 }{ d\omega } = \frac{ q^2 \omega  \mu_1 L }{ 8\pi \varepsilon_0 c^2 } \left( 1 - \frac{ 1 }{ \beta^2n_1^2 } \right) \nonumber\\
&& \times \left\{ 1+ \frac{1}{2} \left( R_{\mathrm{TM,TM}}^{11} \right)^2 + \left( R_{\mathrm{TE,TM}}^{11} \right)^2  - 2  R_{\mathrm{TM,TM}}^{11} \right. \nonumber\\
&& \left. \times \left[ \frac{ J_1\left(2k_1 x_0 \sin\theta_1^{C} \right) }{ 2k_1 x_0 \sin\theta_1^{C} }  - J_2\left(2k_1 x_0 \sin\theta_1^{C} \right) \right] \right\} , \label{TREUH LC 1}
\end{eqnarray}
where $J_1$ and $J_2$ are Bessel functions. If we consider that the topological parameter $\Delta_\Theta$ is smaller than 1, we may approximate Eq.~(\ref{TREUH LC 1}) at order $\Delta_\Theta^2$ as follows 
\begin{eqnarray}
&&\frac{ d \mathcal{E}_1 }{ d\omega } = \frac{ q^2 \omega  \mu_1 L }{ 8\pi \varepsilon_0 c^2 } \left( 1 - \frac{ 1 }{ \beta^2n_1^2 } \right) \nonumber\\
&& \times \left\{ 1+ \frac{ \Delta_\Theta^2 }{ 2 ( \Delta_\Theta^2 + 2\varepsilon\mu^3 ) } - \frac{ 2\Delta_\Theta^2 }{ \Delta_\Theta^2 + 4\varepsilon\mu^3 } \right. \nonumber\\
&& \left. \times  \left[ \frac{ J_1\left(2k_1 x_0 \sin\theta_1^{C} \right) }{ 2k_1 x_0 \sin\theta_1^{C} } - J_2\left(2k_1 x_0 \sin\theta_1^{C} \right) \right] \right\} , \quad \label{TREUH LC 2}
\end{eqnarray}
where we observe how the interference is encoded in the term given by the Bessel functions and is purely modulated by the topological parameter although the permittivities and permeabilities of both media are the same.  Let us complete the analysis of this radiated energy by studying its limit for distances very close to the interface $x_0\rightarrow0$,   
\begin{equation}\label{TREUH LC 3}
\frac{ d \mathcal{E}_1 }{ d\omega } = \frac{ q^2 \omega  \mu_1 L }{ 8\pi \varepsilon_0 c^2 } \left( 1 - \frac{ 1 }{ \beta^2n_1^2 } \right) = \frac{ d \mathcal{E}_1^{(1)} }{ d\omega} \;,
\end{equation}
where terms of order higher than $\Delta_\Theta^2$ were consistently discarded and we observe that this result coincides with Eq.~(\ref{TREUH 1}).
A quick comparison between Eqs.~(\ref{TREUH LC 2}) and (\ref{TREUH LC 3}) shows that only non-negligible distances to the interface $x_0$ lead to new effects.
%
%

On the other hand, after solving the integrals (\ref{TRELH 2}) and (\ref{TRELH 3}), we write the radiated energy at the lower hemisphere (\ref{TRELH}) in the following fashion:
\begin{eqnarray}
\frac{ d \mathcal{E}_2 }{ d\omega } &=&  \frac{ q^2 \omega  \mu_1 L }{ 8\pi^2 \varepsilon_0 c^2 } \left( 1 - \frac{ 1 }{ \beta^2n_1^2 } \right) \nonumber\\ 
&& \times \left[ \frac{1}{2} \left( T_{\mathrm{TM,TM}}^{11} \right)^2 + \left( T_{\mathrm{TE,TM}}^{11} \right)^2 \right] \;, \nonumber\\
&\simeq&\frac{1}{2}  \frac{ q^2 \omega  \mu_1 L }{ 8\pi^2 \varepsilon_0 c^2 } \left( 1 - \frac{ 1 }{ \beta^2n_1^2 } \right) 
= \frac{1}{2} \frac{ d \mathcal{E}_1 }{ d\omega }\;, \label{TRELH LC 1}
\end{eqnarray}
where terms of order $\Delta_\Theta^4$ have been neglected in the second step.
We appreciate that Eq.~(\ref{TRELH LC 1}) is almost the remaining half of the well-known result of the standard VC radiation energy \cite{Panofsky}, which is reasonable due to our integration over the other half of the whole space. Nonetheless, there is a missing half of the energy at the lower hemisphere, because in the limit $\Delta_\Theta=0$, the sum of Eqs.~(\ref{TREUH LC 1}) and (\ref{TRELH LC 1}) must reproduce the standard VC radiated energy \cite{Panofsky}. 
Despite the absence of cylindrical and conical surface waves due to constant Fresnel transmission coefficients, the integrals of the electric field (\ref{integral I7 napp}), (\ref{integral I8 napp}), (\ref{integral I10 napp}) and (\ref{integral I14 napp}) exhibit a pole when rewritten using the Sommerfeld identity (\ref{Sommerfeld}), owing to the presence of $k_{x,1}$ in the denominator. This pole is associated with axially symmetric surface waves that carry the missing energy, similar to the surface waves emitted by an electric point-like dipole near a pure $\Delta_\Theta$ interface \cite{OJF-LFU}. Another way to understand this is by recalling the behavior of dipolar radiation in front of a dielectric--dielectric interface. As shown in Ref.~\cite{Novotny}, the total energy dissipation rate in such interface is 
\begin{equation}
P=P^\uparrow + P^\downarrow_a + P^\downarrow_f + P_m +P_i \;,
\end{equation}
where $P^\uparrow$ is the power radiated into the upper half-space, $P^\downarrow_a$ and $P^\downarrow_f$ stand for the allowed and forbidden zones in the lower half-space, respectively. $P_m$ accounts for power dissipated by surface modes, waveguides, or thermal losses. In our case, the analogs of $P^\uparrow, P^\downarrow_a$ and $P^\downarrow_f$ are obtained by integrating the distributions (\ref{TREUH}) and (\ref{TRELH}) over the corresponding angles. Assuming no intrinsic losses $P_i=0$, $P_m$can be computed as
\begin{equation}
P_m = P - \left(P^\uparrow + P^\downarrow_a + P^\downarrow_f\right)\;.
\end{equation}
For a lossless layered medium with dielectric--dielectric interface and without surface modes, $P_m= 0$. But for a pure $\Delta_\Theta$ interface, the Fresnel coefficients yield a non-zero $P_m$ proportional to $\Delta_\Theta$, indicating the presence of surface modes. This aligns with the findings of Ref.~\cite{OJF-LFU}, supporting the existence of surface waves carrying the missing energy. We conclude this section remarking that the radiated energies (\ref{TREUH LC 1}) and (\ref{TRELH LC 1}) in both hemispheres are independent of $\Delta_\Theta$. This shows that for a traveling particle close to the interface, the generated radiation is pure VC radiation unaffected by the presence of pure $\Delta_\Theta$ interface. This resembles the situation found in Ref.~\cite{OJF-LFU-ORT}.

\section{Retarding force}\label{Friction Force}

\begin{figure*}[ht]
\centering
\subfloat[]{
\label{g:17}
\includegraphics[width=0.485\textwidth]{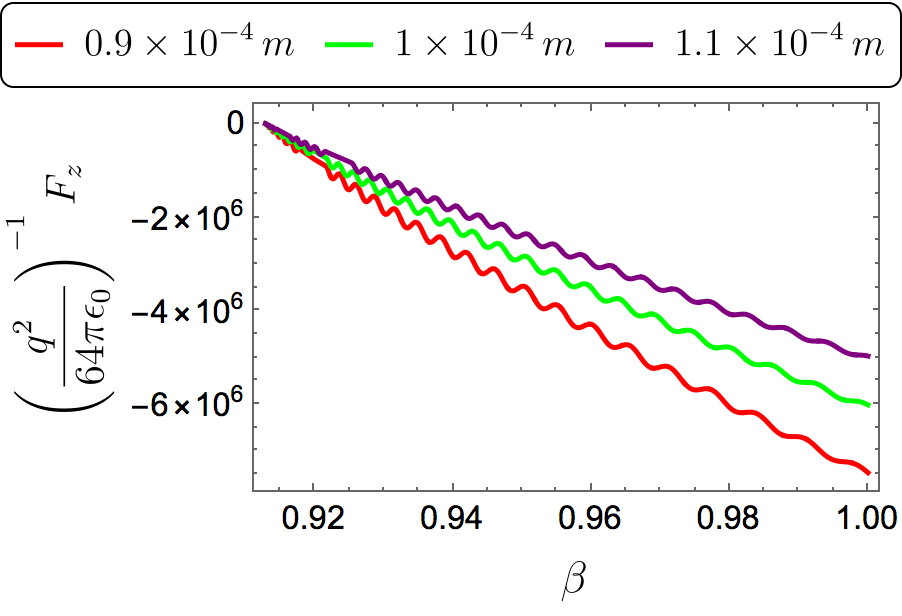}}
\subfloat[]{
\label{g:18}
\includegraphics[width=0.485\textwidth]{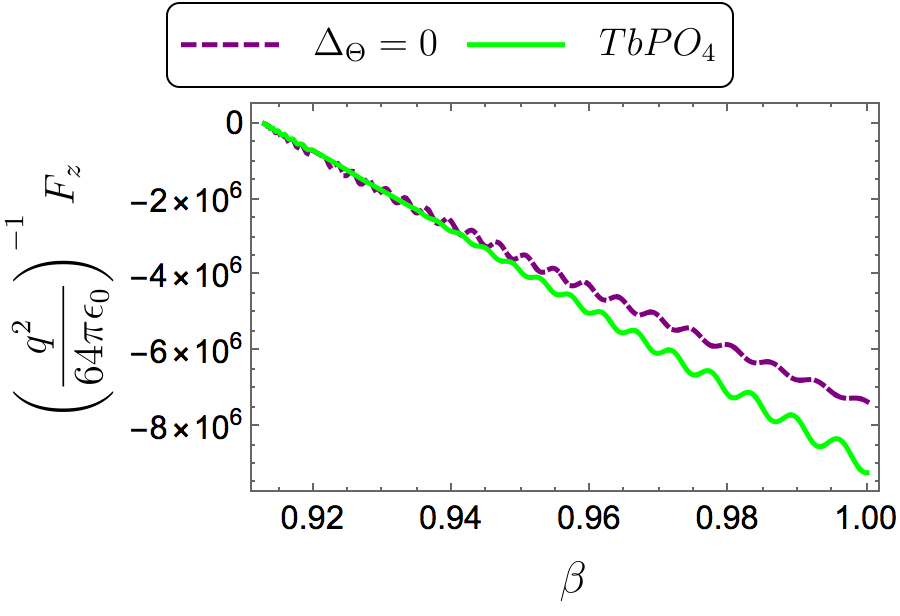}}

\caption{ Plots of the retarding force $F_z$ as function of $\beta$ due to parallel VC radiation in the reflection zone experienced by a particle with single frequency $\omega= 6.0450\times10^{14}\,\, \mathrm{s}^{-1}$. (a) Retarding force for a dielectric--dielectric interface ($\Delta_\Theta=0$) having $\varepsilon_1=1.2$ at the upper layer and the topological insulator TlBiSe$_2$ with $\varepsilon_2=4$ and $\Delta_\Theta=\alpha$ at the lower layer for different values of the height $x_0$ to the interface. (b) Comparison between two scenarios for the retarding force at the same height $x_0=1\times10^{-4}\,\mathrm{m}$. The purple line corresponds to a dielectric--dielectric interface ($\Delta_\Theta=0$) whose permittivities are $\varepsilon_1=1.2$ and $\varepsilon_2=3.4969$. The green solid line represents the same quantity but now with the dielectric $\varepsilon_1=1.2$ at the upper layer and the magnetoelectric TbPO$_4$ with $\varepsilon_2=3.4969$ and $\Delta_\Theta=0.22$ located at the lower layer. The scale is in arbitrary dimensions and all materials are non magnetic. }
\label{PLOTS 3}
\end{figure*}

Another quantity that is of interest to analyze is the 
retarding force experienced by the charged particle when it moves parallel to the interface between a dielectric and a nonmagnetic strong 3D TI. The retarding force, which is calculated through the energy loss per unit length \cite{Bolotovskii,Schwinger} reads
\begin{equation}\label{Force}
F_z = - \frac{ d \mathcal{E}_1 }{ dL }= \sum_{j=1}^{5} F_z^{(j)} = - \sum_{j=1}^{5} \frac{ d \mathcal{E}_1^{(j)} }{ dL } \;.
\end{equation}

Due to practical reasons, we will focus on the reflection part of the force which is in principle accessible to measurement. To do so, we first integrate Eq.~(\ref{Angular distribution UH}) or equivalently Eqs.~(\ref{ADUH 1})--(\ref{ADUH 5}) over $\theta$ in this context. Then we differentiate it with respect to the total length $L$ and solve the integral over $\omega\in[0,\omega_{\mathrm{cut}}]$, where $\omega_{\mathrm{cut}}$ is given by Eq.~(\ref{omega cut}). To carry out the frequency integral, we estimate a cutoff frequency by allowing wavelengths for which the exponential factor appearing in Eq.~(\ref{Phi}) can be set equal to unity, resulting in $\omega_{\mathrm{cut}}\approx c/2x_0\sqrt{\varepsilon_2 - \varepsilon_1}$ \cite{Bolotovskii}. The last step is to integrate over $\phi_1$ obtaining finally $d \mathcal{E}_1/dL$. So, after applying this procedure the retarding force is 
\begin{equation}\label{Force1}
F_z^{(1)} = - \frac{ q^2 }{ 64\pi\varepsilon_0 x_0^2 (\varepsilon_2 - \varepsilon_1) } \left( 1 - \frac{ 1 }{ \beta^2n_1^2 } \right) \; ,
\end{equation}
\begin{eqnarray}
F_z^{(2)} &=& + \frac{ q^2 }{ 32\pi^2\varepsilon_0 x_0^2 (\varepsilon_2 - \varepsilon_1) } \left( 1 - \frac{ 1 }{ \beta^2n_1^2 } \right) \nonumber\\
&& \times \int_{-\pi/2}^{\pi/2} d\phi_1 \cos\left( 2k_1 x_0 \sin\theta_1^{C} \cos\phi_1 \right) \nonumber\\
&& \times \mathrm{Re}\left[ R_{\mathrm{TM,TM}}^{12}(\theta_1^{C},\phi_1,\Delta_\Theta) \right] \cos^2\phi_1 \; , \label{Force2}
\end{eqnarray}
\begin{equation}\label{Force3} 
F_z^{(3)} = 0 \;,
\end{equation}
\begin{eqnarray}
F_z^{(4)} &=& - \frac{ q^2 }{ 64\pi^2\varepsilon_0 x_0^2 (\varepsilon_2 - \varepsilon_1) } \left( 1 - \frac{ 1 }{ \beta^2n_1^2 } \right) \nonumber\\
&&  \times \int_{-\pi/2}^{\pi/2} d\phi_1 \cos^2\phi_1 \big| R_{\mathrm{TM,TM}}^{12}(\theta_1^{C},\phi_1,\Delta_\Theta) \big|^2 \; , \nonumber\\ \label{Force4}
\end{eqnarray}
\begin{eqnarray}
F_z^{(5)} &=& - \frac{ q^2 }{ 64\pi^2\varepsilon_0 x_0^2 (\varepsilon_2 - \varepsilon_1) } \left( 1 - \frac{ 1 }{ \beta^2n_1^2 } \right) \nonumber\\
&&  \times \int_{-\pi/2}^{\pi/2} d\phi_1 \big| R_{\mathrm{TE,TM}}^{12}(\theta_1^{C},\phi_1,\Delta_\Theta) \big|^2 \; , \label{Force5}
\end{eqnarray}
where again the interval of integration over $\phi_1$ was changed from $I_{\mathrm{UH}}$ to $[-\pi/2,\pi/2]$ as in Sec.~\ref{ENERGYUH} and Appendix \ref{D}. Although the last two integrals can be solved analytically, that of Eq.~(\ref{Force3}) involving the interference term cannot be analytically solved. For this reason, we opted to solve numerically the angular integrals and provide plots of the retarding force as shown in Fig.~\ref{PLOTS 3}.

From Eqs.~(\ref{Force1})--(\ref{Force4}), we observe that the retarding force has a Coulombian behavior upon the interference effects due to Eq.~(\ref{Force3}): it is an even function in $\Delta_\Theta$ and the characteristic factor $1-1/\beta^2n_1^2$ of the VC radiation appears as a global factor. 

Figure \ref{PLOTS 3} displays the retarding force $F_z$ as function of $\beta$ and shows the repulsive character of the force meaning that it is opposed to the movement of the charge as Bolotovskii found for the dielectric--dielectric interface \cite{Bolotovskii}. In Fig.~\ref{g:17} we plot the retarding force at constant $\Delta_\Theta$ and different heights $x_0$ to a dielectric--TI interface whose dielectric has $\varepsilon_1=1.2$ and $\Delta_\Theta=0$ and TlBiSe$_2$ as its strong 3D TI with $\varepsilon_2=4$ and $\Delta_\Theta=\alpha$ \cite{TlBiSe2}. This plot shows an increase of the force magnitude for small heights $x_0$. Meanwhile, in Fig.~\ref{g:18} we compare the retarding force at constant height $x_0=1\times10^{-4}\,\mathrm{m}$ from a dielectric--dielectric interface and a dielectric--TI one. The dielectric--dielectric case (dashed purple line) has a dielectric with $\varepsilon_1=1.2$ and $\Delta_\Theta=0$ in its upper layer and a dielectric with $\varepsilon_2=3.4969$ in the lower one. Aiming to illustrate the differences of an increasing topological parameter, the dielectric--TI case (green line) has the dielectric $\varepsilon_1=1.2$ at the upper layer and the magnetoelectric TbPO$_4$ with $\varepsilon_2=3.4969$ and $\Delta_\Theta=0.22$ at the lower one \cite{TbPO4,Rivera}. Here we clearly appreciate an enhancement in the retarding force in the ultra-relativistic regime due to the magnetoelectric TbPO$_4$ resembling the behavior found for transition radiation in the same regime \cite{OJF-SYB}.


\section{Conclusions} \label{CONCLUSIONS}

%

%

%
We have analyzed the parallel Vavilov-Cherenkov radiation emitted by a charged particle propagating with constant velocity $v$ along the $z$ axis parallel to the interface between two generic magnetoelectric media with special emphasis on strong three-dimensional TIs through the Green's function method. By means of the far-field approximation together with the steepest descent method, we were able to obtain analytical expressions for the electromagnetic field, which is due to a superposition of spherical waves and lateral waves with contributions of both kind associated to a purely topological origin. 

In the radiation zone, we study the electric field and the angular distribution of the radiation, whose behaviors for different kinds of interface were illustrated in Figs.~\ref{Electric Field Patterns 1}--\ref{PLOTS 2} respectively. The analysis of these quantities show that the main characteristics of the parallel Vavilov-Cherenkov radiation modified by strong three-dimensional TIs are the following. (i) The Vavilov-Cherenkov angle remains the same and only varies depending on which side of interface the observer is located as described first by Ref.~\cite{Bolotovskii} for a dielectric--dielectric interface. (ii) Although this radiation is also confined to the Vavilov-Cherenkov cone, it is emitted in an inhomogeneous and asymmetric way. Remarkably, the asymmetry of this radiation comes from a term proportional to the topological parameter $\Delta_\Theta$, which is sensitive to the sign of it and could serve to measure it indirectly. (iii) As in standard electrodynamics the height $x_0$ of the source to the interface results crucial to enhance (suppress) the interference effects in the reflected region and the transmitted radiation. 
 
Our expressions at order $\Delta_\Theta^2$ for the radiated energy of this radiation at both sides of the interface show an algebraic dependence on the velocity but are not sensitive to changes in the sign of the topological parameter $\Delta_\Theta$ due to the angular integration. However, the topological parameter contributes with extra terms that are absent in a dielectric--dielectric interface. 

General expressions for the reaction force experienced by the charge were also derived and after numerical integration, we found that it is repulsive and has a Coulomb-type behavior proportional to the squared inverse distance of the charge to the interface as Fig.~\ref{g:17} shows. Despite the lack of sensitiveness to changes in the sign of the topological parameter $\Delta_\Theta$ by the retarding force, it receives a significant increase in the ultra-relativistic regime due to the additional terms arising from the topological parameter as can be appreciated in Fig.~\ref{g:18}. It is hoped that this classical treatment of the retarding force provides insights and motivates further studies on its quantum analog, the quantum friction \cite{QF-Scheel-Buhmann,Pieplow-Henkel}. 

The perpendicular and parallel Vavilov-Cherenkov radiation differ significantly. In the perpendicular configuration of Ref.~\cite{OJF-LFU-ORT}, radiation occurs in both forward and reverse directions, with energy evenly distributed along the Vavilov-Cherenkov cone, showing no sensitivity to the sign of $\Delta_\theta$ and no surface waves. In contrast, the parallel configuration produces only forward radiation, distributed inhomogeneously along the cone and it is asymmetric based on the sign of $\Delta_\theta$, and becomes sharper in the lower hemisphere as shown in Fig.~\ref{PLOTS 2} and discussed in Sec.~\ref{ANGULAR LH}. Surface waves are present in this case, even for a pure $\Delta_\Theta$ interface ($\varepsilon_1=\varepsilon_2$ and $\Delta_\Theta\neq0$), as discussed in Sec.~\ref{ENERGY Pure Delta}. Unlike the perpendicular case, surface waves play a crucial role in the parallel configuration, warranting further study as in Ref.~\cite{Hu-Lin-Wong} for isotropic and uniaxial birefringent media.

 
Finally, we address the question of measuring these effects. In practice, Cherenkov detectors actually measure the number of photons radiated per unit length and per unit frequency, which can be obtained from Eqs.~(\ref{TREUH}) and (\ref{TRELH}). To probe this quantity, it is preferred an experimental setup with an interface having $\varepsilon_1\neq\varepsilon_2$ otherwise it will result difficult to distinguish a new effect as our Eqs.~(\ref{TREUH LC 1}) and (\ref{TREUH LC 2}) for the pure $\Delta_\Theta$ case show.

We note that these findings provide an indirect measurement of the topological parameter $\Delta_\Theta$, aligning with propositions in existing literature \cite{Wu, Dziom, Okada}. Notably, similar effects can manifest in axion insulators \cite{AXIs}, sharing structural characteristics with topological insulators such as gapped bulk and surface states, albeit with topological properties protected by inversion symmetry rather than time-reversal symmetry. It is hoped that the closed form expressions for the reflected parallel Vavilov-Cherenkov radiation will be beneficial for developing waveguides in magnetoelectric media and in tailored graphene-TI heterostructures enhancing the topological effects \cite{Heterostructures1,Heterostructures2,Heterostructures3,Heterostructures4,Heterostructures5}. Moreover, this study holds significance for ongoing investigations into dark matter detection, where axions emerge as a promising candidate \cite{Sikivie}. In this pursuit, condensed matter physics offers novel avenues through TIs \cite{Nenno et al}, antiferromagnetically doped TIs \cite{Marsh et al, Jan Schuette}, or multiferroics \cite{Roising et al}, leveraging the magnetoelectric effect which underpins our results.

\section{Acknowledgments}
O. J. F. has been supported by the postdoctoral fellowship CONACYT-800966 and by the German Research Foundation (DFG, Project No. 328961117-SFB 1319 ELCH).

\appendix

\section{Green's Function components relevant for the electric field} \label{A}
In this appendix, we provide the necessary details of the Green's tensor $\mathbb{G}(\mathbf{r},\mathbf{r}';\omega)$ required to obtain the electric field via Eq.~(\ref{E GF j}) for this configuration. First, the free space part $\mathbb{G}^{(0)}(\mathbf{r},\mathbf{r}';\omega)$ of the Green's tensor, proportional to $\mathds{1}$, is given by 
\begin{eqnarray}
\mathbb{G}^{(0)}(\mathbf{r},\mathbf{r}';\omega) &=& \mu(\omega) \frac{ e^{ \mathrm{i}kR } }{ 4\pi R } \left[ \left( 1 + \frac{ \mathrm{i}kR - 1 }{ k^2 R^2 } \right) \mathds{1} \right. \nonumber\\
&& \left. + \frac{ 3 - 3\mathrm{i}kR - k^2R^2 }{ k^2 R^2 } \frac{ \mathbf{R} \otimes \mathbf{R} }{ R^2 } \right] \;, \label{GF free}
\end{eqnarray}
where $\mathbf{R} = \mathbf{r} - \mathbf{r}'$.

Then the reflective part of the Green's tensor $\mathbb{G}^{(1)}_{\mathbf{r},\mathbf{r}'>0}$ of Eq.~(\ref{GF split}) can be rewritten as 
\begin{equation} \label{GF reflex}
\mathbb{G}^{(1)\; ij}(\mathbf{r},\mathbf{r}';\omega)=\int \frac{ d^2\mathbf{k}_\parallel }{ (2\pi)^2 } e^{ \mathrm{i}\mathbf{k}_\parallel\cdot\mathbf{R}_\parallel } R^{ ij } \left( x,x';\mathbf{k}_\parallel,\omega \right) \;,
\end{equation}
with $\mathbf{R}_\parallel=(y-y',z-z')$, $\mathbf{k}_\parallel=(k_y,k_z)$ is the transversal wavevector (parallel to the interface), and whose only required components in this work are 
\begin{eqnarray}
&& R^{\, xz } \left( x,x';\mathbf{k}_\parallel,\omega \right) = \frac{ \mathrm{i} \mu(\omega) }{ 2k_{x} }e^{ \mathrm{i}k_{x}(|x|+|x'|) } \nonumber\\
&& \times \left[  \mathrm{sgn}(x') \frac{ k_z k_{x} }{ k^2 } R_{\mathrm{TM,TM}} + \frac{ k_y }{ k } R_{\mathrm{TM,TE}} \right] \;,
\end{eqnarray}
\begin{eqnarray}
&& R^{\, yz } \left( x,x';\mathbf{k}_\parallel,\omega \right) = \frac{ \mathrm{i} \mu(\omega) }{ 2k_{x} }e^{ \mathrm{i}k_{x}(|x|+|x'|) } \nonumber\\
&&  \times \left[ - \frac{ k_z k_y }{ k_\parallel^2 }R_{\mathrm{TE,TE}} - \frac{ k_z k_y k_{x}^2 }{ k_\parallel^2 k^2 } R_{\mathrm{TM,TM}} \right.  \nonumber\\
&& \left.  - \mathrm{sgn}(x') \frac{ k_{x} }{ k^2_\parallel k } \left( k_z^2 R_{\mathrm{TE,TM}} + k_y^2 R_{\mathrm{TM,TE}} \right) \right] \;,
\end{eqnarray}
\begin{eqnarray}
&& R^{\, zz } \left( x,x';\mathbf{k}_\parallel,\omega \right) \nonumber\\
&& =  \frac{ \mathrm{i} \mu(\omega) }{ 2k_{x} }e^{ \mathrm{i}k_{x}(|x|+|x'|) }  \left[ \frac{ k_y^2 }{ k_\parallel^2 }R_{\mathrm{TE,TE}} - \frac{ k_z^2 k_{x}^2 }{ k_\parallel^2 k^2 } R_{\mathrm{TM,TM}} \right. \nonumber\\
&& \left.  + \mathrm{sgn}(x') \frac{ k_{x} }{ k^2_\parallel k } \left( k_z k_y R_{\mathrm{TE,TM}} - k_z k_y R_{\mathrm{TM,TE}} \right) \right] , 
\end{eqnarray}
where $\mu(\omega)$ is the permeability of the upper medium, $\mathrm{sgn}$ stands for the sign function and we recall that $k_{x}=\sqrt{k^2-\mathbf{k}_\parallel^2}$. \\

Similarly, the transmissive part of the Green's function $\mathbb{G}^{(1)}_{\mathbf{r},\mathbf{r}'<0}$ of Eq.~(\ref{GF split}) can be rewritten as 
\begin{equation}\label{GF trans}
\mathbb{G}^{(1)\; ij}(\mathbf{r},\mathbf{r}';\omega)=\int \frac{ d^2\mathbf{k}_\parallel }{ (2\pi)^2 } e^{ \mathrm{i}\mathbf{k}_\parallel\cdot\mathbf{R}_\parallel } T^{ ij } \left( x,x';\mathbf{k}_\parallel,\omega \right) \;,
\end{equation}
whose only required components in the present work are 
\begin{eqnarray}
&& T^{\, xz } \left( x,x';\mathbf{k}_\parallel,\omega \right) = \frac{ \mathrm{i} \mu'(\omega) }{ 2k_{x'} }e^{ \mathrm{i}k_{x}|x|+ik_{x'}|x'| } \nonumber\\
&& \times \left[  \mathrm{sgn}(x') \frac{ k_z k_{x'} }{ k k' } T_{\mathrm{TM,TM}} + \frac{ k_y }{ k } T_{\mathrm{TM,TE}} \right] \;,
\end{eqnarray}
\begin{eqnarray}
&& T^{\, yz } \left( x,x';\mathbf{k}_\parallel,\omega \right) = \frac{ \mathrm{i} \mu'(\omega) }{ 2k_{x'} }e^{ \mathrm{i}k_{x}|x|+\mathrm{i}k_{x'}|x'| }  \nonumber\\
&& \times \left[ - \frac{ k_z k_y }{ k_\parallel^2 }T_{\mathrm{TE,TE}} + \frac{ k_z k_y k_{x}k_{x'} }{ k_\parallel^2 k k' } T_{\mathrm{TM,TM}} \right.  \nonumber\\
&& \left.  -  \frac{ \mathrm{sgn}(x') }{ k^2_\parallel } \left( k_z^2 \frac{ k_{x'} }{ k' } T_{\mathrm{TE,TM}} - k_y^2 \frac{ k_{x} }{ k } T_{\mathrm{TM,TE}} \right) \right] \;, \nonumber\\
\end{eqnarray}
\begin{eqnarray}
&& T^{\, zz } \left( x,x';\mathbf{k}_\parallel,\omega \right) \nonumber\\
&& =  \frac{ \mathrm{i} \mu'(\omega) }{ 2k_{x'} }e^{ \mathrm{i}k_{x}|x|+\mathrm{i}k_{x'}|x'| }  \left[ \frac{ k_y^2 }{ k_\parallel^2 }T_{\mathrm{TE,TE}} + \frac{ k_z^2 k_{x}k_{x'} }{ k_\parallel^2 kk^{'} } T_{\mathrm{TM,TM}} \right. \nonumber\\
&& \left.  + \mathrm{sgn}(x') \frac{ k_z k_y }{ k^2_\parallel } \left(  \frac{ k_{x'} }{ k' } T_{\mathrm{TE,TM}} + \frac{ k_{x} }{ k } T_{\mathrm{TM,TE}} \right) \right] , 
\end{eqnarray}
where $k_{x'}=\sqrt{k^{' \, 2} - \mathbf{k}_\parallel^2}$, and $k'$ and $\mu^{ \, \prime }(\omega)$ denote the wave number and the permeability of the source medium. 

Once the labels for the upper and lower media are designed, through Eqs.~(\ref{GF free}), (\ref{GF reflex}), and (\ref{GF trans}) we obtain the four possible cases for a two-layer configuration:  observer and source at the upper medium, observer and source at the lower medium, observer at the upper medium and source at the lower one, and vice versa.

\section{Approximating the required integrals for the electric field} \label{B}
In this appendix, we utilize the steepest descent method to derive the far-field approximation of the $k_\parallel$ integrals pertaining to both reflective (\ref{Ex1 int})--(\ref{Ez1 int}) and transmissive electric fields (\ref{Ex2 int})--(\ref{Ez2 int}). For conciseness, we will solely illustrate the calculations for the $x$ component, as expressed in Eq.~(\ref{Ex1 int}). Similar procedures are employed to obtain the remaining relevant components. Our approach closely adheres to Ref.~\cite{Chew}, which furnishes a meticulous explanation of the general procedure for handling the steepest descent method concerning this class of integrals. 

The three $k_\parallel$ integrals that need to be determined in $E^{\,x}_1(\mathbf{r};\omega)$ are 
\begin{equation}
\mathcal{I}_0 = \int_{-\zeta}^{\zeta}dz' \frac{e^{\mathrm{i}k_1R}}{4\pi R} \left[ - \frac{ (x-x_0)(z-z') }{ R^2 }\right] e^{\mathrm{i}\frac{\omega z'}{v}} \;,
\end{equation}
\begin{equation}
\mathcal{I}_1 =  \int_0^\infty dk_\parallel k_\parallel R_{\mathrm{TM,TM}}^{12}(k_\parallel) J_0(k_\parallel R_\parallel) e^{\mathrm{i}k_{x,1}(x+x_0)} \;,
\end{equation}
\begin{equation}\label{integral I2 App}
\mathcal{I}_2 = \int_0^\infty \frac{ dk_\parallel k_\parallel }{ k_{x,1} }  R_{\mathrm{TM,TE}}^{12}(k_\parallel) J_0(k_\parallel R_\parallel) e^{\mathrm{i}k_{x,1}(x+x_0)} \;,
\end{equation}
where $\mathcal{I}_1$ and $\mathcal{I}_2$ were defined in Eqs.~(\ref{integral I1 napp}) and (\ref{integral I2 napp}).

Let us begin with $\mathcal{I}_0$. Recalling that $R_\parallel=\sqrt{ y^2 + (z-z')^2 } $, $R^2= (x-x_0)^2 + R_\parallel^2 $ and by considering the coordinate conditions 
\begin{equation}\label{FF coordinates}
x\pm x_0\rightarrow\infty \; , \;\; y\rightarrow\infty \; , \;\; z'\ll z\rightarrow\infty \; , 
\end{equation}
we write $R$ in the far-field approximation as $R\simeq R_0-zz'/R_0$ with $R_0=\sqrt{ (x-x_0)^2 + \rho^2 }$ and $\rho=\sqrt{ y^2 + z^2 }$. Therefore, we find 
\begin{equation}
\mathcal{I}_0 = - \frac{e^{ \mathrm{i}k_1R_0  }}{4\pi R_0}  \frac{ (x-x_0)z }{ R_0^2 } \mathcal{K}_1(z/R_0,\omega)  \;, 
\end{equation}
where the function (\ref{Function K1}) was used. This expression coincides with the $x$ component of the free-space contribution of the field (\ref{E reflex}).

Now, our focus shifts to computing the integral $\mathcal{I}_1$. First, we go to the complex plane by expressing the Bessel function $J_0$ in terms of the Hankel function $H_0^{(1)}(k_\parallel R_\parallel)$ and using the reflection formula \cite{Arfken}. This enables us to extend the integration interval to $-\infty$. The result is as follows: 
\begin{equation}
\mathcal{I}_1 = \frac{1}{2} \oint_{C_{S}} dk_\parallel k_\parallel R_{\mathrm{TM,TM}}^{12}(k_\parallel) H_0^{(1)}(k_\parallel R_\parallel) e^{\mathrm{i}k_{x,1}(|x|+x_0)} ,
\end{equation}
where $C_{S}$ is the Sommerfeld integration path depicted in Fig.~2.6.4 of Ref.~\cite{Chew}. Next, we use the far-field conditions (\ref{FF coordinates}) and implement the asymptotic expansion of the Hankel function \cite{Abramowitz}, yielding 
\begin{eqnarray}\label{J1 integral FF}
\mathcal{I}_1 &\sim& e^{ -\mathrm{i}\pi/4 } \sqrt{ \frac{ 1 }{ 2\pi R_\parallel } } \oint_{C_{S}} dk_\parallel \sqrt{k_\parallel} R_{\mathrm{TM,TM}}^{12}(k_\parallel)  \nonumber\\
&& \times e^{ \mathrm{i}k_\parallel R_\parallel } e^{\mathrm{i}k_{x,1}(x+x_0)} \;.
\end{eqnarray}
Then, by choosing the stationary phase as $\mathrm{i}k_\parallel R_\parallel+\mathrm{i}k_{x,1}(x+x_0)$ the saddle-point of $\mathcal{I}_1$ is 
\begin{equation}
k_{\parallel s} =  \frac{ k_1 R_\parallel }{ \tilde{R}_1 } \Rightarrow k_{x,1\,s}= k_1^2 - k_{\parallel s}^2 \;. 
\end{equation}
where $\tilde{R}_1=\sqrt{ (x+x_0)^2 + R_\parallel^2 }$. In this way, around the saddle-point we can write 
\begin{eqnarray}
\mathcal{I}_1 & \sim & \frac{1}{2} R_{\mathrm{TM,TM}}^{12}(k_{\parallel s}) \oint_{C_{S}} dk_\parallel k_\parallel H_0^{(1)}(k_\parallel R_\parallel) e^{\mathrm{i}k_{x,1}(x+x_0)} \;, \nonumber \\
&=& - R_{\mathrm{TM,TM}}^{12}(k_{\parallel s}) \frac{ \partial }{ \partial(x+x_0) }  \left( \frac{ e^{ \mathrm{i}k_1 \tilde{R}_1 } }{ \tilde{R}_1 } \right) \;.
\end{eqnarray}
The latter equality is a consequence of the Sommerfeld identity \cite{Chew}
\begin{equation}\label{Sommerfeld}
\frac{ e^{ \mathrm{i}k_1 \tilde{R}_1 } }{ \tilde{R}_1 } = \frac{ \mathrm{i} }{2} \oint_{C_{S}}  \frac{ dk_\parallel k_\parallel }{ k_{x,1} } H_0^{(1)}(k_\parallel R_\parallel) e^{\mathrm{i}k_{x,1}(x+x_0)} \;.
\end{equation}
After carrying out the partial derivative, the final form of $\mathcal{I}_1$ results
\begin{equation}\label{J1 integral result 1}
\mathcal{I}_1 \sim - \mathrm{i} R_{\mathrm{TM,TM}}^{12}(k_1 R_\parallel /\tilde{R}_1 ) \frac{ k_1(x+x_0) e^{ \mathrm{i}k_1 \tilde{R}_1 } }{  \tilde{R}_1^2 }\;,
\end{equation}
where terms of order $\tilde{R}_1^3$ were discarded.

However, $\mathcal{I}_1$ [Eq.~(\ref{J1 integral FF})] may have poles when the denominator of the Fresnel coefficient $R_{\mathrm{TM,TM}}^{12}$ becomes zero. From Eq.~(\ref{Fresnel 5}) we read that the poles are given by the roots of the following polynomial:
\begin{equation}\label{polynomial}
(\varepsilon_{2}k_{x,1}+\varepsilon_{1}k_{x,2})\mu_1 \mu_2 ( k_{x,1}\mu_2 + k_{x,2}\mu_1) + \Delta_\Theta^{2}k_{x,1}k_{x,2}=0 \;. 
\end{equation}
By recalling the definition $k_{x,j}=\sqrt{k^2_j-\mathbf{k}_\parallel^2}$, we note that it is a fourth-degree polynomial in $k_\parallel$ and is common for all involved integrals of the electric field as Eqs.~(\ref{Fresnel 1})--(\ref{Fresnel 8}) show. The analysis of the four roots of (\ref{polynomial}) shows that only the root $\varkappa$ is positive and reproduces correctly the corresponding root of a dielectric--dielectric interface ($\Delta_\Theta=0$). After computing the residue of the integrand at $\varkappa$ via the Proposition 4.1.2 of Ref.~\cite{Marsden-Tromba} and applying the residue theorem to the integral (\ref{J1 integral FF}), we obtain
\begin{eqnarray}\label{J1 integral general result 2}
\mathcal{I}_{1,\mathrm{pole}} &=& \sqrt{ \frac{ 2\pi\mathrm{i} \varkappa }{ R_\parallel } } \mathrm{Res}\left(R_{\mathrm{TM,TM}}^{12};\varkappa\right)  \nonumber\\
&& \times e^{ \mathrm{i}\varkappa R_\parallel + \mathrm{i}\sqrt{k_1^2 - \varkappa^2 }(x+x_0) } \;.
\end{eqnarray}
Although there are explicit expressions for $\varkappa$ and the other three roots of the polynomial (\ref{polynomial}), they are unwieldy to use. For this reason, we restrict ourselves to study an interface without magnetic properties, i.e., $\mu_1=\mu_2=1$. At order $\Delta_\Theta^2$ the four roots are 
\begin{eqnarray}
\varkappa &\equiv& \varkappa_1= -\varkappa_2 = k_1 \sin\tilde{\theta}_{B} \;, \\
\varkappa_3 &=& -\varkappa_4 = \left[ \frac{ \varepsilon_1 - \varepsilon_2 }{ \sqrt{2}\Delta_\Theta } + \frac{ 3\Delta_\Theta (\varepsilon_1 + \varepsilon_2) }{ 4 \sqrt{2}(\varepsilon_1 - \varepsilon_2) } \right] \frac{\omega}{c} \,,\;\; 
\end{eqnarray}
where 
\begin{equation}
\sin\tilde{\theta}_{B} = \sqrt{\frac{\varepsilon_2 }{ \varepsilon_1 + \varepsilon_2 }} \left[ 1 + \frac{ \Delta_\Theta^2 \varepsilon_1 \varepsilon_2 }{ (\varepsilon_1-\varepsilon_2)^2 (\varepsilon_1+ \varepsilon_2) } \right] .
\end{equation}
Only $\varkappa$ is physically admissible for the same reasons mentioned above. The next step is to determine the residue of the integrand at $\varkappa$. By applying the Proposition 4.1.2 of Ref.~\cite{Marsden-Tromba} and the residue theorem, we find
\begin{eqnarray}\label{J1 integral result 2}
\mathcal{I}_{1,\mathrm{pole}} &=& \sqrt{ \frac{ 2\pi\mathrm{i} \omega }{ R_\parallel c } }  e^{ \mathrm{i}\varkappa R_\parallel + \mathrm{i}\sqrt{k_1^2 - \varkappa^2 }(x+x_0) } \frac{ \mathcal{N}(R_{\mathrm{TM,TM}}^{12}) }{  \mathcal{D}'(R_{\mathrm{TM,TM}}^{12}) } \Bigg|_\varkappa \,, \nonumber\\
&\simeq&\sqrt{ \frac{ 2\pi\mathrm{i} \omega }{ R_\parallel c } }  e^{ \mathrm{i}\varkappa R_\parallel + \mathrm{i}\sqrt{k_1^2 - \varkappa^2 }(x+x_0) } \mathcal{F}_1(\varepsilon_1,\varepsilon_2,\Delta_\Theta) \;,  \nonumber\\
\end{eqnarray}
where the prime means derivative with respect to $k_\parallel$, $\mathcal{N}$ and $\mathcal{D}$ stand for the numerator and denominator of $R_{\mathrm{TM,TM}}^{12}$, respectively. The function $\mathcal{F}_1$ is explicitly written in Appendix {C}. In the last line only terms at order $\Delta_\Theta^2$ were considered. After adding the results (\ref{J1 integral result 1}) and (\ref{J1 integral general result 2}) or for a non magnetic interface (\ref{J1 integral result 1}) and (\ref{J1 integral result 2}) the calculation of $\mathcal{I}_1$ is completed.

By adapting this procedure to the remaining integral $\mathcal{I}_2$ given by Eq. (\ref{integral I2 App}), we obtain 
%
%
\begin{equation}\label{J2 integral result 1}
\mathcal{I}_2  \sim  R_{\mathrm{TM,TE}}^{12}(k_1 R_\parallel /\tilde{R}_1) \frac{ e^{ \mathrm{i}k_1 \tilde{R}_1 } }{ \mathrm{i} \tilde{R}_1 } \;, 
%
%
\end{equation}
\begin{equation}\label{J2 integral general result 2}
\mathcal{I}_{2,\mathrm{pole}} = \sqrt{ \frac{ 2\pi\mathrm{i} \varkappa }{ R_\parallel } } e^{ \mathrm{i}\varkappa R_\parallel + \mathrm{i}\sqrt{k_1^2 - \varkappa^2 }(x+x_0) }  \mathrm{Res}\left(R_{\mathrm{TE,TM}}^{12};\varkappa\right) \,, 
\end{equation}
\begin{equation}\label{J2 integral result 2}
\mathcal{I}_{2,\mathrm{pole}} = \sqrt{ \frac{ 2\pi\mathrm{i} \omega }{ R_\parallel c } } e^{ \mathrm{i}\varkappa R_\parallel + \mathrm{i}\sqrt{k_1^2 - \varkappa^2 }(x+x_0) } \frac{ \mathcal{F}_2(\varepsilon_1,\varepsilon_2,\Delta_\Theta) }{ \sqrt{k_1^2 - \varkappa^2 } } \;,
\end{equation}
where $\mathcal{F}_2$ is explicitly written in Appendix {C}. Again, the sum of Eqs.~(\ref{J2 integral result 1}) and (\ref{J2 integral general result 2}) or for a nonmagnetic interface (\ref{J2 integral result 1}) and (\ref{J2 integral result 2}) finishes the calculation of $\mathcal{I}_2$.

Finally, by substituting $\mathcal{I}_1$ and $\mathcal{I}_2$ into Eq.~(\ref{Ex1 int}), utilizing the far-field approximation (\ref{FF coordinates}) by writing  $\tilde{R}_1\simeq R_1-zz'/R_1$ with $R_1=\sqrt{(x+x_0)^2 + \rho^2}$ and carrying out the integral over $z'$, one arrives at the $x$ component of Eq.~(\ref{E reflex}). For the case of a non magnetic interface, the $x$ component of the surface-wave contribution is given by Eq.~(\ref{E reflex sw}). 

To derive the transmissive components of the electric field as described in Eq.~(\ref{E trans}), we follow a similar procedure. However, the corresponding stationary phase yields a fourth-degree polynomial, resulting in solutions that are cumbersome and analytically challenging. Given the significant suppression of transmitted radiation, in this study, we focus on distances where $x_0\ll x$, enabling us to derive analytical expressions. 





\section{Functions $\mathcal{F}_j(\varepsilon_1,\varepsilon_2,\Delta_\Theta)$ with $j=1,2$ for the surface waves}\label{C}
In this appendix, we write the explicit form of the functions $\mathcal{F}_j(\varepsilon_1,\varepsilon_2,\Delta_\Theta)$ with $j=1,2$ of the reflected electric field (\ref{E reflex sw}) and the transmitted one (\ref{E trans sw}), when non-magnetic media $\mu_1=\mu_2=1$ are considered. 
\begin{equation}
\mathcal{F}_1(\varepsilon_1,\varepsilon_2,\Delta_\Theta) = \frac{ \Delta_\Theta^2 \alpha_1^{3/2} \alpha_2 }{ \alpha_3 } \;,  \label{curlyF1}
\end{equation}
\begin{equation}\label{curlyF2}
\mathcal{F}_2(\varepsilon_1,\varepsilon_2,\Delta_\Theta) = \frac{ \alpha_1^{3/2} \alpha_4 \Delta }{ \alpha_3 \sqrt{ \varepsilon_1 - \alpha_1^2 } } \;,
\end{equation}
where we introduced 
\begin{eqnarray}
\alpha_1 &=& \sqrt{ \frac{ \varepsilon_1\varepsilon_2 }{ \varepsilon_1 + \varepsilon_2 } } \;, \\ 
\alpha_2 &=& \frac{ 2 \alpha_1^4 }{ \varepsilon_2 (\varepsilon_1 - \varepsilon_2) } \left[ \varepsilon_2 + \alpha_1^2 \left( \varepsilon_1 + 1 \right) \right] \;, \\ 
\alpha_3 &=& - \frac{ (\varepsilon_1 + \varepsilon_2)^2 }{ \alpha_1 } \;, \\
\alpha_4 &=& - 2 \sqrt{ \varepsilon_1 } \alpha_1^2 \;.
\end{eqnarray}

\section{Solving the radiated energy integrals}\label{D}

In this appendix, we solve the required integrals to obtain the radiated energy of Eq.~(\ref{TREUH}). Let us first calculate $d\mathcal{E}_1^1/d\omega$ being the radiated energy of the free term because it is the easiest one to integrate. This will serve to illustrate the procedure based on Refs.~\cite{OJF-LFU-ORT,Panofsky}, which  will be applied repeatedly in this appendix. Integrating the expression (\ref{ADUH 1}) with respect to the corresponding solid angle we obtain
\vspace{0.5cm}
\begin{eqnarray}
\frac{ d \mathcal{E}_1^{(1)} }{ d\omega} &=& \sqrt{ \frac{ \varepsilon_0 \varepsilon_1 }{ \mu_0 \mu_1 } } \frac{ q^2 \omega^2  \mu^2_1 }{ 16\pi^2 \varepsilon_0^2 c^4} \left( 1 - \frac{ 1 }{ \beta^2n_1^2 } \right) \nonumber\\
&&\times \int_{0}^1 d(\cos\theta)  \mathcal{K}_1^2(\theta,\omega) .
\end{eqnarray}
The delta-like behavior of the integrand in the limit $\zeta \gg \omega/v $ shows that the radiation is sharply localized in a main lobe around the VC angle $\theta_1^{C}$ given by Eq.~(\ref{Cherenkov reflex}), producing a half-cone in the forward direction as Figs.~\ref{Electric Field Patterns 1} and \ref{Electric Field Patterns 2} show. Consequently, after employing the change of variable $u=\omega\zeta(1-vn_1\cos\theta)/v$, we are allowed to extend the integration limits of the integral to $\pm \infty$ as long as we include the maximum of the lobe situated at $u=0$. In this way, it results that
\begin{eqnarray}
\frac{ d \mathcal{E}_1^{(1)} }{ d\omega } &=& \sqrt{ \frac{ \varepsilon_0 \varepsilon_1 }{ \mu_0 \mu_1 } } \frac{ q^2 \omega  \mu^2_1 \zeta }{ 4\pi^2 \varepsilon_0^2 c^3 n_1 } \left( 1 - \frac{ 1 }{ \beta^2n_1^2 } \right) \int_{-\infty}^{\infty} du \frac{\sin^2 u }{ u^2 } \;, \nonumber\\
&=& \frac{ q^2 \omega  \mu_1 L }{ 8\pi \varepsilon_0 c^2 } \left( 1 - \frac{ 1 }{ \beta^2n_1^2 } \right) \;, 
\end{eqnarray}
where we introduced the total length $L=2\zeta$ traveled by the particle.  

The second term comes from the angular integration of Eq.~(\ref{ADUH 2}):
\begin{eqnarray}
&&\frac{ d \mathcal{E}_1^{(2)} }{ d\omega } = - 2 \sqrt{ \frac{ \varepsilon_0 \varepsilon_1 }{ \mu_0 \mu_1 } } \frac{ q^2 \omega^2  \mu^2_1 }{ 16\pi^3 \varepsilon_0^2 c^4 } \left( 1 - \frac{ 1 }{ \beta^2n_1^2 } \right) \nonumber\\
&& \times \int d\Omega_1 \mathcal{K}_1^2(\theta,\omega) \mathrm{Re}\left[ R_{\mathrm{TM,TM}}^{12}(\theta_1^{C},\phi_1,\Delta_\Theta) \right] \cos^2\phi_1 \nonumber\\
&& \times \cos\left( 2k_1 x_0 \sin\theta_1^{C} \cos\phi_1 \right) .
\end{eqnarray}
All the $\theta$ dependency lies only on the $\mathcal{K}_1$ function enabling us to repeat the above procedure. So upon  carrying out the integral over the azimuthal angle $\phi_1$, we find 
\begin{eqnarray}
\frac{ d \mathcal{E}_1^{(2)} }{ d\omega } &=& - \frac{2}{\pi} \frac{ d \mathcal{E}_1^{(1)} }{ d\omega } \int_{I_{\mathrm{UH}}} d\phi_1 \cos\left( 2k_1 x_0 \sin\theta_1^{C} \cos\phi_1 \right)  \nonumber\\
&& \times \mathrm{Re}\left[ R_{\mathrm{TM,TM}}^{12}(\theta_1^{C},\phi_1,\Delta_\Theta) \right] \cos^2\phi_1 \;. 
\end{eqnarray}

The calculation of $d\mathcal{E}_1^{(3)}/d\omega$ starts from Eq.~(\ref{ADUH 3}), which after applying the same procedure can be written in the following fashion:
\begin{equation}
\begin{aligned}
\frac{ d \mathcal{E}_1^{(3)} }{ d\omega }& = \frac{ 2 }{ \pi \beta n_1 } \frac{ d \mathcal{E}_1^{(1)} }{ d\omega } \int_{I_{\mathrm{UH}}} d\phi_1  \cos\phi_1\sin\phi_1 \\
& \times \mathrm{Re}\left[ R_{\mathrm{TM,TM}}^{12}(\theta_1^{C},\phi_1,\Delta_\Theta) R_{\mathrm{TE,TM}}^{12\,*}(\theta_1^{C},\phi_1,\Delta_\Theta) \right], \\
&= 0 \;.
\end{aligned}
\end{equation}

This integral vanishes by noting that the integrand is an odd function and after a change of variable one can transform the interval of integration $I_{\mathrm{UH}}$ to $[-\pi/2,\pi/2]$.

Finally, along the same lines and beginning with Eqs.~(\ref{ADUH 4}) and (\ref{ADUH 5}), one finds that the final expressions for $d\mathcal{E}_1^{(4)}/d\omega$ and $d\mathcal{E}_1^{(5)}/d\omega$ are just Eqs.~(\ref{TREUH 4}) and (\ref{TREUH 5}).


\end{document}